\documentclass[twocolumn]{aastex7}

\begin{document}

\title{Effect of Matter Accretion on Lithium Enhancement of Giants}

\author[orcid=0000-0001-9291-4261]{Xue-Feng Li}
\altaffiliation{University of Chinese Academy of Sciences}
\affiliation{Yunnan Observatories, Chinese Academy of Sciences, P.O. Box110, Kunming 650216, China}
\email[show]{lixuefeng@ynao.ac.cn}  

\author[0000-0002-0349-7839]{Jian-Rong Shi}
\affiliation{CAS Key Laboratory of Optical Astronomy, National Astronomical Observatories, Beijing 100101, China}
\affiliation{University of Chinese Academy of Sciences, Beijing 100049, China}
\email{sjr@bao.ac.cn}  

\author[0000-0002-1424-3164]{Yan Li}
\affiliation{Yunnan Observatories, Chinese Academy of Sciences, P.O. Box110, Kunming 650216, China}
\affiliation{University of Chinese Academy of Sciences, Beijing 100049, China}
\affiliation{Key Laboratory for Structure and Evolution of Celestial Objects, Chinese   Academy of Sciences, P.O. Box110, Kunming 650216, China}
\affiliation{Center for Astronomical Mega-Science, Chinese Academy of Sciences, Beijing 100012, China}
\email{ly@ynao.ac.cn}  

\author[0000-0002-8609-3599]{Hong-Liang Yan}
\affiliation{CAS Key Laboratory of Optical Astronomy, National Astronomical Observatories, Beijing 100101, China}
\affiliation{University of Chinese Academy of Sciences, Beijing 100049, China}
\affiliation{Institute for Frontiers in Astronomy and Astrophysics, Beijing Normal University,  Beijing 102206, China}
\email{hlyan@nao.cas.cn}  

\author[0000-0002-2510-6931]{Jing-Hua Zhang}
\affiliation{South-Western Institute for Astronomy Research, Yunnan University, Chenggong District, Kunming 650500, China}
\email{zhang_jh@ynu.edu.cn}  

\author[0000-0002-7072-2522]{Fei Guo}
\affiliation{Yunnan Normal University, Kunming 650500, China}
\email{guofei@ynnu.edu.cn}  


\begin{abstract}

A subset of low-mass giants ($<2.2\,M_{\odot}$) exhibit anomalous lithium enhancement behavior, which is still an open topic.
Given that more massive giants retain more surface lithium, increasing mass by accreting circumstellar matter could be a channel to enrich lithium. We evaluate this process in the current work.
Using MESA, we construct a model of matter accretion, including mass loss, that evolves a star from the main sequence turnoff to the red giant branch tip. The mean accretion rate is estimated from the upper limit of the accreted mass and the evolutionary time of the star during this period, and a grid of accretion rates is constructed. We separately consider their effects on the lithium enhancement of giants, both in terms of the mass and the composition of accretion.
Accreting matter with higher lithium abundances has a promoting effect on the lithium enhancement of giants. The accreted matter with excess lithium alleviates the dilution of lithium in the convective envelope during the first dredge-up. The added mass results in lower temperatures at the bottom of the convective envelope, which likewise weakens the depletion of surface lithium.
Weak accretion of circumstellar matter is a possible route to lithium enhancement for giants, and it predicts an upper limit on the lithium abundance of $\rm \sim 2.5\,dex$. However, the mass increment it requires poses a potential challenge to real astrophysical environments. Such accretion suppresses lithium dilution and depletion of the star during the first dredge-up, thus exhibiting lithium enhancement behavior.

\end{abstract}



\section{Introduction}\label{sect1}

The phenomenon of anomalous lithium (Li) enhancement in low-mass giants ($<2.2\,M_{\odot}$) has been an open topic \citep[e.g,][]{2000A&A...358L..49D, 2003ApJ...593..509D, 2010ApJ...723L.103C, 2024MNRAS.535.1243D, 2024ApJ...964...42S}. 
Li enhancement is quantified relative to standard stellar evolution predictions, where model-derived Li abundances, $A(\rm Li)$\footnote{$A(\rm Li)$=log($N_{\rm Li}$/$N_{\rm H}$)+12, where $N_{\rm Li}$ and $N_{\rm H}$ are the number densities of Li and H, respectively.}, serve as the reference baseline for a given mass and evolutionary stage. The theoretical threshold for low-mass stars is $\rm 1.5\,dex$ \citep[e.g.][]{2000A&A...359..563C}. Stars above this value are regarded as Li-rich. At present, the highest Li abundance can reach $\sim 6.0\,\rm dex$ \citep{2024ApJ...973..125K}. Since \citet{1982ApJ...255..577W}'s pioneering study, theoretical efforts to reproduce Li enhancement in giants have continued, but the problem has grown increasingly complex with advancing observational capabilities.
Growing sample of observations can give us a glimpse of the physical attribution behind them \citep[e.g.,][]{1989ApJS...71..293B, 2000A&A...359..563C, 2011ApJ...730L..12K, 2016MNRAS.461.3336C, 2019MNRAS.484.2000D, 2019ApJS..245...33G, 2019ApJ...877..104Z, 2020MNRAS.494.1348D, 2021A&A...651A..84M, 2021MNRAS.505.5340M, 2021ApJ...919L...3Z, 2022ApJ...931..136Z, 2023AJ....165...52C, 2024ApJS..271...58D}. The enhancement of Li varies across diverse evolutionary stages \citep[e.g.][]{2019MNRAS.484.2000D}. The number of Li-rich stars is very small \citep{2019ApJS..245...33G} and decreases with increasing Li abundance \citep{2023AJ....165...52C}. The formation of Li enhancement giants appears to be associated with rotation velocity \citep[e.g.][]{2021A&A...651A..84M} and the presence of companion stars \citep[e.g.][]{2016MNRAS.461.3336C}. Additionally, a central challenge is that Li enhancement exhibits no clear systematic dependence on fundamental stellar parameters \citep[see Section 1 of][]{2025ApJ...982L...4L}.

An important point is that asteroseismology has enabled a clear distinction between red giant branch (RGB) stars and clump giants (core He burning). Combined with spectroscopic analysis, subsequent observations confirm that the majority of Li-rich giants are in the clump phase. Notably, \citet{2020NatAs...4.1059K} reported a high occurrence of Li enhancement in clump giants (believed that all low-mass clump giants are Li-rich), suggesting it may be common in this evolutionary stage. However, standard stellar evolution models, which neglect Li loss \footnote{Here "Li loss" generally refers to the decrease in Li abundance (involving various factors), while "Li depletion" represents the reduction in Li abundance caused by participation in the thermonuclear reaction.} during pre-main sequence and main sequence (MS) stages, still predict low Li abundances at the RGB and clump stages \citep[e.g.][]{2023ApJ...943..115L,2024MNRAS.529.1423L}. Combined with the observed distribution characteristics \citep[e.g.][]{2020NatAs...4.1059K,2021MNRAS.505.5340M}, this means that Li enhancement\footnote{Li enhancement is different from Li-rich. The scarcity of Li-rich giants is indicated by observations, while Li enhancement can be quantified in a simple method. Here, abundances exceeding standard model predictions are generally considered evidence of Li enhancement behavior—though not necessarily meeting Li-rich criteria.} is the normalcy for giants.

Currently, Li-rich giants (defined as $A(\rm Li)>1.5\,dex$) are observationally rare, occurring in only $\sim 1-2\%$ of low-mass giants \citep[e.g.][]{2016MNRAS.461.3336C}. However, the high incidence of Li enhancement behavior suggests the existence of extra physical processes in these stars. To accumulate Li on the surface of a star, it is necessary to analyze the source of Li. The beryllium produced by the H burning inside a star can decay into Li. However, due to the difference in reaction temperatures between the two, a suitable channel needs to be established, that is, the mixing process. Another type is Li pollution formed from external sources. Theoretical research on the Li enhancement is also alive and ongoing \citep[e.g.,][]{2019ApJ...880..125C, 2020ApJ...901L..18S, 2021MNRAS.503.2746M, 2022A&A...668A.126G, 2023ApJ...943..115L, 2024MNRAS.529.1423L, 2025ApJ...982L...4L}, and a more comprehensive introduction can be found in \citet{2016MNRAS.461.3336C}, \citet{2022AcASn..63....2Y}, \citet{2024MNRAS.535.1243D}, and \citet{2024ApJ...964...42S} and their references.

Although there is no significant difference in the mass distribution of Li-rich giants \citep{2022ApJ...931..136Z}, the Li abundances of some normal stars exhibit a rather special connection to mass. It appears that some higher-mass giants retain more surface Li \citep[e.g.,][]{2016A&A...587A..66D, 2023AJ....166...60T}. In addition, the samples summarised in \citet{2023ApJ...943..115L} show an increasing trend in Li abundance with increasing mass. A similar trend is predicted by the standard convection model \citep{2022ApJ...933...58C}. This implies a possible pattern for Li enhancement in low-mass giants, which could be achieved by increasing mass. Although some work suggests that Li enhancement may not be strongly correlated with mass \citep[e.g.,][]{2020MNRAS.494.1348D}, there have been many attempts to induce Li enhancement by increasing stellar mass, such as engulfment \citep[e.g.,][]{1999MNRAS.308.1133S,2009ApJ...705L..81V, 2016ApJ...829..127A}, mergers \citep{2020ApJ...889...33Z}, etc. \citet{2016ApJ...829..127A} showed that Li enhancement is closely related to the Li component of the matter added by a star. \citet{2020ApJ...889...33Z} stated that the Li abundance of a star formed by the merger of a He white dwarf and a RGB star is determined by the mass of the He white dwarf progenitor. Matter accretion is one scenario that can increase the mass of stars. Possible sources of matter include mass lost by asymptotic giant branch (AGB) stars \citep{1990ApJ...361L..69S,1995ApJ...441..735S} and massive stars, as well as mass transfer between binaries. In this paper, we will assess the impact of accreting circumstellar matter on the Li abundance of low-mass giants.

The paper is structured as follows: $\rm Sect.\,\ref{sect2}$ describes the settings of the accretion model; $\rm Sect.\,\ref{sect3}$ presents the modeling results from the mass and composition of the accreted matter and analyzes the source and composition of accreted matter; in $\rm Sect.\,\ref{sect4},$ we discuss the possibility of matter accretion based on observations, and in $\rm Sect.\,\ref{sect5},$ we list the main conclusions.

	\section{Method}\label{sect2}
\subsection{Inputs}\label{sect21}

First, we construct a basic convection model with the help of the Modules for Experiments in Stellar Astrophysics (MESA; r11701 \citep{2011ApJS..192....3P, 2013ApJS..208....4P, 2015ApJS..220...15P, 2018ApJS..234...34P, 2019ApJS..243...10P}). Our models refer to the test suite provided by the MESA, and the specific path is `.../.../mesa-r11701/star/test\_suite/7M\_prems\_to\_AGB'. The key elements of model construction can be found in $\rm Table\,\ref{t3}$ of Appendix \ref{AA}. The convective boundaries are used with the Schwarzschild criterion and the treatment of the convection is based on the mixing length theory \citep{1968pss..book.....C}. The mixing length coefficient $\alpha_{\rm MLT}$ uses the default value given by the above MESA template, i.e., $\alpha_{\rm MLT}=2.0$. Convective mixing is the solely mixing process, and the convection model does not involve other processes such as rotation, diffusion, and thermohaline mixing, etc. To ensure that the elements in the convective envelope are completely and evenly mixed, we add a strong surface convective overshooting to it (the overshooting coefficient $f_{\rm ov}$ is 0.80), and none at other convective boundaries.

For the input physics to the model, the chemical composition is chosen as \texttt{GS98} \citep{1998SSRv...85..161G}, the equation of state is given by the results of \citet{2002ApJ...576.1064R}, and the OPAL opacity tables follow the results of 
\citet{1993ApJ...412..752I,1996ApJ...464..943I}. The nuclear reaction network is \texttt{pp\_extras.net}, including 12 isotopes: $\rm ^{1,2}H$, $\rm ^{3,4}He$, $\rm ^{7}Li$, $\rm ^{7}Be$, $\rm ^{8}B$, $\rm ^{12}C$, $\rm ^{14}N$, $\rm ^{16}O$, $\rm ^{20}Ne$, and $\rm ^{24}Mg$. 

The stellar parameters of the basic model are $1.2\,M_{\odot}$ \footnote{It should be noted that the $1.2\,M_\odot$ models are situated right near the edge of the Li dip in the MS stage. Here, it is chosen as the input parameter because it corresponds to the peak in the count-mass relation \citep[see e.g.][]{2024MNRAS.529.1423L}.} and $Z=0.02$ (the solar metallicity). The two values quantify key distribution characteristics of the observation samples \citep[see][]{2022ApJ...931..136Z, 2021MNRAS.505.5340M}.
In $\rm Sect.\,\ref{sect3}$, we appropriately extend  the range of stellar parameters to enable a more precise and detailed analysis of the results.
Referring to the meteoritic abundance \citep[e.g.,][]{1998SSRv...85..161G,2009ARA&A..47..481A}, we set up an input Li abundance of $\rm 3.3\,dex$ at the zero-age main sequence (ZAMS). Despite the presence of the Li loss during the pre MS \citep[see e.g.,][]{1965ApJ...141..993I}, our current work does not include this component, so we uniformly give initial Li abundance at the ZAMS. Our stellar models evolve from the ZAMS all the way to the RGB tip.

\subsection{Matter Accretion and Mass Loss} \label{sect22}

We then add matter accretion to the convection model developed in $\rm Sect.\,\ref{sect21}$. The standard convection model predicts that more massive giants retain more surface Li \citep[e.g.,][]{2022ApJ...933...58C,2023ApJ...943..115L}, and in the absence of extra Li loss/enhancement, the dilution of elements by the convective envelope during the first dredge-up (FDU) is the only process affecting surface Li abundance \citep[see e.g.,][]{1967ApJ...147..624I}. Obviously, this requires increasing the mass of the star before the Li loss ceases. Considering that the current work is an attempt to enhance surface Li of giants, and in the absence of certainty as to when matter accretion begins, we will initiate the accretion process just after the MS turnoff, and take the mass fraction of the central H at this moment as $10^{-9}$ \citep[see e.g.,][]{2024MNRAS.529.1423L}.

This study primarily examines the impact of accreted matter on Li enhancement in low-mass stars during their RGB expansion phase. The accretion rate is estimated based on the upper limit of the maximum accretable mass and the RGB evolutionary timescale. Subsequently, we construct an accretion rate grid using this framework. The typical evolutionary timescales for Sun-like stars in the FDU phase are on the order of $10^8$ yr \citep[e.g.,][]{1998MNRAS.298..525P, 2000MNRAS.315..543H}, so for an upper limit on accretion mass of $\sim 1.0\,M_{\odot}$, the corresponding mean rate of matter accretion should be less than $10^{-8}\,M_{\odot}\,\rm yr^{-1}$. On the other hand, assuming the composition of the accretion is similar to that of the convective envelope of star, and is uniform.  Then, using the surface density of the Sun, about $\rm 10^{-7}\,g\,cm^{-3}$ \citep[e.g.,][]{2001ApJ...555..990B}, the stellar radius enlarges nearly 100 times during the RGB stage, so the volume expands to $10^6$ times of its original size, and the accreted mass is $\sim 0.1\,M_{\odot}$. So in this case, the average accretion rate is about $10^{-9}\,M_{\odot}\,\rm yr^{-1}$. Such accretion rate is much lower than the calculated accretion rate of \cite{1999MNRAS.308.1133S}, i.e., $\sim 10^{-5}\,M_{\odot}\,\rm yr^{-1}$, when increasing stellar mass by engulfing planets or brown dwarfs. Whereas the low accretion rate in our models ensures that the accreted matter is quickly dissolved into the convective envelope (detailed information can be found in Appendix \ref{AC}). 

Assuming there is a steady supply of circumstellar matter, it can be expected that the rate of matter accretion is related to the stellar mass and size throughout the stellar evolution. Since it is difficult to establish a clear relationship between the gravitational effects and stellar expansion due to uncertain mass/density distributions of accreted matter and distance from the star, we use a grid of constant accretion rates to investigate the role of matter accretion. The advantage is that this grid can reflect the increment in mass of stars due to accretion, thereby facilitating a more precise evaluation of how circumstellar matter accretion impacts Li enhancement. In our accretion models, we adopt `\texttt{mass\_change = 1d-10}', representing a matter accretion rate is $1\times10^{-10}\,M_{\odot}\,\rm yr^{-1}$. Here, \texttt{1d-10} is just a value chosen for our subsequent analysis.

To test the effect of matter accretion on the abundance of Li, two aspects must be considered: the mass of accreted matter and its composition.  In order to isolate the effect of the mass on Li abundance, the composition of accretion should be eliminated. Therefore, we maintain consistency in the composition of the accreted matter and the stellar convective envelope throughout evolution. Specific to model setting, `\texttt{accrete\_same\_as\_surface = .true.}' is assumed.

In order to investigate the effect of accretion composition on the Li abundance of giants, conversely, we will input diverse compositions of accretion, while controlling the same accretion rate. For a detailed description of the composition setup, see Appendix \ref{AB}.

In the MESA code, matter accretion or loss needs to run with the `\texttt{wind\_scheme=' '}' open. Mass loss is only incorporated into accretion models; it is not included in non-accretion models. During the RGB and AGB stages, the coefficients of mass loss caused by stellar wind are 0.5 \citep{1975MSRSL...8..369R} \footnote{Here we adopt the widely used Reimers' law, but how to regulate the RGB mass loss is still under extensive research \citep[e.g.][]{2025arXiv250512794L}.} and 0.1 \citep{1995A&A...297..727B}, respectively. In addition, we also refer to the `.../.../mesa-r11701/star/defaults/controls.defaults' file and list the stellar wind settings here. 
\[
\begin{array}{lp{0.8\linewidth}}
	\texttt{cool\_wind\_RGB\_scheme = 'Reimers' }    \\
	\texttt{Reimers\_scaling\_factor = 0.5}    \\
	\texttt{cool\_wind\_AGB\_scheme = 'Blocker'}   \\
	\texttt{Blocker\_scaling\_factor = 0.1 }      \\
	\texttt{RGB\_to\_AGB\_wind\_switch = 1d-4}\\
\end{array}
\]
According to our tests, the above settings do not affect the surface Li content of the RGB stars. We focus only on the behavior of surface Li in the RGB phase, while the subsequent AGB phase is outside the scope of current work. The  mass loss due to the stellar winds during the RGB is included  in our accretion models. Changing the scaling factor has a minor effect on the Li abundance, as \citet{1975MSRSL...8..369R} suggested that the scaling factor should not exceed three. About the empirical formulae for \citet{1975MSRSL...8..369R}, there is $\dot{M}\propto\eta L R /M$, where $L$, $R$, and $M$ are the stellar luminosity, radius, and mass, respectively. $\dot{M}$ is the mass loss rate and $\eta$ is the scaling factor. 

Based on our model parameters, the rate of mass loss $\dot{M}$ can be estimated to be about $10^{-13\sim -12}\,M_{\odot}\,\rm yr^{-1}$ during the FDU and $\sim 10^{-8}\,M_{\odot}\,\rm yr^{-1}$ in the late RGB phase. While we expect the Li enhancement of giants by matter accretion to occur in the FDU, the mass loss in this stage is much lower than the matter accretion rate we tested (i.e., $\sim 10^{-10}\,M_{\odot}\,\rm yr^{-1}$). Predictably, our accretion models will have a net mass gain. In the late RGB phase, on the other hand, one would expect the mass of the star to decrease as mass loss exceeds matter accretion. The mass of the convective envelope exceeds $0.7\,M_{\odot}$ in the late RGB phase, however, the mass loss at this stage is on the order of $0.1M_{\odot}$ when $\dot{M}$ is $\sim 10^{-8}\,M_{\odot}\,\rm yr^{-1}$. Although there is an obvious mass loss, it is predicted to have a negligible effect on the Li abundance in the homogenised convective envelope.

We list here the symbols that are covered in the follow-up.		
\[
\begin{array}{lp{0.8\linewidth}}
	M_{\odot}  & solar mass     \\
	L_{\odot}   & solar luminosity     \\
	R_{\odot} & solar radius \\
	R_{\rm acc}    &    matter accretion rate \\
	\dot{M}    &    mass loss rate \\
	M_{\rm e}  & mass of the convective envelope                   \\
	M_{\rm acc}   & mass of matter accreted           \\
	C_{\rm e}        &  $N_{\rm Li}/N_{\rm H}$ of the convective envelope            \\
	C_{\rm acc}    &  $N_{\rm Li}/N_{\rm H}$ of matter accreted            \\
	T_{\rm acc}    &  timescale of matter accretion            \\
\end{array}
\]

\subsection{Analysis for Default Values in MESA Modeling}
Recently, \citet{2025arXiv250114867C} had shown that many different modeling choices, whether physical or numerical, can affect surface chemistry, mixing, and the FDU in the RGB. Different modeling choices significantly affect the prediction of surface element abundance and the mixing process by altering the depth of the convection zone, nuclear reaction efficiency and the evolution of chemical abundance. Physical parameters (such as initial abundance, nuclear reaction rate, and mixing mechanism) dominate the trend differences, while numerical selection mainly affects the calculation accuracy.

They analyzed the interpolation methods, time steps and spatial resolution, and found that the adjustments in these numerical methods only produced extremely low prediction errors. Therefore, these values are selected with default settings when we execute the MESA simulations.

$\alpha_{\rm MLT}$ serves as a free parameter in stellar models. Increasing 
$\alpha_{\rm MLT}$ elevates the base temperature of the surface convection zone and intensifies Li destruction. However, calibrating $\alpha_{\rm MLT}$ requires compromise among $\rm log\,$g$-T_{\rm eff}$, the position of RGB bump, and the abundance of Li \citep{2025arXiv250114867C}. We mainly focus on the enhancement behavior of Li (abundance increment), and the choice of $\alpha_{\rm MLT}$ has little influence on the increment. Similarly, we also adopt the default values. Similar situations apply to the selection of the boundaries of convective zones.

The differences in the initial abundances of C and N directly affect the prediction of the surface [C/N] after the FDU. For example, the lower metallicities of \texttt{A09} ($Z_\odot\sim 0.014$, see \citet{2009ARA&A..47..481A}) may lead to a slightly smaller mixing amplitude during the FDU process, but the overall difference is limited (approximately 0.01 dex, see \citet{2025arXiv250114867C}). The abundance variation of Li is also affected by opacity (or metallicity), and we have made a coarse grid calculation for it in the following text. Furthermore, we uniformly input the Li abundance as $\rm 3.3\,dex$ for all the models. So, we choose \texttt{GS98}.

In addition, the comparative selection of the opacity table and the nuclear reaction rate will also have different effects on the prediction of the element abundance value. The focus of this article is to explore the Li enhancement effect of matter accretion. Therefore, we ignore their influences and perform calculations using the default numerical table.

\subsection{Ignored Physics Processes}
This paper focuses on the Li enhancement effect of matter accretion. To evaluate its role alone, we have ignored many physics processes. Matter accretion and mass loss are considered only on the basis of the standard model. A brief introduction to some physics processes is as follows: 

1. During the pre MS stage, a star has not yet begun to burn H in its core and is in the process of contraction and warming up. We do not evaluate the evolution of Li abundance in this stage when constructing the models. The behavior of Li at this stage can be influenced by several key physics processes: 1) Temperature-dependent destruction: Pre MS stars usually have a deep convection zone, especially in low-mass stars. Convection will bring the Li inside to the surface and simultaneously carry the substances on the surface into the high-temperature area, resulting in the destruction of Li \citep[e.g.][]{1997ARA&A..35..557P}. When the bottom temperature of the convective envelope exceeds the critical temperature for Li nuclear reaction (approximately $2.6\times 10^6\rm \,K)$, Li will be destroyed through proton capture reactions (such as $\rm ^7Li(p,\alpha)^4He)$. This process depends on the mass and age of the star, as stars with greater mass reach higher temperatures more quickly.
2) Rotation and Magnetic Field: The rotation of stars may affect the internal mixing efficiency, thereby altering the distribution of Li \citep[e.g.][]{2012A&A...539A..70E}. Magnetic fields may suppress convection, thereby reducing the loss of Li \citep[e.g.][]{2014MNRAS.445.4306J, 2015ApJ...807..174S}. 
3) Accretion and Peristellar disks: Young stars may acquire matter through accretion disks, temporarily altering the Li abundance on their surfaces, but the impact of this process may be relatively short-lived \citep[e.g.][]{2010A&A...521A..44B}.

2. The phenomenon of Li loss is the norm in the MS stage. Numerous physics processes can cause the attenuation of Li in the MS. For instance, overshooting \citep[e.g.][]{2017ApJ...845L...6B} that extends the convection zone and increases the temperature at the bottom; diffusion processes \citep[e.g.][]{1998ApJ...504..559T, 2015A&A...579A.122A} that disrupt the homogeneous characteristics of the convection zone;  rotation mixing (and its synergistic effect with the magnetic field) \citep[e.g.][]{1981ApJ...243..625E, 1992A&A...255..191C, 2005Sci...309.2189C, 2016ApJ...829...32S} triggered at the radiative zone; and mixing drived by internal gravity waves \citep[e.g.][]{1991ApJ...377..268G}, etc.

3. The evolution of Li in the giant phase has always been the focus of our attention. We ignore the mixing process that has been considered so far. Thermohaline mixing can predict the Li evolution trend of RGB stars \citep[e.g.][]{2020NatAs...4.1059K}, but it is limited by the selection effect of the aspect ratio of `salt fingers' and the enhancement of the intensity of the mixing by introducing diffusion \citep[see][]{2024MNRAS.529.1423L}. Internal gravity waves can cause Li enhancement, but it do not match the observation characteristics of RGB stars \citep{2023ApJ...943..115L}. Meridional circulation can promote the formation of clump giants with Li enhancemnt \citep{2025ApJ...982L...4L}. \citet{2021MNRAS.503.2746M} introduced the energy loss of neutrinos to explore the evolution of Li. Some external celestial bodies can also have an impact on the Li of the giants \citep[e.g.][]{2019ApJ...880..125C}. In addition, there are some processes that can cause Li depletion, such as overshooting.

\section{Result and Analysis} \label{sect3}
\subsection{Mass Effect} \label{sect31}

\begin{figure}
	\centering
	\includegraphics[width=8.2cm]{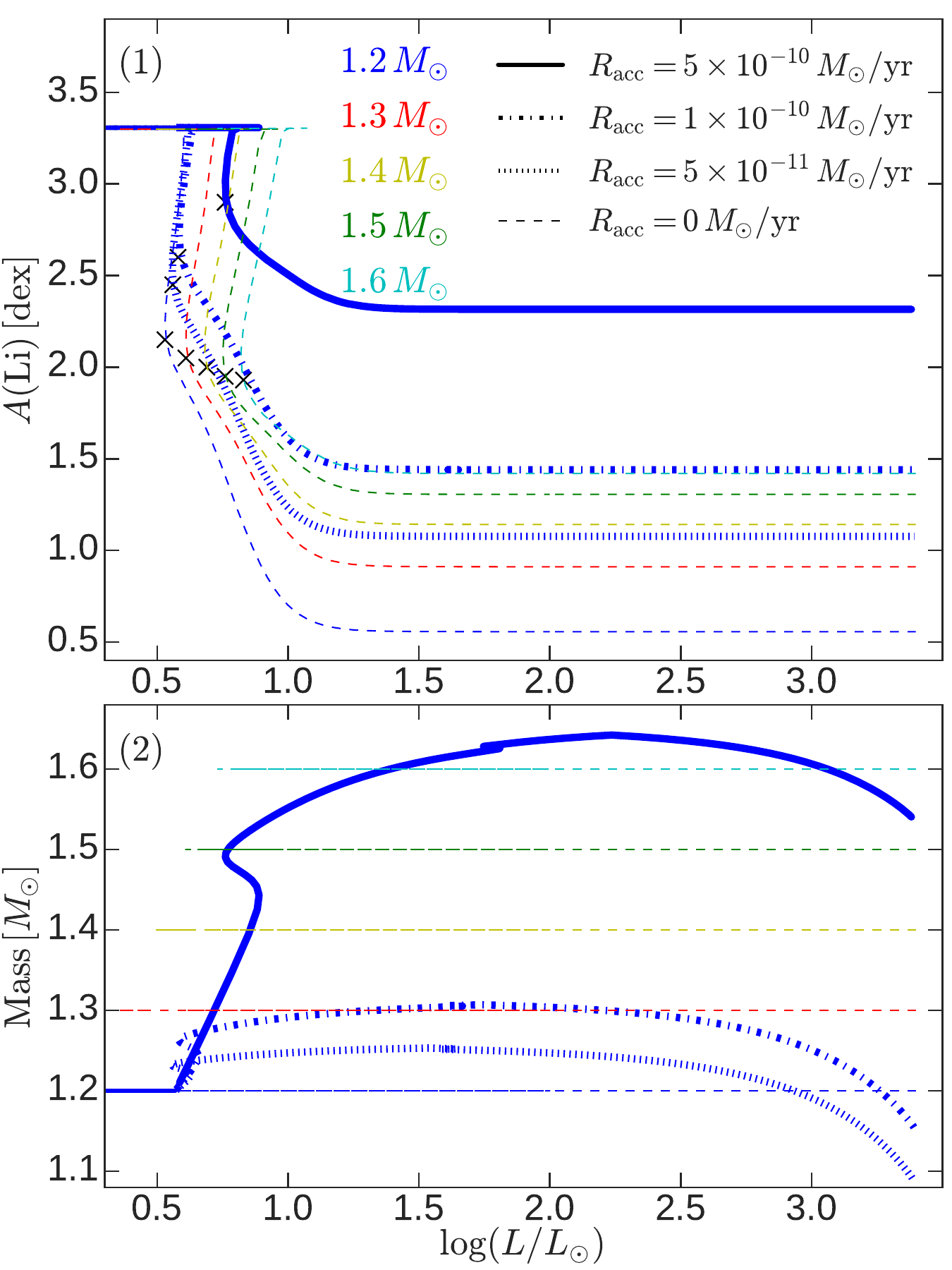}
	\caption{Evolutionary trajectories of Li abundance and stellar mass with luminosity from the ZAMS to the RGB tip. Panel (1): $A(\rm Li)$ vs. Luminosity. Panel (2): Stellar mass vs. Luminosity. The rate of accretion is marked by different linestyles, i.e., $R_{\rm acc} = $0 (the dashed lines), $5\times10^{-11}$ (the dotted lines), $1\times10^{-10}$ (the dot-dashed lines), and $5\times10^{-10}$$\,M_{\odot}\rm \, yr^{-1}$ (the solid lines), with a thick line indicating the presence of an accretion process and a thin line indicating the absence of one. Stars with different initial masses are indicated by separate colours, and $Z=0.02$ for all models. We mark the luminosity inflection points during the FDU phase with symbol `$\times$'. A corresponding Hertzsprung-Russell diagram is shown in $\rm Fig.\,\ref{a0}$ of Appendix \ref{AC}.}
	\label{f1}%
\end{figure}

\begin{figure*}
	\centering
	\includegraphics[width=18cm]{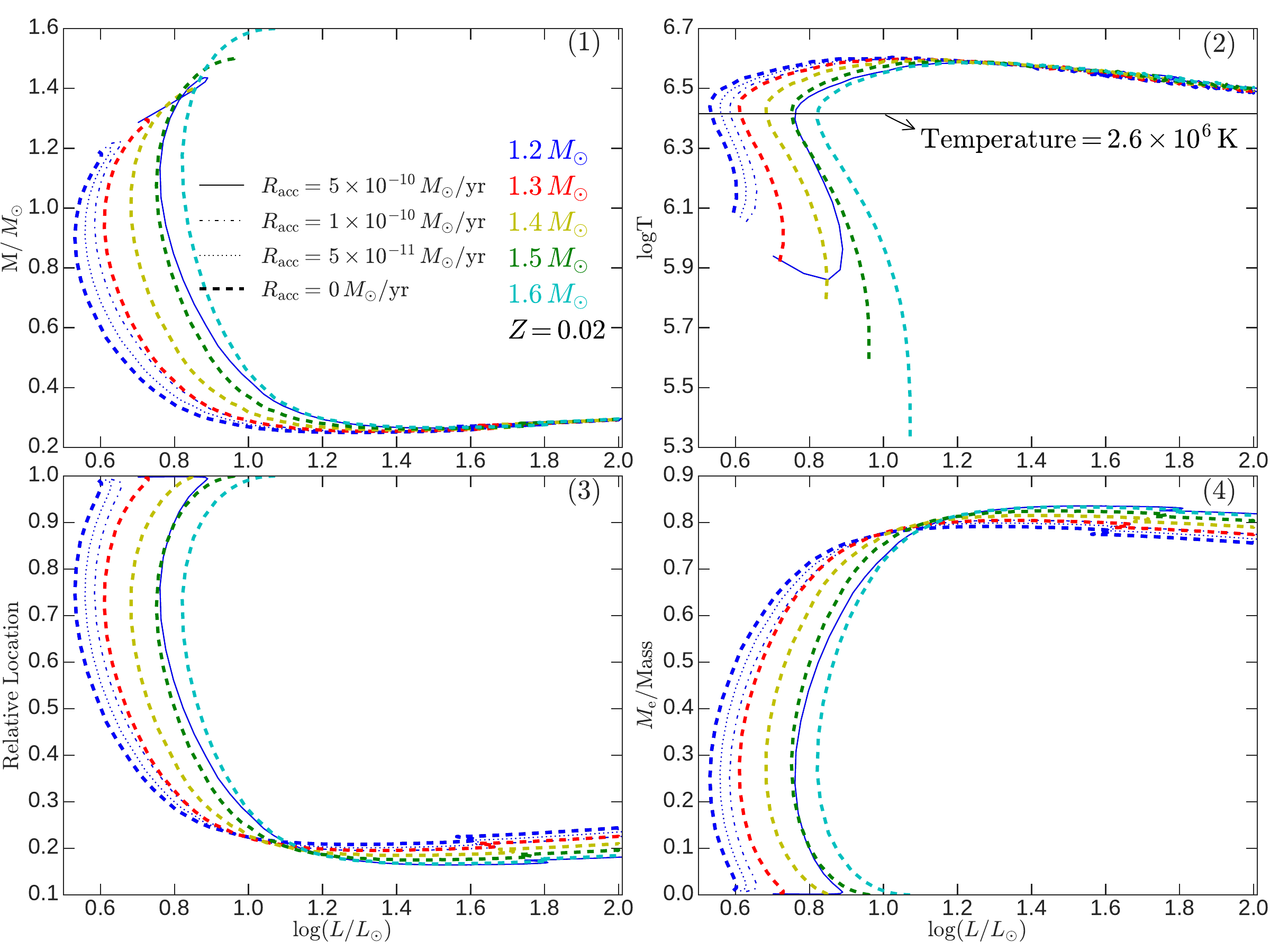}
	\caption{Structural diagram corresponding to the stellar models of $\rm Fig.\, \ref{f1}$ between the MS turnoff and luminosity ascends to $100\,L_{\odot}$. The labelling is similar to that of $\rm Fig.\, \ref{f1}$, but in order to present the differences in structure more clearly, in contrast to $\rm Fig.\, \ref{f1}$, the thin lines indicate the presence of accretion, while the thick lines do not. Panel (1) shows the position of the lower boundary of the convective envelope in mass label (i.e., the mass embeded), and panel (2) shows the corresponding temperature evolution. In panel (2), we mark the temperature at which Li is destroyed with a black horizontal line. In panel (3), the y-axis is the relative location of the bottom of the convective envelope (i.e., mass covered by a sphere with radius at the bottom of the convective envelope/total mass of star). The ratio of the mass of the convective envelope to the total mass of star is the y-axis in panel (4). }
	\label{f2}%
\end{figure*}

\begin{figure*}
	\centering
	\includegraphics[width=5.36 cm]{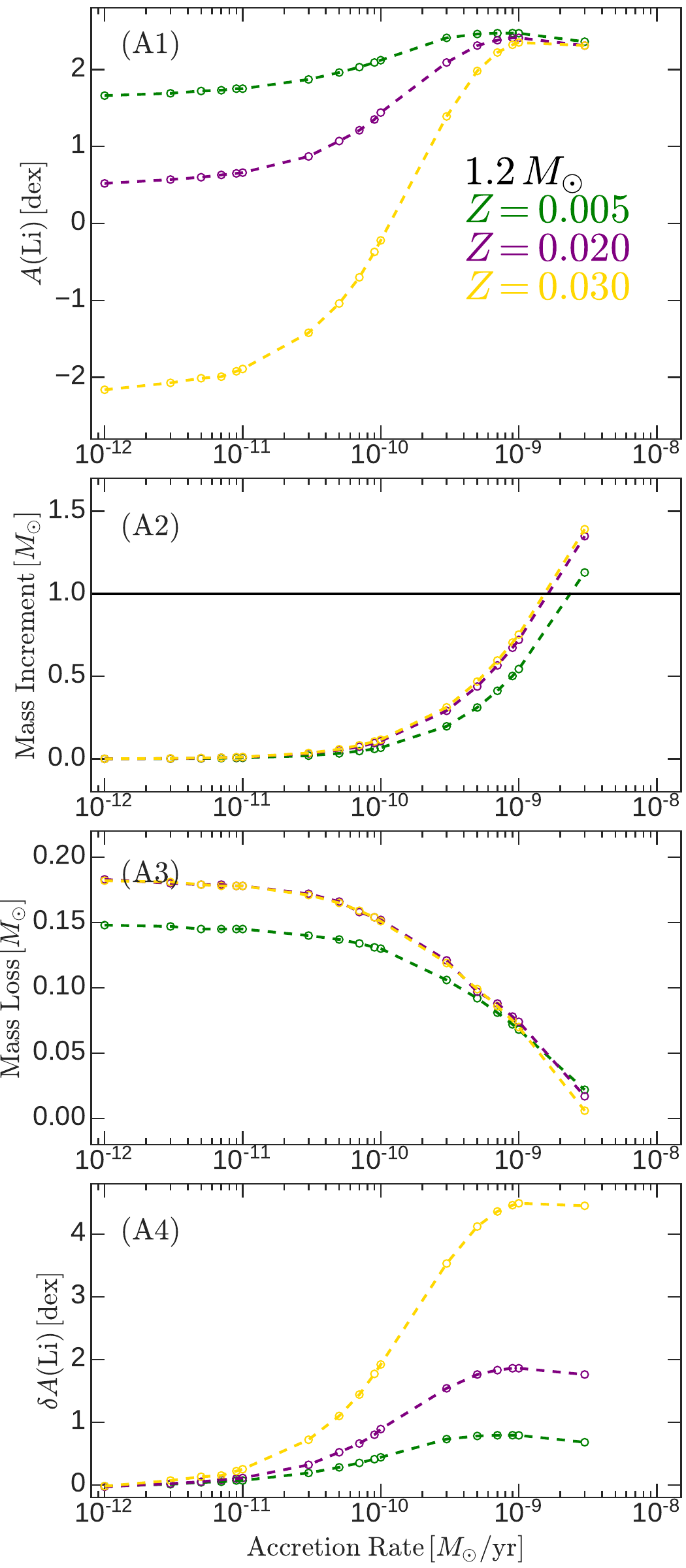}
	\includegraphics[width=5.20 cm]{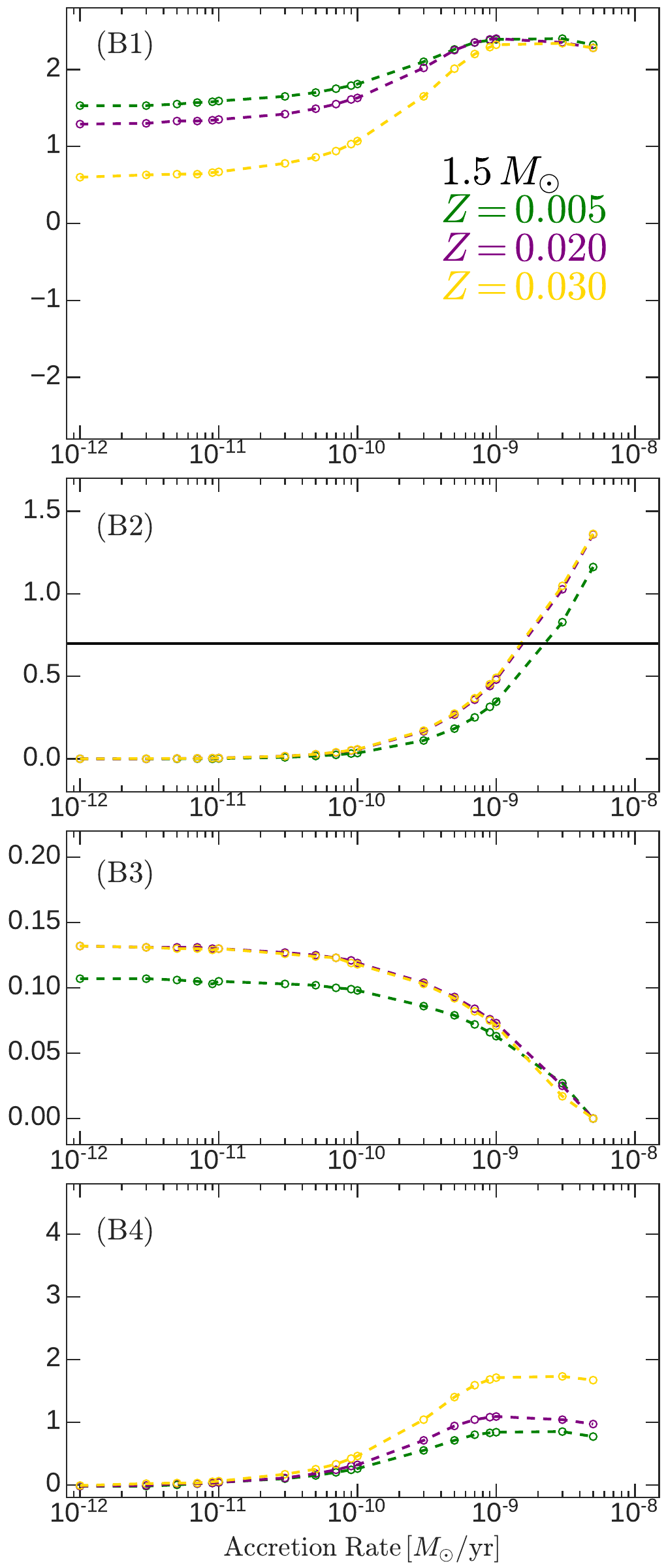}
	\includegraphics[width=5.20 cm]{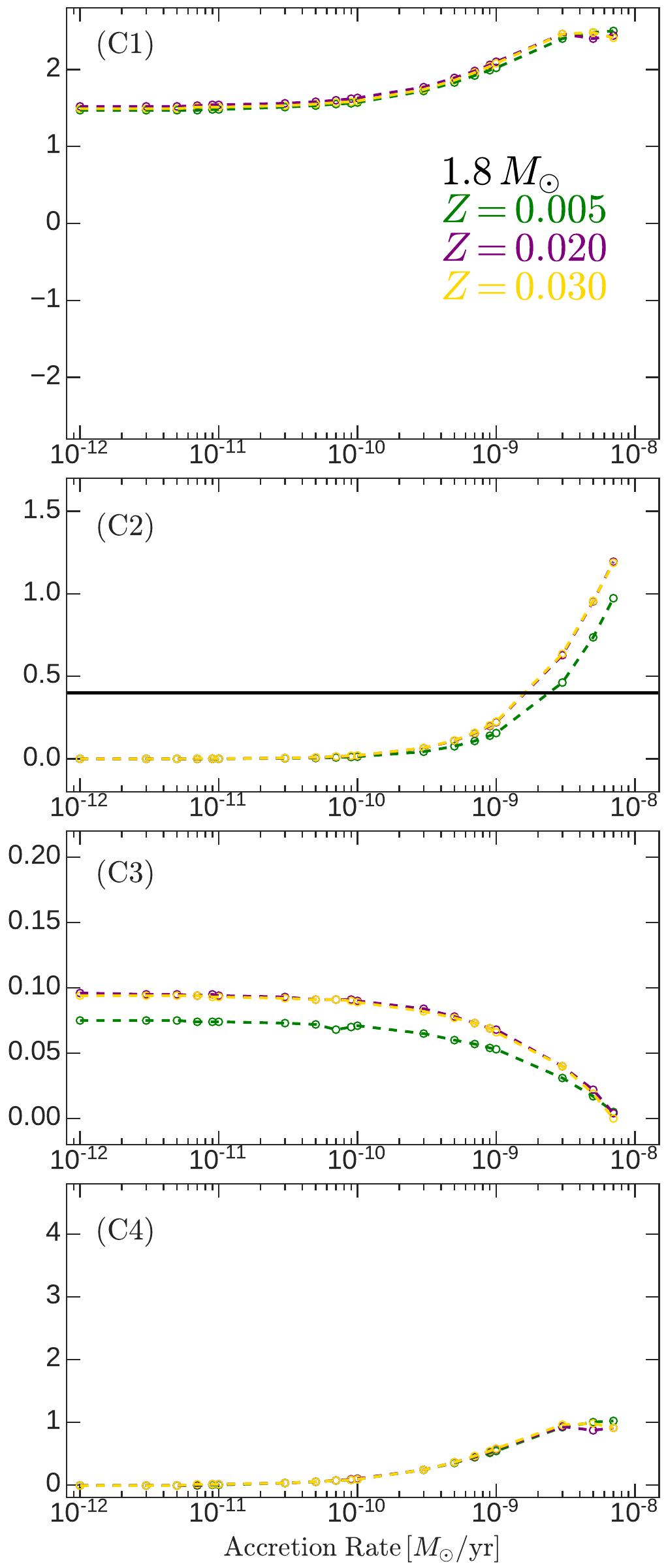}
	\caption{Li abundance, mass increment, mass loss, and Li abundance increment from low accretion rate grid scanning models obtained. Panels (A): $1.2M_{\odot}$; (B): $1.5M_{\odot}$;  (C): $1.8M_{\odot}$. \texttt{1}: Li abundances at the RGB tip; \texttt{2}:  masses increment (maximum mass gained during evolution subtracts initial mass); \texttt{3}: mass losses (maximum mass gained during evolution subtracts stellar mass at the RGB tip); \texttt{4}: difference in Li abundance between models with and without accretion evolving to the RGB tip. The black solid lines in 2nd row panels indicate the upper limit of the accretion mass for the low-mass stars with 1.2, 1.5, and $1.8\,M_{\odot}$, i.e, $1.0$, $0.7$, and $0.4\,M_{\odot}$.}
	\label{f3}%
\end{figure*}

\begin{figure}
	\centering
	\includegraphics[width=8.2cm]{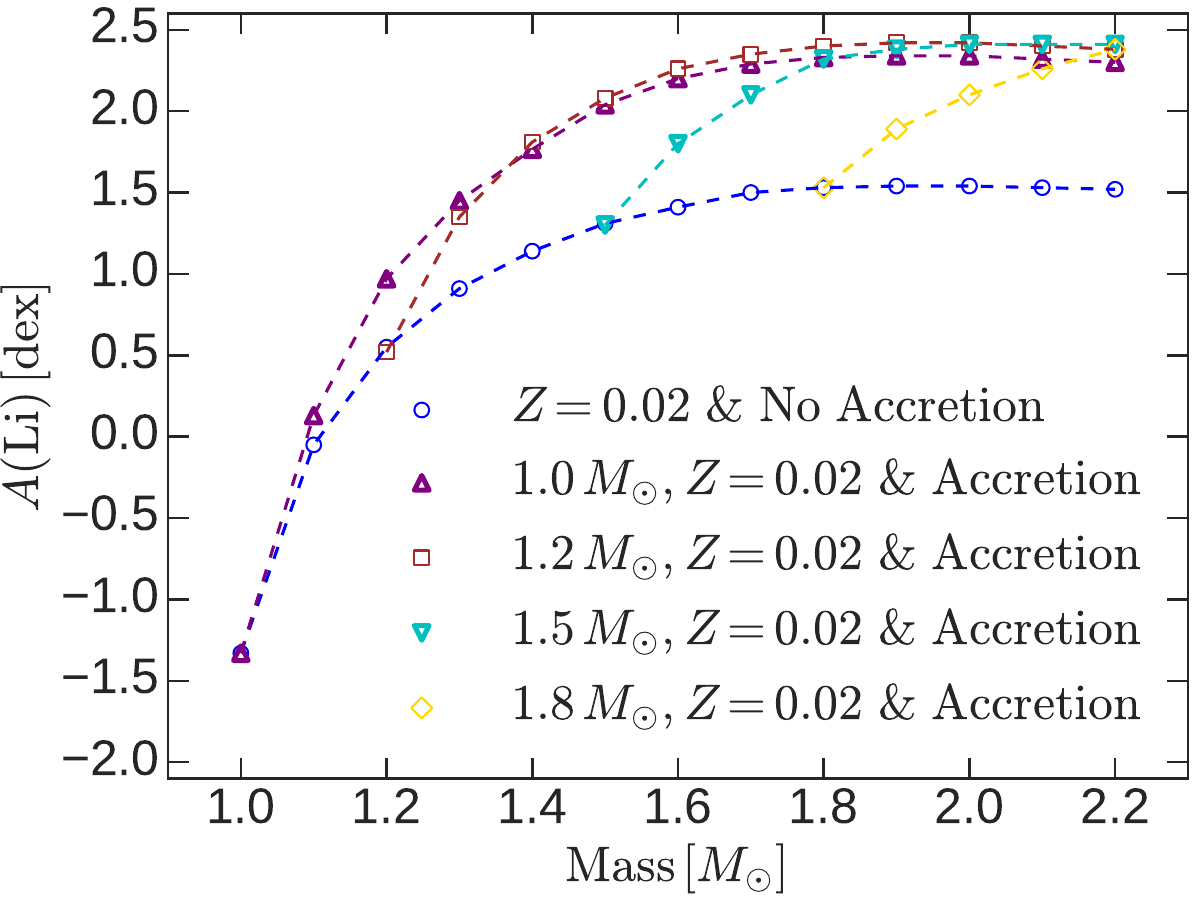}
	\caption{Li abundances predicted by the accretion model for different initial masses when the maximum mass increment is a positive integer multiple of $0.1\,M_{\odot}$, respectively. Open circles are the Li abundances predicted by models with different initial masses at the RGB tip in the absence of accretion. The initial masses represented by the triangles, squares, inverted triangles and rhombuses are 1.0, 1.2, 1.5, and $1.8\,M_{\odot}$ respectively.}
	\label{f4}%
\end{figure}

$\rm Fig.\,\ref{f1}$ shows the evolution of Li abundance and mass with luminosity for the stellar models with different accretion rates. $\rm Fig.\,\ref{f1}\,(1)$ presents that the Li abundance remains essentially unchanged during the MS and beyond the RGB bump. 
One by one, $\rm Fig.\,\ref{f1}\,(2)$ displays the evolution of mass with luminosity for the stellar models in $\rm Fig.\,\ref{f1}\,(1)$.
There is a clear decline of surface Li abundance during the FDU. For $1.2\,M_{\odot}$ stars with $Z=0.02$, the higher the matter accretion rate, the lower the Li abundance decay is, i.e., the higher Li abundances can be maintained. In particular, for $1.2\,M_{\odot}$ stars with $Z=0.02$, the Li abundances corresponding to the four accretion rates (i.e., $R_{\rm acc}=0, 5\times10^{-11}, 1\times10^{-10},\rm \ and\  5\times10^{-10}$$\,M_{\odot}\rm \, yr^{-1}$) at the RGB tip are 0.55, 1.08, 1.44, and $\rm 2.32\,dex$, respectively. Furthermore, in the case of $R_{\rm acc}=5\times10^{-10}\,M_{\odot}\rm \, yr^{-1}$, the Li abundance of a $1.2\,M_{\odot}$ star is almost an order of magnitude higher than that of the star with $1. 6\,M_{\odot}$ and $R_{\rm acc}=0\,M_{\odot}\rm \, yr^{-1}$ as they evolve into the vicinity of the RGB bump, at that stage the mass of the star with an initial mass of $1.2\,M_{\odot}$ increases to $\sim 1.6\,M_{\odot}$. In $\rm Fig.\,\ref{f1}\,(2)$, the mass loss dominates as the luminosity rises to $\sim 100\,L_{\odot}$ (i.e., $\dot{M}>R_{\rm acc}$). As we predicted in $\rm Sect.\,\ref{sect22}$, the mass loss does not cause the change in Li abundance. This is due to, for a $1.2\,M_{\odot}$ star with $Z=0.02$ and $R_{\rm acc}= 1\times10^{-10}$$\,M_{\odot}\rm \, yr^{-1}$ in the late RGB phase, the mass loss is about $0.15\,M_{\odot}$, which is less than the convective envelope mass at this point ($\sim 0.8\,M_{\odot}$). This does not cause significant stellar structure changes.

As discussed above, matter accretion can significantly weaken the Li loss during the FDU phase, which then manifests itself as `Li enhancement'. Recently, \citet{2024MNRAS.529.1423L} probed the effect of convective boundary selection on the Li abundance of giants and showed a similar behavior.

It is also clear, compared to a star with originally higher masses,  the star increases mass by matter accretion will show a signiﬁcantly higher Li abundances. Such a discrepancy is clearly related to structural changes inside star, and we present in $\rm Fig.\,\ref{f2}$ the detailed structures of the stellar models involved in $\rm Fig.\,\ref{f1}$. The Li abundances do not change significantly after the RGB bump, and furthermore, $\rm Fig.\,\ref{f1}\,(2)$ indicates that the mass loss starts to dominate in the case of the luminosity exceeds $100\,L_{\odot}$. Therefore, we focus solely on the structural changes during the period from the MS turnoff until the luminosity ascends to $100\,L_{\odot}$. Additional structural details can be found in $\rm Figs.\,\ref{a1}$ and \ref{a2} of Appendix \ref{AC}.
$\rm Fig.\,\ref{f2}$ shows the structural details concerning the  convective envelope, with luminosity in the horizontal coordinates and mass, temperature, relative location (i.e., mass covered by a sphere with radius at the bottom of the convective envelope/total mass of the star), and mass of the convective envelope/total mass of the star on the y-axis, respectively. Overall, the ingression depth of the convective envelope is kept negatively correlated with the mass of the star during the FDU, so that higher mass stars have a relatively lower surface Li loss in this phase.

In detail, the change in Li abundance occurs in two periods, 
bounded by the luminosity inflection point (marked with `$\times$' in $\rm Fig.\,\ref{f1}\,(1)$), before which the convective envelope begins to expand, so that the Li contained in the previously thin convective envelope will be diluted, leading to a decrease in Li abundance (referred to as Li dilution).
Expanding until the temperature at the lower boundary of the convective envelope reaches $\sim 2.6\times10^6\,\rm K$, i.e., near the luminosity inflection point (see $\rm Fig.\,\ref{f2}\,(2)$), and the Li in the envelope begins to participate in the reaction, $\rm  ^{7}Li(p,\alpha)\rm ^{4}He$, and therefore a decrease in surface Li ensues (recorded as Li depletion). 

In the non-accretion models, as can be seen in $\rm Fig.\,\ref{f2}\,(1)$, the bottoms of the convective envelopes of stars with different masses near the luminosity inflection point are all near $0.8\,M_{\odot}$ in mass coordinates, so that more massive stars develop larger convective envelopes, leading to a more pronounced dilution of Li.
As seen in $\rm Fig.\,\ref{f1}\,(1)$, the Li abundance decreases more for more massive stars as the luminosity inflection point is reached. 
However, after this inflection point, the degree of the Li depletion decreases instead, since their convective envelopes are not as deep as those of the less massive stars (see $\rm Figs.\,\ref{f2}\,(1)$ and (3)), and the temperature is lower (see $\rm Fig.\,\ref{f2}\,(2)$).

$\rm Figs.\,\ref{f2}\,(2)$  and \ref{a0} demonstrate that accretion promotes the expansion of the convection zone, both inward and outward. This expansion enhances the dilution of Li. However, as the accreted matter itself contains Li, the injection of more Li into the star counteracts this dilution effect to some extent. Consequently, in $\rm Fig.\,\ref{f1}(1)$, accretion models show markedly higher Li abundances than non-accretion models during the Li dilution stage as the accretion rate increases. Specifically, when the accretion starts at the MS turnoff, the dilution effect in the convective envelope will be weakened before the luminosity inflection point because of the injection of matter of the same composition, so the Li loss is not significant. In the case of higher accretion rates (e.g., $R_{\rm acc}= 5\times10^{-10}$$\,M_{\odot}\rm \, yr^{-1}$), more matter is injected into the convective envelope and the dilution effect in the convective envelope is obviously weakened. Whereas, at lower accretion rates (e.g., $R_{\rm acc}= 5\times10^{-11}$$\,M_{\odot}\rm \, yr^{-1}$), the dilution effect is somewhat weaker. Therefore, in the case where accretion is considered, the extent of the Li loss is inversely correlated with the accretion rate before the luminosity inflection point occurs. 

After this inflection point, when the temperature at the bottom of the convective envelope exceeds $2.6\times10^6\,\rm K$, the behavior of surface Li in stars with matter accretion will be consistent with those in the more massive stars without matter accretion. For the model with $1.2\,M_\odot$ and  $R_{\rm acc}= 5\times10^{-10}$$\,M_{\odot}\rm \, yr^{-1}$, the Li abundance decreases from 2.9 to $2.3\,\rm dex$, with an increment of $-0.6\,\rm dex$. This is consistent with the situation in the non-accretion model with $1.6\,M_\odot$, where $A(\rm Li)$ drops from 2.0 to $1.4\,\rm dex$. Similar situations also occurred in models with $1.2\,M_\odot$ \&  $R_{\rm acc}= 1\times10^{-10}$$\,M_{\odot}\rm \, yr^{-1}$ and with $1.3\,M_\odot$ \&  $R_{\rm acc}= 0$$\,M_{\odot}\rm \, yr^{-1}$, but the Li abundance varies more significantly, approximately $1.1\,\rm dex$. The above situation is supported by $\rm Fig.\,\ref{f2}\,(2)$, as the bottom temperature of the convective envelope in the larger mass model is lower, and its destructive effect on Li is weaker.

As can be seen in $\rm Fig.\,\ref{f1}$, the higher matter accretion rates induce a stronger `Li enhancement', even if it experiences the surface Li loss process during the FDU phase. 
It is clear that the Li abundances of giants in our models cannot exceed the input value $\rm 3.3\,dex$. 
We inspect the strength of `Li enhancement' for the stellar models within a grid of accretion rates. 
The Li abundances (1st row), mass increments (2nd row), mass losses (3rd row), and the increment of Li abundance (4th row) for stellar models with various accretion rates during the RGB stage are shown in $\rm Fig.\,\ref{f3}$. The model masses for panels (A), (B), and (C) are 1.2, 1.5, and $1.8\,M_{\odot}$, respectively.

\citet{2016ApJ...829..127A} stated that the upper limit on the Li abundance that the RGB stars can achieve by engulfing planets or brown dwarfs is $\rm 2.2\,dex$. Asteroseismological and spectroscopic analysis of Li-rich giants \citep[$A(\rm Li)>1.5\,dex$, see e.g.,][]{2000A&A...359..563C} also show that the RGB stars have an upper limit on Li abundance of $\sim \rm 2.6\,dex$ \citep{2021NatAs...5...86Y}. Our accretion models likewise reveal a similar upper limit. From $\rm Fig.\,\ref{f3}$ (A1), an upper limit on the Li abundance is found to be $\rm 2.5\,dex$, which corresponds to an accretion rate of about $10^{-9}\,M_{\odot}\rm\, yr^{-1}$. It is independent of the metallicity, but from $\rm Fig.\,\ref{f3}$ (A4) it is clear that the increment in Li abundance is positively correlated with the metallicity, and the highest increment can be up to 5 orders of magnitude at $Z=0.03$. Comparing $\rm Figs.\,\ref{f3}$ (A1), (B1), and (C1), it can be seen that the stars with different masses present the same upper limit in the presence of accretion. It is clear from $\rm Figs.\,\ref{f3}$ (A4), (B4), and (C4) that the degree of the `Li enhancement' due to accretion decreases with increasing stellar mass in the case of higher accretion rates ($R_{\rm acc}>10^{-11}$$\,M_{\odot}\rm \, yr^{-1}$),
and the differences in enhancement due to metallicity are  eliminated in the meantime. When $R_{\rm acc}<10^{-11}$$\,M_{\odot}\rm \, yr^{-1}$; however, the above situation does not occur since $M_{\rm acc}$ is very small ($<0.01\,M_{\odot}$). In addition, there is a strong correlation between the mass increase of a star and the extent of `Li enhancement'. For a $1.2\,M_{\odot}$ star with $Z=0.02$, the Li abundance increases by two orders of magnitude for $M_{\rm acc}$$ = 0.5\,M_{\odot}$, while it is about $\rm 1.0\,dex$ when $M_{\rm acc}$$ = 0.1\,M_{\odot}$. It is apparent that this feature is similar to $\rm Fig.\,\ref{f1}\,(1)$, where higher mass stars maintain higher Li abundances. Although the mass loss has no effect on the Li abundance, there is a more pronounced mass loss in the low accretion rate model (see $ \rm Figs.\,\ref{f3}$ (A3), (B3), and (C3)), which is due to the fact that the mass increment due to accretion is inherently small (see $ \rm Figs.\,\ref{f3}$ (A2), (B2), and (C2)). Such a scenario would result in higher Li abundances for more massive stars, while the opposite for less massive stars. This further corroborates our original envisage in $\rm Sect.\,\ref{sect1}$.

$\rm Fig.\,\ref{f4}$ shows the disparity in Li abundance induced by mass addition caused by accretion for stars with different initial masses. It can be seen that the Li abundances are analogous ($\sim \rm 2.5\,dex$) when the mass of all the stars with initial masses of 1.0, 1.2, 1.5, and $1.8\,M_{\odot}$ is increased by accretion to $2.2\,M_{\odot}$, showing an enhancement of Li abundance by about $\rm 1.0\, dex$ compared to that of a star with an initial mass of $2.2\,M_{\odot}$.
This increase is caused by the inhibition of the diluting effect of the convective envelope by accreting matter. The `Li enhancement', brought about by accreting mass and then losing again, is more pronounced than the distribution of Li abundance caused by differences in the mass of a star itself. For example, a $1.2\,M_{\odot}$ star with $Z=0.02$ evolves back to its mass of $1.2\,M_{\odot}$ in this way having a Li abundance of $\rm 1.63\,dex$, whereas by natural evolution alone, it has a $A(\rm Li)$ of only $\rm 0.55\,dex$. The discrepancy is due to that the mass added by accretion in such a way makes the giants with higher Li abundances compared to a star with a higher mass from the time of its formation. 

\subsection{Source of Accreted Matter}\label{sect32}
Li enhanced giants are ubiquitous, e.g., in the Milky Way: bulge \citep[see e.g.,][]{2009A&A...508..289G}, halo \citep[e.g.,][]{2013MNRAS.430..611M}, thick disk \citep[see e.g.,][]{2011A&A...529A..90M}, and thin disk \citep[e.g.,][]{2016A&A...589A..57R}. Our models require the matter source, and it may be the case that the star impedes the dispersed matter ejected by other stars and causes this matter to surround it. Possible sources of matter can be mass loss of stellar wind from massive stars, AGB stars, supernova remnants, and so on \citep[e.g.,][]{1995ApJS..101..181W}. We will estimate each of these possible sources below.

\subsubsection{Accretion of Matter Lost by Massive Stars and AGB Stars} \label{sect411}
The lost matter of massive stars via stellar winds can be in the form of clumping \citep[e.g.,][]{1988ApJ...335..914O, 2013MNRAS.428.1837S}. \citet{2006ApJ...637.1025F} gathered a collection of O-type star samples, which has enabled us to estimate the average density, denoted as \( \bar{\rho} \), of the stellar wind clumping. Upon reviewing Table 1 from \citet{2006ApJ...637.1025F}, we designate the stellar radius as the boundary, \(r\), where the stellar wind extends and take \(v_\infty\) as the terminal velocity of the wind. 
Please note that  for the stellar wind the actual location should exceed the stellar radius and the velocity should be less than \( v_\infty \). Considering the mass loss rate for O-type stars is approximately \( 10^{-7}- 10^{-4}\,M_{\odot}\,\rm yr^{-1}\)\citep{2006ApJ...637.1025F} and \(\dot{M}\propto r^2\,\bar{\rho}\,v_{\infty}\) (see \citet{2014ARA&A..52..487S}), the resultant density range is forecast to be between \(10^{-11}\) and \(10^{-6}\rm \,g\,cm^{-3}\).

The scales of the clumping are less than the stellar radius \citep{2014ARA&A..52..487S}, hence, assuming a clumping size of \(10^{10}\rm \,cm\), the mass of these clumps can be estimated to range from \(10^{-14}\) to \(10^{-9}\,M_\odot\). Subject to gravitational forces, the target star will accumulate these clumps in its vicinity. In a scenario where the overlap of clumps is neglected, the star's radius will expand to \(\sim 100\,R_{\odot}\) during the RGB phase, potentially incorporating approximately \(10^8\) clumps. This accumulation would lead to a mass increase within the order of \(10^{-6}\) to \(10^{-1}\,M_\odot\). Combined with \(\rm Fig.\,\ref{f3}\), the mass loss of massive stars can bring a `Li enhancement' effect to our model. However, since the predicted accretion mass is highest in the order of $0.1\,M_\odot$, referring to \(\rm Fig.\,\ref{f1}\), this means that `Li enhancement' does exist, but manifesting as Li-rich characteristics seems to be rather difficult. Like O-type stars, the matter loss from Wolf-Rayet stars also occurs through the aggregation of matter into clumps \citep{1988ApJ...334.1038M}. In this scenario, the density of the clumps is critical, and a higher \(\dot{M}\) means that the clumps maintain a higher density, while Wolf-Rayet stars currently have mass loss rates that exceed those of O-type stars by \(1-2\) orders of magnitude \citep{2007ARA&A..45..177C}. This may provide a significant mass gain for our model and facilitate the formation of Li-rich giants.

Massive stars have very short evolutionary timescales, while low-mass giants are old stars. If the `Li enhancement' giants are formed by accretion of massive stars, then old stars would be in a young stellar population, i.e., the multi-population problem in star clusters. One possible explanation is that star clusters might have trapped gas clouds from the outside through their gravitational potential wells, which then formed new stars, and subsequent observations confirmed this \citep{2016Natur.529..502L}. The number of Li-rich giants is small in globular clusters \citep{2016ApJ...819..135K}, it should be noted, which means that the overall extent of `Li enhancement' may be low. On the other hand, the ejected matter clumping is bound around the star, then the star will accrete the matter through changes in its own scale, which will be comparable to the star's evolution timescale in the RGB stage. The average accretion rate thus converted would be within the grid in \(\rm Fig.\,\ref{f3}\).

From the perspective of accretion mass alone, massive stars are suitable donors. However, the phenomenon of low-mass stars accreting massive stars may not occur easily. Collecting matter lost by nearby but binary companion massive stars or AGB stars has been discussed extensively in the context of globular cluster second generation formation. Current case is that there are many lines of argument (energy, velocity, density, timescale, etc) to suggest that that it does not happen, especially for stars not currently in dense cluster environments \citep[e.g.][]{2017MNRAS.470..977L, 2018ARA&A..56...83B}. 
The role of AGB stars in star clusters faces limitations, primarily in the following aspects:
	1). Mass Supply: Due to their low velocity and relatively low energy deposition rate, AGB stellar winds are more likely to be retained by globular clusters and contribute to second-generation star formation. However, the mass they provide might be insufficient.
	2). Timescale Mismatch: The active phase of AGB stellar winds is relatively long, lasting approximately $10^8\,\rm yr$. In contrast, the timescale for massive stellar winds to form the second generation of stars in globular clusters is only $\sim 10^6\,\rm yr$.
While these factors may prevent AGB stars from being the primary source for second-generation star formation, they can still serve as excellent matter donors during accretion processes, as the ejected matter has a low velocity and a long existence time.

In addition, AGB stars are a more common scene than the massive stars as accretion sources. According to the initial mass function, the prevalence of AGB stars significantly outnumbers that of massive stars, and their distribution area is more extensive. The circumstellar envelopes of AGB stars may exist in the form of clumping \citep[e.g.,][]{1998A&A...333L..51W, 2000A&A...357..169O}. In addition, the mass loss rate is about \(10^{-8\sim -4}\,M_\odot\,\rm yr^{-1}\) and \(v_\infty\) is \(\sim 10^6\, \rm cm\,s^{-1}\) \citep{2009A&A...499..515R}, thus the acceptor star can share the increment of mass from O-type stars when it gains mass from the AGB stars (about the order of $10^{-1}\,M_\odot$). 

\subsubsection{Accreting the Supernova Remnants}
The mass lost by the star can be as high as \(\sim 10^{34}-10^{35}\,\rm g\), and the loss rate is usually $10^{-3}-10^{-8}\,M_\odot\,\rm yr^{-1}$ \citep{2000ARA&A..38..573W, 2014ARA&A..52..487S}, that is, the timescale is much smaller than the accretion time of our model (\(\sim 10^8\,\rm yr\)). Therefore, a predictable routine scenario is that acceptor stars tend to be distributed in a diffuse region of lost matter, where the density of matter around the star is relatively low, making it difficult for the star to accrete sufficient matter through its own expansion.
Take type II supernovae as an example, the products of supernova explosion will be distributed throughout interstellar space in a very short time (\(\sim 10^5\,\rm s\)), and the escape velocity is about $10^8\,\rm cm\,s^{-1}$ \citep{2003ApJ...582..905H}. 
Thus, the matter is distributed on a scale of approximately $10^{13}\,\rm cm$, resulting in an average density of about $10^{-5}-10^{-4}\,\rm g\,cm^{-3}$. However, the acceptor star cannot sustain itself at such a short distance from the supernova.  Taking the age of the supernova remnants in the current Large Magellanic Cloud (\(\sim 10^{11}\,\rm s\), see \citet{2017ApJS..230....2B}), the spatial scale of the supernova remnant is about $10^{19}\,\rm cm$, and the average density of matter will be very low, which is obviously not conducive to the considerable increase of the mass of the star.  Because the dispersion time of matter is much shorter than the accretion time of the model. The resulting `Li enhancement' would be very weak for most of our models of stars. 
In a region of dispersed matter, if the star is a rapid rotating rotor, it will enhance the rate of accretion and potentially lead to a relatively considerable increase in mass.

Unlike type II supernovae, for type Ia supernovae, they may leave behind a remnant companion star after explosion \citep{2004Natur.431.1069R}, and the companion star may be contaminated by heavy elements \citep{2010ApJ...715...78P, 2012ApJ...750..151P}. 
As a result, the distance between the acceptor star and the supernova progenitor can be short, potentially increasing the mass. For type Ia supernovae, however, the ejection velocity and current remnant age in the Large Magellanic Cloud are similar to type II supernovae \citep{2017ApJS..230....2B}, but with lower ejection mass (\(\sim 10^{33}\,\rm g\)). Thus, unlike the case of the clumping in \(\rm Sect.\,\ref{sect411}\), the ejected matter rapidly disperses and forms a remnant within a short period of time. It can then be assumed that the acceptor star, with a radius of about \(10^{11-12}\,\rm cm\), is \(10^{13}\,\rm cm\) away from the supernova progenitor. Since the acceptor star is expanding very slowly and far beyond the time the remnants were formed. It can be assumed that all matter within an azimuth, $\phi \sim \rm sin\, \phi \approx 10^{-2} - 10^{-1} \rm \,rad$, corresponding to the volume of the acceptor star, is accreted by the acceptor star. The received mass is only about \(10^{-3}-10^{-2}\, M_\odot\). Clearly, the mass increment is not high, and `Li enhancement' is also not significant. Nevertheless, the acceptor star may also be enriched in Li by contamination of its elements by Ia supernova remnants. As shown in \(\rm Fig.\,\ref{f5}\,(1)\), even if the mass of accretion is low, the composition of the accretion has a high Li content, and the `Li enhancement' effect is also significant. Unfortunately, to date, no statistical samples of Li-rich giants have been found in supernova remnants. For the time being, it is not possible to determine the proportion of giant samples that are Li enhanced due to mass or contaminated by Li, which requires follow-up observation support.

In summary, considering the incremental mass of accretion matter from supernova remnants and the current statistical sample of Li-rich giants in the remnants, accreting the supernova remnants is not a major source of matter for the current work.

\subsubsection{Accretion between Binaries}
In order to achieve the mass increment of \(10^{-1}\,M_\odot\) orders of magnitude through accretion, there are several necessary conditions, one is that the velocity of the lost matter is not too fast to facilitate the star to capture, and the other is that the star cannot be too far away from the matter source. Assuming that the outflow of matter maintains a uniform composition, to attain a significant mass increase on the order of \(10^{-1}\,M_\odot\), the star's distance from the matter source should not exceed approximately \(10^{13}\,\rm cm\). This condition is more readily achievable in binary systems, particularly close binaries. Such as the potential donor, an AGB star, the typical spatial scale at which matter accretion occurs in the scenario described by \citet{2018A&ARv..26....1H} is about \(10^{13}\,\rm cm\). In addition, the accretion rates that we explore in this work are all well within the range of accretion rates characteristic of various types of close binaries \citep[see e.g.,][]{1984ApJS...54..443P, 1988A&A...202...93R, 2009MNRAS.395.1127C}.

From the MS turnoff to the RGB tip (the timescale is \(\sim10^9\,\rm yr\) for our $1.2\,M_{\odot}$ and $Z=0.02$ model, here, with the long duration is attributed to the fact that the mass fraction of the central hydrogen from $10^{-9}$ to $0$ takes about $\sim 6\times 10^8\,\rm yr$), all model initiates the accretion process, but mass loss dominates in the late RGB phase. As detailed in $\rm Sect.\,\ref{sect31}$, `Li enhancement' due to mass effects is primarily observed during the FDU, which lasts about \(10^8\,\rm yr\), a timescale we refer to as the `effective accretion'. This phase, albeit long, is significantly shorter than the whole period after the FDU, which extends to roughly \(4 \times 10^8\,\rm yr\). Consequently, the aforementioned `effective accretion' is feasible within the context of a binary system.

It should be noted here that the timescale of the actual matter transfer between binary stars is not long.   We evaluate the accretion situation within a relatively short period of time, as shown in $\rm Fig.\,\ref{a3}$ of Appendix \ref{AC}. It can be found that within a relatively short period of time, when $0.1\,M_\odot$ is accreted, the Li abundance of the star increases from 0.55 to $\rm 1.31 \, dex$. The stable Li abundance of a star with an initial mass of $1.3\,M_\odot$ in the giant stage is less than $1.0\,\rm dex$ (see $\rm Fig.\,\ref{f1}$). During the FDU, the `Li enhancement' caused by a $0.1\,M_\odot$ increase in stellar mass does not seem to be strongly related to the required time or accretion rate.   Therefore, as long as a star accrets matter in a relatively short period of time during the FDU stage, it can also promote the retention of Li. Four points regarding this situation should be noted: 1) Unlike the accretion of matter by the expansion of stars themselves, the simulation of matter transfer between binary stars is more complex, and some more realistic scenarios should be considered, such as \citet{2021ApJ...923..277R} and \citet{2023ApJ...944...89S}. 2) Accreting a relatively high mass in a short period of time may cause significant changes in the internal structure. 3) When the accretion timescale is much smaller than that of the FDU, it can be seen from $\rm Fig.\,\ref{f1}\,(1)$ that effective accretion needs to occur before the luminosity inflection point. The rapid transfer of matter between binary stars also needs to meet this condition in order to cause significant Li enhancement. 4) Since the timescale of matter transfer is much smaller than that of the FDU, when accretion begins also needs to be evaluated. In addition, whether accretion continues (or episodic) also requires further analysis. Therefore, in this case, the average accretion rate estimated by dividing the accretion mass by the timescale of the FDU may not be a suitable choice. Research on the mode of matter transfer between binary stars will draw attention to the above-mentioned issues. This paper mainly focuses on the situation of weak accretion that does not significantly affect the structure of stars. Therefore, we only roughly analyze the candidates of this matter donor.

Recently, \citet{2024A&A...690A.367C} cast a shadow on this scenario by finding that the binary fraction among Li-rich giants is the same as that for Li-normal giants. At this point, our model results show that mass increase can lead to `Li enhancement', but it does not necessarily lead to the formation of Li-rich giants. Given that our model focuses on the mass transfer outcomes between binaries without delving into the intricate mass transfer details, it currently does not provide the formation probability of the Li-rich giant within binary systems. Determining this probability will necessitate future large-scale, detailed computational modeling.

In conclusion, AGB stars and binary systems are possible matter donors. However, the mass increment provided by the former is difficult to reach $1.0\,M_{\odot}$, which means that the expected $2.5\,\rm dex$ of the weak accretion model is a relatively ideal value. The latter can provide sufficient matter, but the relationships among its accretion timescale, accretion rate, and accretion stages still need to be further analyzed.

\subsection{Effect of the Composition of Accreted Matter on Lithium Enhancement of Giants} \label{sect33}

\begin{table}
	\caption{Predicted Li abundances at varying accretion compositions.} 
	\label{t1}
	\centering          
	\begin{tabular}{l l l c c c} 
		\hline \hline \noalign{\smallskip}
		& $C_{\rm acc}$   &   $C_{\rm e}$ &  $M_{\rm e}$ & $M_{\rm acc}$ & $A(\rm Li)$\\
		&    & &  [$M_{\odot}$]& [$M_{\odot}$] &  [$\rm dex$]\\
		\hline \noalign{\smallskip}
		& $10^{-8}$ & $10^{-10.5}$  & 0.9& 0.001&  1.63\\
		& $10^{-9}$ &  $10^{-10.5}$ & 0.9& 0.001&  1.52\\
		Case 1	& $10^{-10}$ & $10^{-10.5}$   & 0.9& 0.001&  1.50\\
		& $10^{-11}$ & $10^{-10.5}$   & 0.9& 0.001&  1.50\\
		&  $10^{-12}$ & $10^{-10.5}$   & 0.9& 0.001&  1.50\\
		\hline \noalign{\smallskip}
		&$10^{-8}$ & $10^{-10.5}$  & 0.9& 0.01&  2.15\\
		&$10^{-9}$ &  $10^{-10.5}$ & 0.9& 0.01&  1.63\\
		Case 2 &$10^{-10}$ & $10^{-10.5}$   & 0.9& 0.01&  1.51\\
		&$10^{-11}$ & $10^{-10.5}$   & 0.9& 0.01&  1.50\\
		&$10^{-12}$ & $10^{-10.5}$   & 0.9& 0.01&  1.50\\
		\hline \noalign{\smallskip}
		&$10^{-8}$ & $10^{-10.5}$  & 0.9& 0.05&  2.75\\
		&$10^{-9}$ &  $10^{-10.5}$  & 0.9& 0.05&  1.92\\
		Case 3 &$10^{-10}$ & $10^{-10.5}$  & 0.9& 0.05&  1.55\\
		&$10^{-11}$ & $10^{-10.5}$  & 0.9& 0.05&  1.48\\
		&$10^{-12}$ &  $10^{-10.5}$  & 0.9& 0.05&  1.48\\
		\hline \noalign{\smallskip}
		&$10^{-8}$ & $10^{-10.5} $ & 0.9& 0.10&  3.01\\
		&$10^{-9}$ &  $10^{-10.5} $ & 0.9& 0.10&  2.11\\
		Case 4 &$10^{-10}$ & $10^{-10.5}$  & 0.9& 0.10&  1.59\\
		&$10^{-11}$ & $10^{-10.5} $ & 0.9& 0.10&  1.47\\
		&$10^{-12}$ &  $10^{-10.5}$  & 0.9& 0.10&  1.46\\
		\hline \noalign{\smallskip}
		&$10^{-8}$ & $10^{-10.5} $ & 0.9& 0.50&  3.56\\
		&$10^{-9}$ &  $10^{-10.5} $ & 0.9& 0.50&  2.58\\
		Case 5 &$10^{-10}$ & $10^{-10.5}$  & 0.9& 0.50&  1.75\\
		&$10^{-11}$ & $10^{-10.5} $ & 0.9& 0.50&  1.38\\
		&$10^{-12}$ &  $10^{-10.5}$  & 0.9& 0.50&  1.32\\
		\hline \noalign{\smallskip}
		&$10^{-8}$ & $10^{-10.5} $ & 0.9& 0.90&  3.70\\
		&$10^{-9}$ &  $10^{-10.5} $ & 0.9& 0.90&  2.71\\
		Case 6 &$10^{-10}$ & $10^{-10.5}$  & 0.9& 0.90&  1.82\\
		&$10^{-11}$ & $10^{-10.5} $ & 0.9& 0.90&  1.32\\
		&$10^{-12}$ &  $10^{-10.5}$  & 0.9& 0.90&  1.21\\
		\noalign{\smallskip}
		\hline
	\end{tabular} 
\end{table}

\begin{figure}
	\centering
	\includegraphics[width=8.2cm]{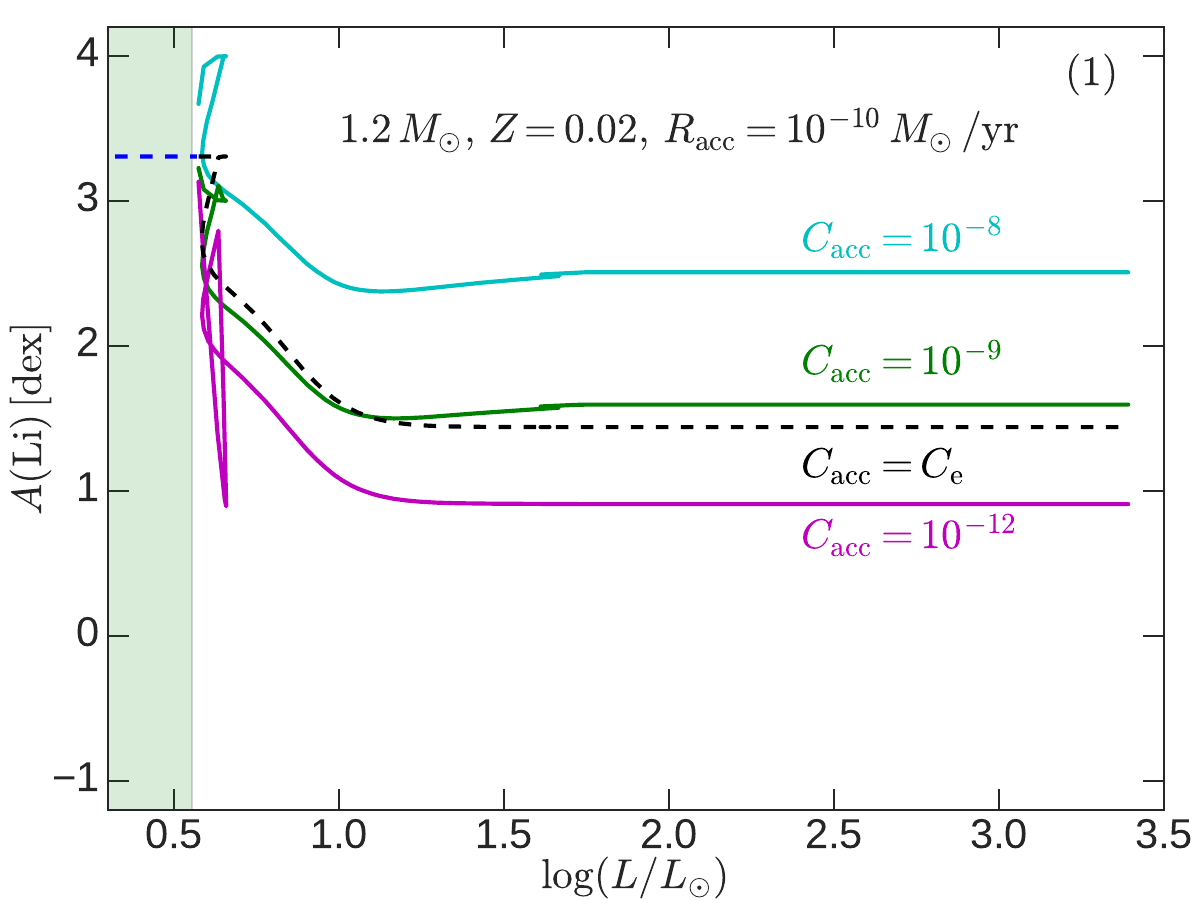}
	\includegraphics[width=8.2cm]{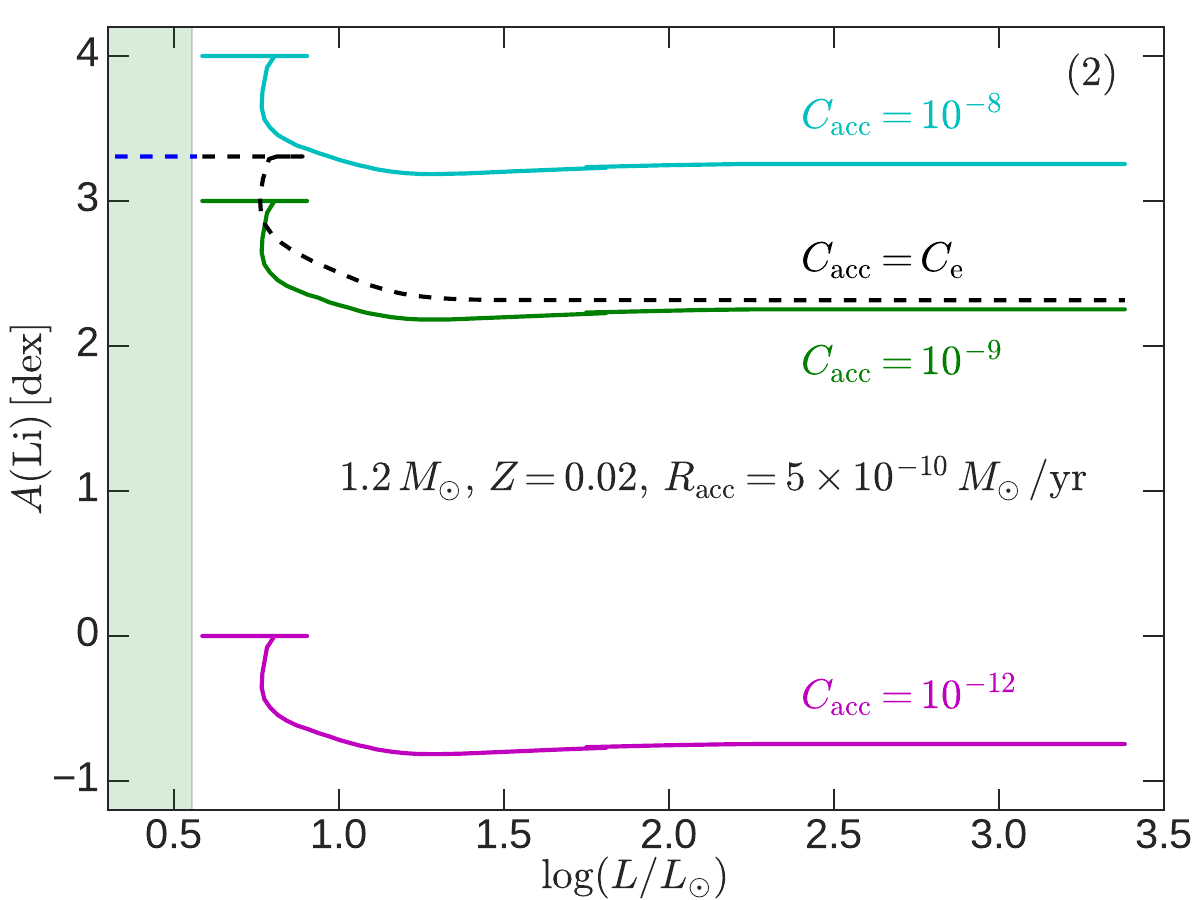}
	\caption{$A(\rm Li)$ vs. Luminosity. Accretion starts at the MS turnoff up to the RGB tip. The shaded areas are the MS. The black dashed lines are the evolutionary trace in $\rm Fig.\,\ref{f1}\,(1)$ with $R_{\rm acc} = 10^{-10}$ and $5\times10^{-10}\,M_{\odot}\,\rm yr^{-1}$.}
	\label{f5}%
\end{figure}

In $\rm Sect.\,\ref{sect31}$, to verify the effect of the mass increment caused by matter accretion on the Li enhancement for giants, we ensure that $C_{\rm acc}$ and $C_{\rm e}$ are the same. In this subsection, we will explore the influence of the composition of the accreted matter on the effect of giants Li enhancement. Here, the composition of the accreted matter specifically refers to the ratio of the number densities of Li to hydrogen (i.e., $C_{\rm acc}$). Furthermore, we assume that Li is evenly distributed throughout the circumstellar matter.

In this paper, we focus only on the low-mass stars, so for a $1.2
\,M_{\odot}$ star, the upper limit on $M_{\rm acc}$ is $1.0\,M_{\odot}$. To assess the impact of accretion composition alone, it is necessary to eliminate the influence of mass increase present in $\rm Sect.\,\ref{sect31}$. For the sake of our calculations, we take a $1.2\,M_{\odot}$ star with $Z=0.02$ and $R_{\rm acc}=1\times10^{-10}\,M_{\odot}\rm\, yr^{-1}$ evolving to the RGB bump (i.e., $A(\rm Li) = 1.5\,dex$ and $C_{\rm e}=10^{-10.5}$), at this evolution stage the total mass is about $1.3\,M_{\odot}$, and $M_{\rm e}$ is $\sim 0.9\,M_{\odot}$. In addition, we assume that the $M_{\rm acc}$ of the star at subsequent stages are 0.001, 0.01, 0.05, 0.10, 0.50, and $0.90\,M_{\odot}$, respectively. As shown in $\rm Fig.\,\ref{f2}\,(4)$ (or $\rm Fig.\,\ref{a2}$), the accreted matter is dissolved into the convective envelope, thus in the presence of accretion, the Li abundance of the star can be expressed as: 
\begin{equation}
	A(\mathrm{Li}) = 
	\mathrm{log} (\frac{M_{\mathrm e}}{M_{\mathrm e}+ M_{\mathrm{acc}}}  C_{\mathrm e} + 
	\frac{M_{\mathrm{acc}}}{M_{\mathrm e}+ M_{\mathrm{acc}}}  C_{\mathrm{acc}} ) +12.
\end{equation}
$\rm Table\,\ref{t1}$ shows our test results. It is clear that the Li abundances of the giants are very sensitive to the $C_{\rm acc}$ of the accreted matter. Higher $C_{\rm acc}$ can increase the surface Li abundance of a star. In the weak accretion case, the stellar mass growth is slow. It can be found that the effect of accreting matter on the Li enhancement is positive even when $M_{\rm acc}$ is very small. Such as at $C_{\rm a cc}=10^{-8}$ and $M_{\rm acc}=0.01\,M_{\odot}$, the Li abundance increases form 1.5 to $\rm 2.15\,dex$ (see the Case 2 of $\rm Table\,\ref{t1}$). The greater the accreted mass, the higher the Li abundance. However, the original surface Li of a star is diluted if a larger mass of matter is accreted when $C_{\rm acc}$ is low. For example, for the Cases 3, 4, 5, and 6 of $\rm Table\,\ref{t1}$, $A(\rm Li)$ will decrease when $C_{\rm acc}<10^{-10.5}$. 

To summarise, the role of accreting matter on Li will result in enhancement if the accretion matter contains more Li than the surface of a star. Conversely, it will cause Li loss.

To isolate the effects of composition differences, our model calculations ignore the FDU stage. Next, we will initiate the accretion process at the MS turnoff, which inevitably entrains the mass effects. For this reason, two values of $R_{\rm acc}$ ($10^{-10}$ and $5\times10^{-10}\,M_{\odot}\rm \,yr^{-1}$) are chosen as references to each other. Meanwhile, we consider three compositions with $C_{\rm acc}=10^{-8},\ 10^{-9},\ \rm and\ 10^{-12}$.

$\rm Fig.\,\ref{f5}$ plots the progression of Li abundance with luminosity for various compositions of accreted matter.
During the MS, the Li abundance remains constant at $\rm 3.3\,dex$, i.e., $C_{\rm e} = 10^{-8.7}$. After the MS turnoff, the dilution of Li is modulated by the composition and mass of accreted matter, as the accreted matter is injected into the star. As can be seen from $\rm Figs.\,\ref{f5}\, (1)\ and\ (2)$, Li does not dilute, but rather increases, when  $C_{\rm acc} = 10^{-8}$ > $C_{\rm e}$ (i.e., the cyan lines above the black dashed lines in $\rm Fig.\,\ref{f5}$). In this situation, the higher the mass of the accreted matter, the higher the Li abundance. Conversely, when $C_{\rm acc} = 10^{-12}$, a higher mass of accreted matter exacerbates the Li dilution. In $\rm Fig.\,\ref{f5}\,(2)$, $A(\rm Li)$ is $\rm \sim -0.7\,dex$, while it is about $\rm 1.0\,dex$ in the case of $R_{\rm acc}=10^{-10}\,M_{\odot}\,\rm yr^{-1}$.  Notably, an anomalous dip appears in $\rm Fig.\,\ref{f5}\,(1)$, potentially associated with the blue hook. This feature is transient and does not affect the overall findings.

The behavior of Li at the beginning of accretion is closely related to the strength of expansion of the convective envelope, the mass of the accreted matter, and its composition. Thereafter, there is a Li loss during the FDU due to the dilatation of the convective envelope. After the RGB bump, the evolutionary time of a star during the RGB phase is very short, on the order of $\rm 10^{7}\,yr$, thus, the corresponding mass increment is $\sim 10^{-3}\,M_{\odot}$. This is considerably less than the mass of the convective envelope ($> 0.7\,M_{\odot}$), and even if the composition of Li in the accreted matter exceeds that of its surface at this time, the enhancement of Li is very small (such as the Case 1 of $\rm Table\,\ref{t1}$).

\subsection{Composition of Accreted Matter}\label{sect34}
Oxygen-rich AGB stars in the Magellanic Clouds show Li abundances exceeding $\rm 2.0\,dex$ \citep{1990ApJ...361L..69S,1995ApJ...441..735S}, and their loss of mass may provide $C_{\rm acc}>10^{-10}$ of accreted matter. The Li abundance in the population II is about $\rm 2.2\,dex$ \citep[e.g.,][]{1993A&A...279L...9S}.
In the solar system, the Li abundance is about $\rm 3.3\,dex$ \citep[i.e., $C_{\rm acc}=10^{-8.7}$, see e.g.,][]{1999A&A...352..117R}. In addition, similar compositions are found in the interstellar medium \citep{1984A&A...138..303F, 2003ApJ...586..268K} and in the nearby star-forming region \citep{2006A&A...446..971J}. Each of the above possible scenarios could provide the source of matter accreted with excess Li. 
In terms of the Milky Way, due to the supplementation of AGB stars, nova systems, and cosmic ray effect with Li, the Milky Way is also rich in Li  \citep[e.g.,][]{2018A&A...610A..38F}. Whereas for the giants, the Li enhancement in our accretion models is also related to the Li content. 
In the case of $C_{\rm acc}=10^{-9}$ in $\rm Fig.\,\ref{f5}\,(2)$, the upper limit of the Li abundance is also close to $\rm 2.5\,dex$ as $R_{\rm acc}$ approaches $10^{-9}\,M_{\odot}\,\rm yr^{-1}$. Which is in agreement with the results of the mass effect caused by matter accretion in $\rm Sect.\,\ref{sect31}$. The composition of the accreted matter is consistent with the stellar surface during the FDU phase, i.e., Li is overabundant in the accreted matter, and the above source of circumstellar matter is capable of providing a similar composition.

\section{Discussion}\label{sect4}

\subsection{Validation of Observations}\label{sect41}
\begin{table}[htb]
	\caption{\label{t2}List of Li abundances of part giants with IR excesses.}
	\centering 
	\begin{tabular}{lrr}
		\hline \hline \noalign{\smallskip}
		Object & $A(\rm Li)\,[dex]$ & References	\\ 
		\hline \noalign{\smallskip}
		HD 30834 & 2.4 / 1.8 & 1, 2 \\
		HD 146850 & 2.0 / 1.6 & 1, 2 \\
		HD 157457 & 1.5  & 1 \\
		IRAS 17596-3952 & 2.3 & 3, 4  \\ 
		IRAS 19012-0747 & 2.5 & 5, 6\\
		IRAS 19285-0517 & 2.5 & 4, 7 \\
		TYC 7843-2018-1 & 1.6& 8 \\
		TYC 9112-00430-1 & 2.5 & 9\\
		\hline \noalign{\smallskip}
		HD 21078 & 1.3 & 2 \\
		HD 40359 & 1.3 & 2 \\
		HD 114182 & 1.0 &  2 \\
		HD 129456 & $-0.5$ & 2 \\
		HD 131530 & 1.3 & 2 \\
		HD 152786 & 1.3 & 1 \\
		HD 153687 & 0.2 & 2 \\
		HD 156061 & 0.8 & 2 \\
		HD 169689 & 1.0 & 1 \\
		HD 175492 & 1.3 & 1 \\
		HD 176884 & 1.2 &  1 \\
		\hline \noalign{\smallskip}
		HD 19745 & 3.4 & 3 \\
		HD 219025 & 2.9 / 3.3 & 1, 2\\
		HD 233517 & 4.0 / 4.3 & 10, 11 \\
		IRAS 13313-5838 & 3.1 & 12 \\
		IRAS 13539-4153 & 3.9 & 3 \\
		\hline
	\end{tabular}
\tablecomments{The references are as follows: 
(1) \citet{1999A&A...342..831J}, (2) \citet{1998AJ....116.2466F}, (3) \citet{2005AJ....129.2831R}, (4) \citet{2015A&A...577A..10B}, (5) \citet{2000A&A...364..674C}, (6) \citet{2006A&A...449..211P}, (7) \citet{2002AJ....123.1993R}, (8) \citet{2022MNRAS.511.3741M}, (9) \citet{2011ApJ...743..107R}, (10) \citet{2000ApJ...542..978B}, (11) \citet{2015A&A...574A..31S}, and (12) \citet{2002AJ....123.2703D}.}
\end{table}

Our accretion models provide a pathway for Li enhancement by accreting circumstellar matter at low accretion rates. This requires that the giants with Li enhancement should possess circumstellar matter and/or matter disks. Stars in such environments could exhibit the phenomenon of infrared (IR) excess \citep[e.g.,][]{1999A&A...342..831J, 2003ApJ...582.1032J}. The IR excess has been associated with the giants of Li overabundant \citep[e.g.,][]{1997ApJ...482L..77D,1999A&A...342..831J,2005AJ....129.2831R,2015A&A...577A..10B, 2015AJ....150..123R, 2025A&A...693A..98D}. $\rm Table\,\ref{t2}$ lists the Li abundances of a selection of giants with IR excesses. Based on the Li abundance values we divide them into three fractions, i.e., $<1.5$, $1.5-2.5$, and $\rm >2.5\,dex$. The Li abundances of giants with IR excesses are spread over a wide range, indicating that both significant and slight Li enhancement can lead to the IR excess. As can be seen in $\rm Fig.\,\ref{f3}$, the `Li enhancement' induced by matter accretion can be significant or slight.

However, \citet{1996ApJ...456L.115D} related the Li enhancement of the giants by the mass loss, and stated that the physical processes that cause Li enhancement excite the loss of mass from the giants, resulting to an observed IR excess. Our results align with this perspective but differ in some aspects. Our accretion models suggest that the `Li enhancement' of the giants is driven by the increase in mass during the FDU, whereas the mass loss dominates in the later phase. The increase in mass during the previous period will not directly trigger the mass loss behavior.

Different routes to the same destination, our models also have a significant loss of mass. With the help of the corrected Reimers' law of mass loss \citep[see][]{2005ApJ...630L..73S}, \cite{2015A&A...577A..10B} found the giants with $\dot{M}$ in the range of  $\sim 10^{-9}-10^{-7}\,M_{\odot}\,\rm yr^{-1}$ show IR excesses, which is consistent with our predicted mass loss rate (i.e., $\dot{M}\sim10^{-8}\,M_{\odot}\,\rm yr^{-1}$).
From the above, the giants with IR excess phenomena may be a good corroboration of the accretion model we have developed.

\subsection{Accretion from Main Sequence}\label{sect42}
\begin{figure*}
	\centering
	\includegraphics[width=18cm]{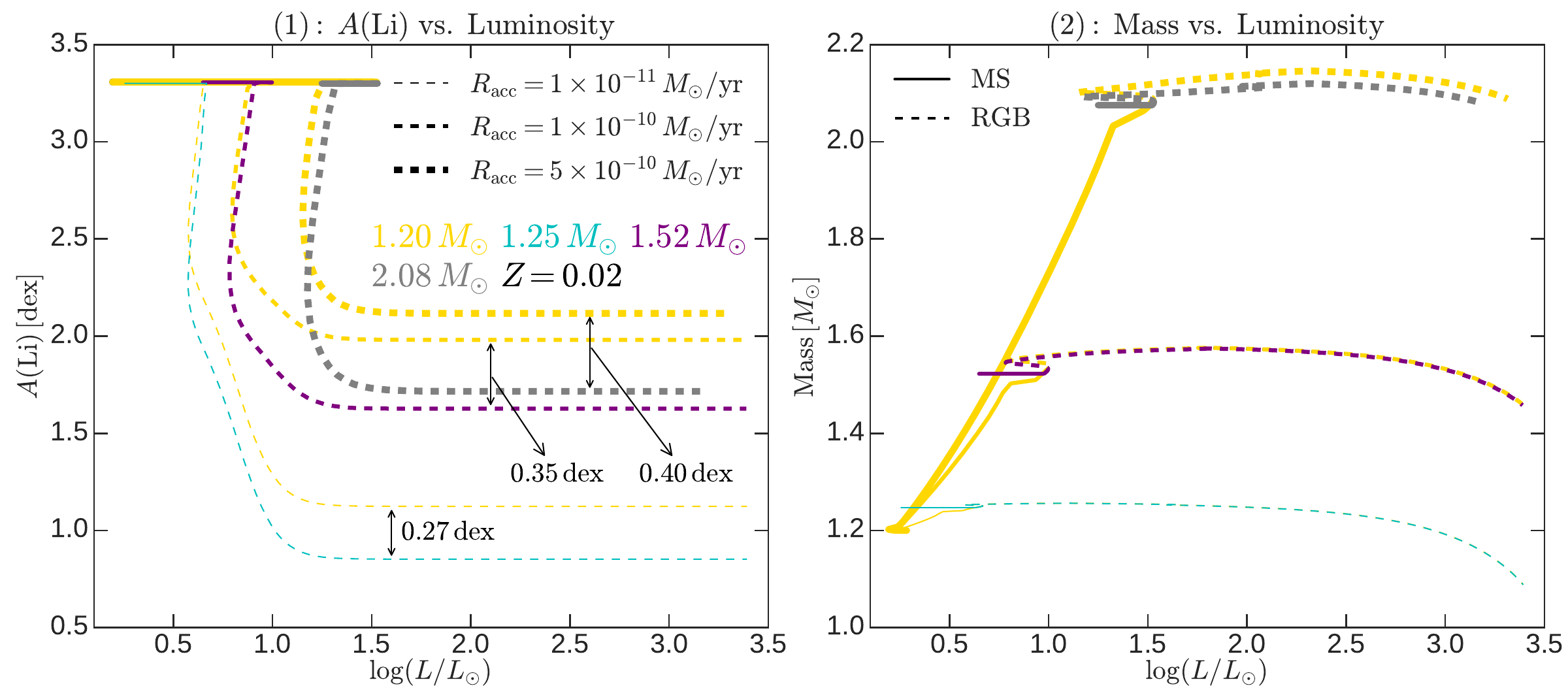}
	\caption{Similar to $\rm Fig.\,\ref{f1}$, but the gold lines indicate that accretion starts from the ZAMS, while the cyan, purple, and gray lines indicate that accretion starts from the MS turnoff. Lines of the same thickness indicate that they have the same accretion rate, i.e., $R_{\rm acc}=$ $10^{-11}$, $10^{-10}$, $5\times10^{-10}\,M_{\odot}\,\rm yr^{-1}$ from thin to thick. For the three accretion rates, a $1.2\,M_{\odot}$ star and $Z=0.02$ increases in mass to 1.25, 1.52, and $2.08\,M_{\odot}$, respectively, during the MS.}
	\label{f6}%
\end{figure*}

As can be seen in $\rm Fig.\,\ref{f1}$ and $\rm Table\,\ref{t1}$, the timing of the onset of matter accretion has different effects on the `Li enhancement' of giants. The positive effect of the accretion process starting from the post-MS on the Li abundance is reflected in the suppression of the dilution effect in the convective envelope, and the increasing stellar mass leading to a weakening of the Li loss during the FDU. It is clear that after the RGB bump, due to the short evolution time, the mass increase induced by the weak accretion does not have a significant effect on the Li abundance, even if the $C_{\rm acc}$ of the accreted matter is relatively high.

This subsection will discuss the effect of matter accretion from the MS onwards on the increase of Li abundance. Here again, we refer to the setup of $\rm  Sect.\,\ref{sect31}$ and take $C_{\rm acc}$ = $C_{\rm e}$. $\rm Fig.\,\ref{f6}$ shows the evolution of Li abundance and mass as a function of  luminosity for the two modes of accretion from the ZAMS (the gold lines) and accretion from the MS turnoff (the non-gold lines). The thickness of the lines indicates the different accretion rates. The $1.2\,M_{\odot}$ and $Z=0.02$ stars in three accretion rates ($R_{\rm acc}=10^{-11}, 10^{-10},\ \rm and\  5\times10^{-10}$$\,M_{\odot}\rm\, yr^{-1}$) increase their mass from the ZAMS up to the MS turnoff, to 1.25, 1.52, and $2.08\,M_{\odot}$, respectively. Accretion starting from the ZAMS can form giants with higher Li abundances than those starting from the MS turnoff, and higher $R_{\rm acc}$ induce stronger `Li enhancement'. The difference between the two patterns is reflected in the varying degrees of Li loss between the MS turnoff and the luminosity inflection point. The accretion starting from the MS turnoff has a more pronounced Li loss. During the MS the star has accrued matter of the same component, and although $M_{\rm acc}$ is relatively large at this stage, $M_{\rm e}$ is still low due to adjustments in the stellar structure. The bulk of the accreted matter is incorporated into the stellar interior, therefore they do not form a convective zone. After the MS turnoff, the dilution effect on the convective envelope due to expansion recedes, because the high Li composition matter is incorporated into the stellar interior close to convective zone during the MS. Whereas, the accretion pattern from the MS turnoff onwards does not suffer from the above situation (see $\rm Fig.\,\ref{f1}\,(1)$), since there is no accretion on the MS. After the luminosity inflection point, the behavior of the surface Li follows an analogous depletion in the two patterns (i.e., accretion from both the MS the MS turnoff).

Overall, in the time dimension, as the onset of accretion becomes later, the gain of matter accretion on the Li abundance for the giants shows signs of abating. During the MS, the larger $M_{\rm acc}$ is, the more it suppresses the Li loss during the FDU. In the subgiant branch, accretion from the luminosity inflection point onwards misses the suppression of dilution of Li before that point. In the late RGB phase, low $M_{\rm acc}$ does not have a striking effect on the Li abundance.

\subsection{Differences between Model Evolution and Observation}

It can be found from $\rm Figs.\,\ref{f1}\ and\ \ref{f5}$ that the accretion models predict an almost constant evolution pattern of Li abundance after the RGB bump. However, at this stage the observed Li abundance shows a downward trend \citep[e.g.,][]{2019MNRAS.484.2000D, 2020NatAs...4.1059K}. Limited by convective mixing, our accretion models cannot reproduce this trend of Li evolution. The results of \citet{2020NatAs...4.1059K} show that from the RGB bump to the RGB tip, the Li abundances of giants are mainly distributed in the range of $\rm +0.7$ to $\rm -0.5\,dex$, i.e., there is a decrease of about one order magnitude. 

Assuming an uneven mass/density distribution of the accreted matter, when a star evolves to the stage beyond the RGB bump, the decline in Li abundance could occur by accreting more matter with low $C_{\rm acc}$ to dilute the convective envelope.  A noteworthy example is shown in $\rm Table\,\ref{t1}$, where the Li abundance decreases by $\rm 0.3\,dex$ when about $1.0\,M_{\odot}$ of matter with $C_{\rm acc}=10^{-12}$ are accreted. Prior to this, the $C_{\rm acc}$ (approximately $10^{-8.7}$ to $10^{-10}$) of the accreted matter is the same as that of the stellar surface. For this situation, the distribution of Li content in the circumstellar matter should exhibit a sharp decline from the interior to the exterior. This assumption may help to explain the observed distribution, but it requires further observations to support it.

\subsection{Mass Parameter}

As has been noted in recent literature, mass is the main parameter determining what should be considered normal and what should be labeled anomalous in the case of lithium by giant stars \citep[e.g.,][]{2022ApJ...933...58C}.
For example, in the case of star clusters, the masses of its member stars are usually determined collectively (from the isochron that best fits the cluster as a whole) rather than independently for each star. Here, a giant star that has experienced a mass increase like the one considered in current work (say mass M2) would actually be incorrectly labeled as mass M1 (M2$>$M1, and M1 is the mass of the red giant in the cluster). If Li is detected on its surface after the FDU, it will likely be labeled as an anomaly (Li-rich). On the other hand, for a field giant that has undergone this work scenario and happens to be the object from Kepler or TESS asteroseismic measurement, the seismic mass may reflect its current mass (M2) and therefore may be considered normal based on its post-FDU Li content. 
 
In our view, the `Li enhancement' mechanism in current work will be applicable for above two cases. 
On the one hand, because the scenario we build is to achieve `Li enhancement' by increasing mass, the improve in surface Li is therefore a natural consequence of the way cluster stars determine their mass.
On the other hand, most or a large fraction of the current observed work on Li-rich giants has been done on field giants, aided by asteroseismic masses and evolutionary states.
For the field giants, the applicability of the above scenario will not be affected by precise determination of mass. 
Within a specific range, there is a positive correlation between the mass increase and the extent of `Li enhancement', potentially leading to the formation of a Li-rich giant.
In \(\rm Fig.\,\ref{f4}\), it is easy to see that stars with masses increasing from the initial mass M0 to M2 show a higher surface Li abundance than stars with masses of M2 that have not been accreted. 
In \(\rm Fig.\,\ref{f1}\), in addition, a star with mass increases to $1.6\,M_\odot$ can differ in Li abundance by about $\rm 1.0\,dex$ from a star with that mass itself, which can be labeled the Li-rich giants. But for a star that increases its mass to \(1.3\,M_{\odot}\), the Li abundance will still be marked as normal (which will be the normalcy of the matter accretion mechanism), despite the `Li enhancement' occurring. The natural consequence is, for field giants whose mass is determined by astroseismology, this `Li enhancement' scenario may increase the Li abundance value of normal stars as a whole, by \(\sim 0.5\,\rm dex\) for lower mass stars and by \(\sim 0.1\,\rm dex\) for higher mass stars (see \(\rm Fig.\,\ref{f3}\)).

Therefore, for the current scenario to be seen as enhancing Li, it will not depend on how we, as observer, determine the mass of the star.

\section{Conclusion}\label{sect5}
The effect of matter accretion on the Li enhancement behavior of low-mass giants is assessed and the conclusions of the current work are as follows:

\begin{enumerate}
	\item Weak accretion of circumstellar matter is a possible channel to `Li enhancement' for giants; this accretion predicts an upper limit on the Li abundance of $\rm \sim 2.5\,dex$. However, this requires the star's mass to increase to around $2.2\,M_\odot$.
	
	\item By analyzing the sources of matter, we have found that AGB stars and binary systems are possible donors. For AGB star donors, the accretion mass is on the order of $0.1\,M_\odot$. Therefore, the upper limit of $2.5\,\rm dex$ predicted by the model may be non-conservative. The amount of matter transfer in a binary system is sufficient.  However, the enhancement effect on Li from rapid accretion of substantial matter requires further evaluation.
	
	\item Matter accretion can suppress the dilution of surface Li for a star during the FDU, thus exhibiting `Li enhancement' behavior.
	
	\item The extent of `Li enhancement' is closely related to the mass and composition of the accreted matter. Accreting more matter causes more pronounced `Li enhancement'. Higher proportions of the number densities of Li to hydrogen in the accretion also inspires stronger `Li enhancement'.
	
	\item The increment in Li abundance increases significantly with metallicity, and our accretion models have a more significant `Li enhancement' effect for stars with lower masses and higher metallicities.
	
	\item The earlier the onset of accretion, the more pronounced the effect on `Li enhancement'.
	
	\item The accretion model we developed can be supported by the observations of giants with IR excesses.
\end{enumerate}

\begin{acknowledgments}
	
The work is supported by National Natural Science Foundation of China (grant Nos: 11973079, 12288102, 12133011, 11973052, 12022304, 12090040, 12090044, 12173080, 12273104, and 12373036), the National Key R\&D Program of China Nos. 2021YFA1600400, 2021YFA1600402, the Natural Science Foundation of Yunnan Province (grant No. 202201AT070158), and the Yunnan Fundamental Research Projects (grant No. 202401AS070045). H.-L. Y. acknowledges support from the Youth  Innovation Promotion Association of the CAS and the NAOC Nebula Talents Program. J.-H. Z. acknowledges support from NSFC grant No. 12103063 and from China Postdoctoral Science Foundation funded project (grant No. 2020M680672). F. G. acknowledges support from the Yunnan Province Foundation (grant No. 202301AU070160).

\end{acknowledgments}

\software{MESA-r11701  \citep{2011ApJS..192....3P, 2013ApJS..208....4P, 2015ApJS..220...15P, 2018ApJS..234...34P, 2019ApJS..243...10P}
          }


\appendix
\section{Modeling Settings}\label{AA}
With the help of MESA-r11701, we construct the matter accretion model. Some basic information is shown in Table $\rm \ref{t3}$, and a more detailed description is provided in $\rm Sect.\,\ref{sect2}$.

\begin{table}[htb]
	\caption{\label{t3}The basic information of the model.}
	\centering 
	\begin{tabular}{lcl}
		\hline \hline \noalign{\smallskip}
		Item & Value & Description	\\ 
		\hline \noalign{\smallskip}
		MESA Version & 11701 & \citet{2011ApJS..192....3P, 2013ApJS..208....4P, 2015ApJS..220...15P, 2018ApJS..234...34P, 2019ApJS..243...10P}\\
		The Equation of State & - & \citet{2002ApJ...576.1064R}\\
		Opacity Table & - & 
		\citet{1993ApJ...412..752I,1996ApJ...464..943I}\\
		Chemistry Composition & \texttt{GS98} & \cite{1998SSRv...85..161G}\\
		Nuclear Reaction Rate & NACRE, CF88 & \citet{1999NuPhA.656....3A}, \citet{1988ADNDT..40..283C} \\
		Nuclear Reaction Net & \texttt{pp.extras.net} & pp chain\\
		Atmospheric Boundary & \texttt{simple\_atmosphere} & $\tau=\frac{2}{3}$ \\
		Convective Zone Boundary & $\nabla = \nabla_{\rm ad}$ & Schwarzschild boundary \\
		\hline \noalign{\smallskip}
		Input Mass  & $1.2\,M_{\odot}$ & See \citet{2022ApJ...931..136Z}\\
		Input Metallicty ($Z$)  & 0.02 & See \citet{2021MNRAS.505.5340M}\\
		Input Li Abundance & $3.3\,\rm dex$ & Meteorite abundance\\
		\hline \noalign{\smallskip}
		Evolution Stage & - & From zero-age main sequence to the RGB tip\\
		Zero-Age Main Sequence & - & MESA setting: \texttt{stop\_near\_zams = .true.}\\
		Main Sequence Turnoff & - & MESA setting: \texttt{xa\_central\_lower\_limit\_species(1) = 'h1'} \\ & & \ \ \ \ \ \ \ \ \ \ \ \ \ \ \ \ \ \ \ \ \ \texttt{xa\_central\_lower\_limit(1) = 1d-9} \\
		Tip of Red Giant Branch & - & MESA setting: \texttt{power\_he\_burn\_upper\_limit = 10}\\
		\hline \noalign{\smallskip}
		Convection Mixing& - &MLT \\
		$\alpha_{\rm MLT}$ &2.00& Default value\\
		Convection Overshooting & - & It operates on the surface of a star and $f_{\rm ov}=0.80$ \\
		Other Mixing Processes & - & - \\
		\hline \noalign{\smallskip}
		Matter Accretion & - & Consider the accretion of circumstellar matter \\
		Preset Accretion Rate & ($2.2\,M_{\odot}-1.2\,M_{\odot}$)/Evolution Timescale & The estimated average accretion rate \\
		Accretion Rate Grid & Based on the Preset Accretion Rate & The grid settings are shown in $\rm Fig.\, \ref{f3}$\\
		Accretion Stage & - & RGB or Main sequence + RGB (see $\rm Sect.\,\ref{sect42}$) \\
		Accretion Composition & - & 1 Composition of a star's surface\\ & & 2 See Appendix \ref{AB} \\
		\hline \noalign{\smallskip}
		Mass Loss & $\dot{M}\propto\eta L R /M$ &  Reimers' empirical formulae \citep{1975MSRSL...8..369R}  \\
		Mass Loss Stage& - &  RGB \\
		\hline
	\end{tabular}
\end{table}

\section{Input Settings for the Composition of Accreted Matter}\label{AB}

Referring to the solar composition\footnote{\url{https://opalopacity.llnl.gov/pub/opal/type1data/GN93/ascii/GN93hz}}, we list  the composition proportions of 23 elements in the accreted matter, including H, He, Li, Be, C, N, O, Ne, Na, Mg, Al, Si, P, S, Cl, Ar, K, Ga, Ti, Cr, Mn, Fe, and Ni.

Inlist file of matter accretion in the RGB stage as follow:

\begin{verbatim}	
	&star_job
	
	show_log_description_at_start = .false.
	
	load_saved_model = .true.
	saved_model_name = 'to_msto.mod'
	
	save_model_when_terminate = .true.
	save_model_filename = 'to_trgb.mod'
	
	kappa_file_prefix = 'gs98'
	
	pgstar_flag = .false.
	
	/ ! end of star_job namelist
	
	&controls
	
	! when to stop
	!log_L_upper_limit = 2.0
	power_he_burn_upper_limit = 10
	
	! overshooting
	overshoot_f0_above_nonburn_shell = 0.004
	overshoot_f_above_nonburn_shell = 0.80 
	
	! accreted material chemical composition
	accrete_same_as_surface = .false.
	accrete_given_mass_fractions = .true.
	num_accretion_species = 23
	accretion_species_id(1) = 'h1'
	accretion_species_xa(1) = 0.700
	accretion_species_id(2) = 'he4'
	accretion_species_xa(2) = 0.280
	accretion_species_id(3) = 'li7'
	accretion_species_xa(3) = 49d-10      !N(Li)/N(H)=1e-9
	accretion_species_id(4) = 'be7'
	accretion_species_xa(4) = 1d-99
	accretion_species_id(5) = 'c12'
	accretion_species_xa(5) = 3.4657d-3   !0.173285*0.02
	accretion_species_id(6) = 'n14'
	accretion_species_xa(6) = 1.06304d-3  !0.053152*0.02
	accretion_species_id(7) = 'o16'
	accretion_species_xa(7) = 9.64544d-3  !0.482272*0.02 !!0.482273
	accretion_species_id(8) = 'ne20'
	accretion_species_xa(8) = 1.97336d-3  !0.098668*0.02
	accretion_species_id(9) = 'na23'
	accretion_species_xa(9) = 3.998d-5    !0.001999*0.02
	accretion_species_id(10) = 'mg24'
	accretion_species_xa(10) = 7.5146d-4  !0.037573*0.02
	accretion_species_id(11) = 'al27'
	accretion_species_xa(11) = 6.476d-5   !0.003238*0.02
	accretion_species_id(12) = 'si28'
	accretion_species_xa(12) = 8.104d-4 !0.040520*0.02
	accretion_species_id(13) = 'p30'
	accretion_species_xa(13) = 7.1d-6   !0.000355*0.02
	accretion_species_id(14) = 's32'
	accretion_species_xa(14) = 4.2284d-4!0.021142*0.02
	accretion_species_id(15) = 'cl35'
	accretion_species_xa(15) = 9.12d-6  !0.000456*0.02
	accretion_species_id(16) = 'ar40'
	accretion_species_xa(16) = 1.0758d-4!0.005379*0.02
	accretion_species_id(17) = 'k39'
	accretion_species_xa(17) = 4.2d-6   !0.000210*0.02
	accretion_species_id(18) = 'ca40'
	accretion_species_xa(18) = 7.468d-5 !0.003734*0.02
	accretion_species_id(19) = 'ti48'
	accretion_species_xa(19) = 4.22d-6  !0.000211*0.02
	accretion_species_id(20) = 'cr52'
	accretion_species_xa(20) = 2.01d-5  !0.001005*0.02
	accretion_species_id(21) = 'mn55'
	accretion_species_xa(21) = 1.096d-5 !0.000548*0.02
	accretion_species_id(22) = 'fe56'
	accretion_species_xa(22) = 1.43588d-3!0.071794*0.02
	accretion_species_id(23) = 'ni59'
	accretion_species_xa(23) = 8.918d-5 !0.004459*0.02
	
	! mass gain or loss
	mass_change = 1d-10 ! unit Msun/year
	cool_wind_RGB_scheme = 'Reimers' 
	Reimers_scaling_factor = 0.5    
	cool_wind_AGB_scheme = 'Blocker' 
	Blocker_scaling_factor = 0.1     
	RGB_to_AGB_wind_switch = 1d-4
	
\end{verbatim}

\section{Supplementary Visual Evidence}\label{AC}

\begin{figure}
	\centering
	\includegraphics[width=9cm]{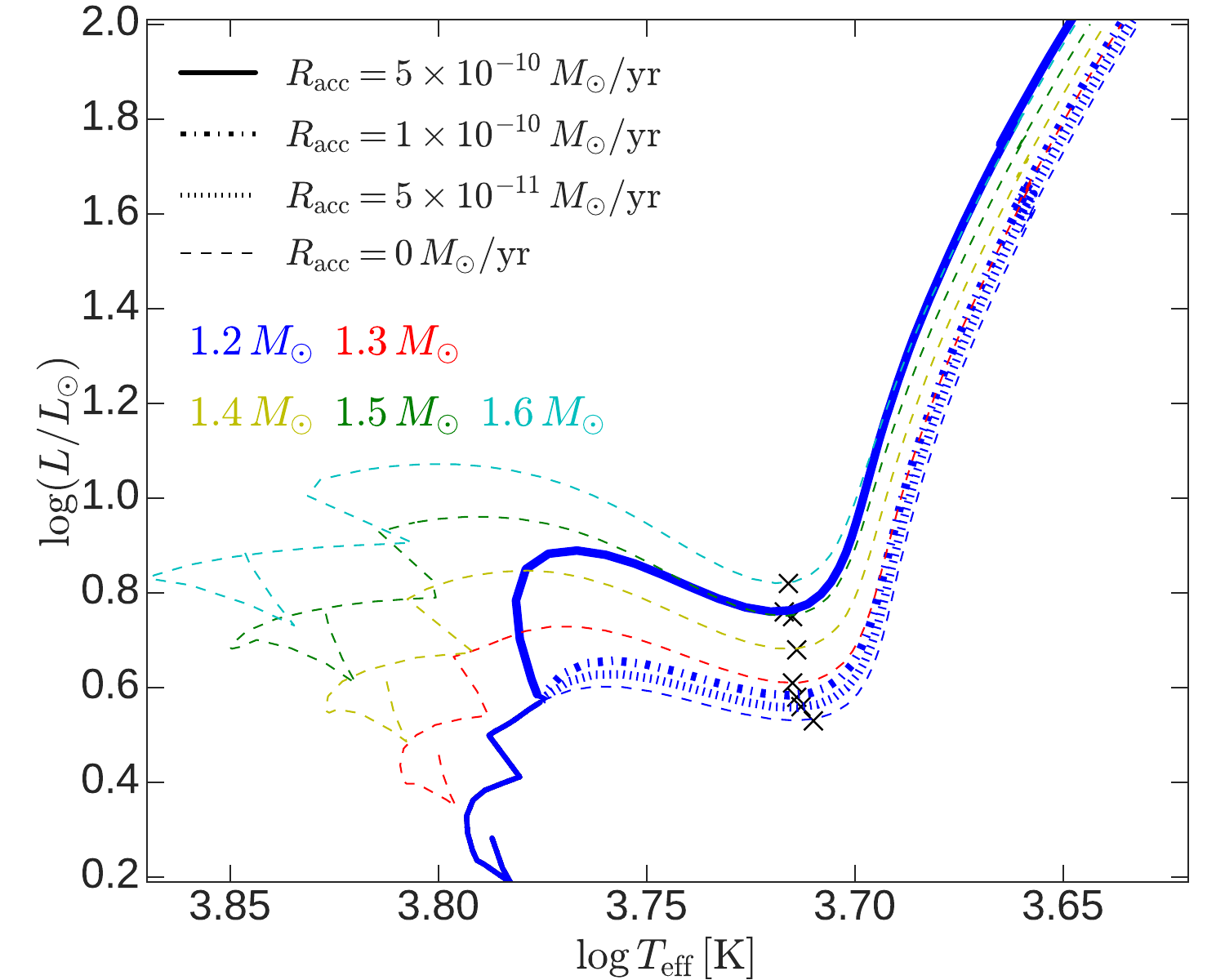}
	\caption{Hertzsprung-Russell diagram from the ZAMS to $100\,L_\odot$. All the marks in the figure are consistent with those in $\rm Fig.\,\ref{f1}$.}
	\label{a0}%
\end{figure}

$\rm Fig. \,\ref{a0}$ displays the Hertzsprung-Russell diagram corresponding to $\rm Fig. \,\ref{f1}$. To enable direct comparison with $\rm Fig.\,\ref{f2}$ in our main analysis, we only show the information from the ZAMS to $100\,L_\odot$.

\begin{figure*}[hbt]
	\centering
	\includegraphics[width=5.5cm]{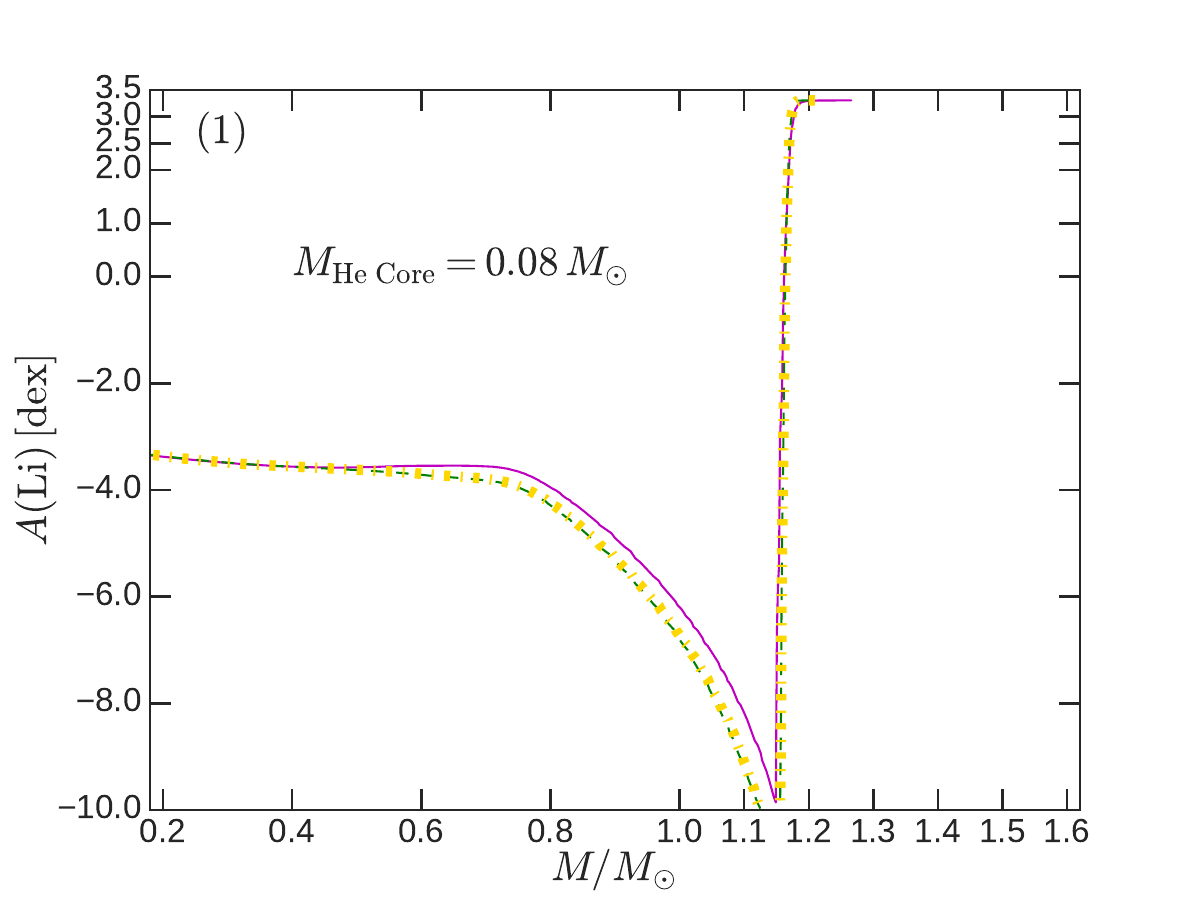}
	\includegraphics[width=5.5cm]{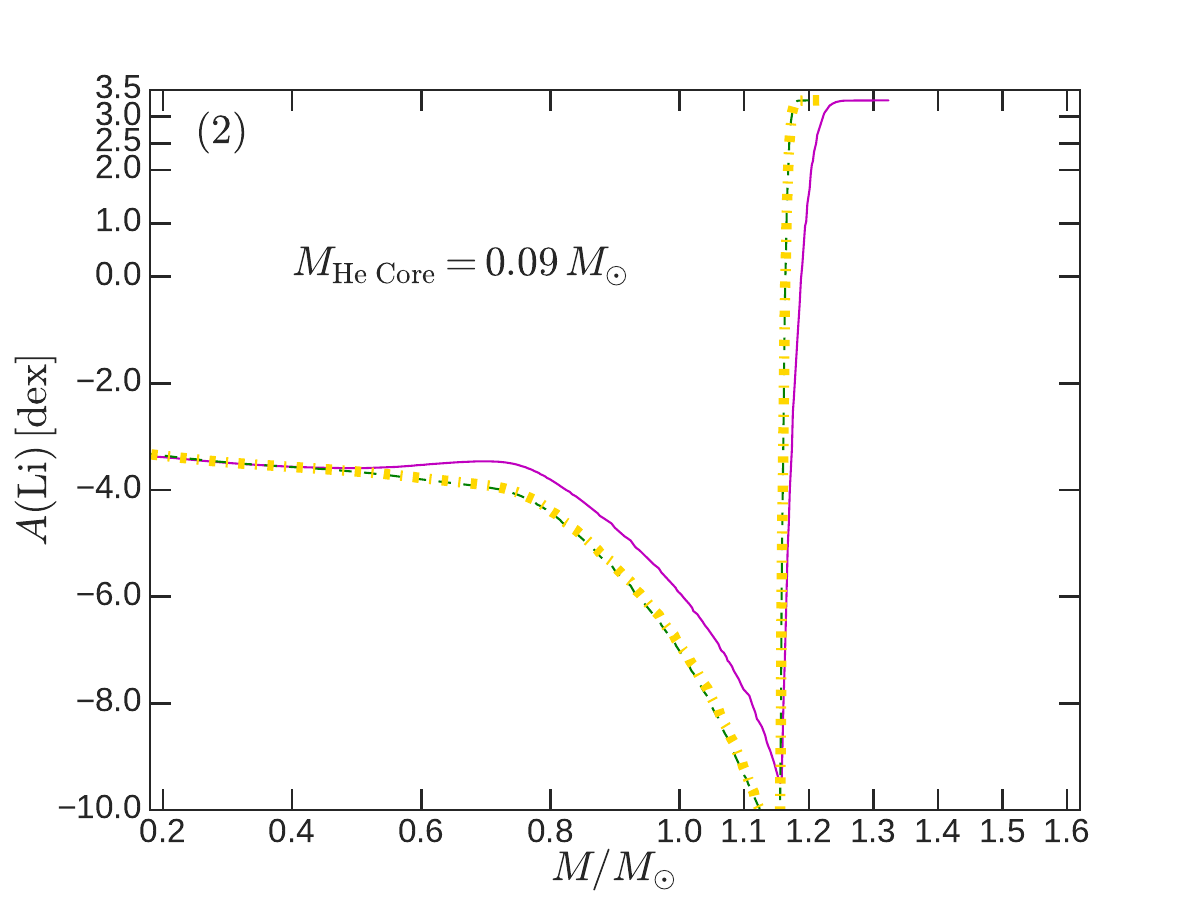}
	\includegraphics[width=5.5cm]{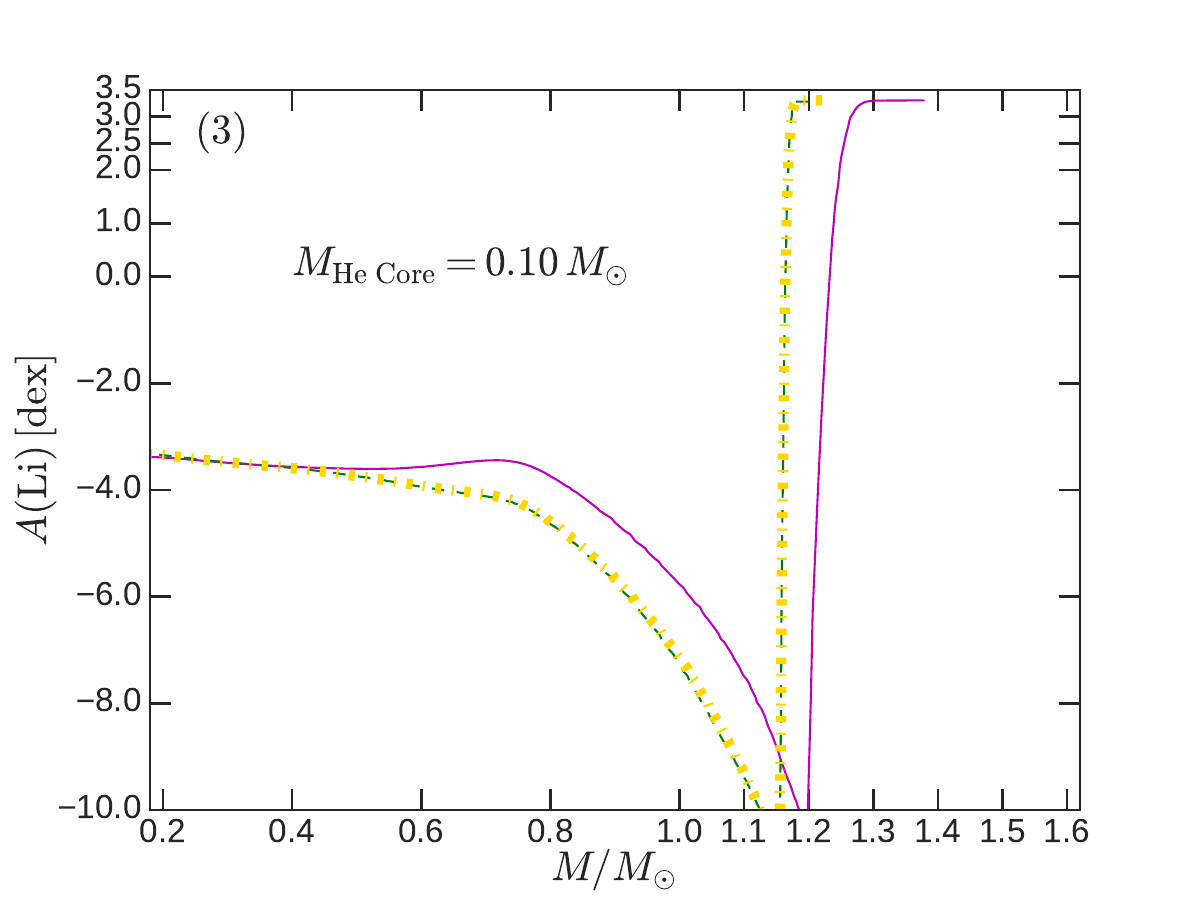}
	\includegraphics[width=5.5cm]{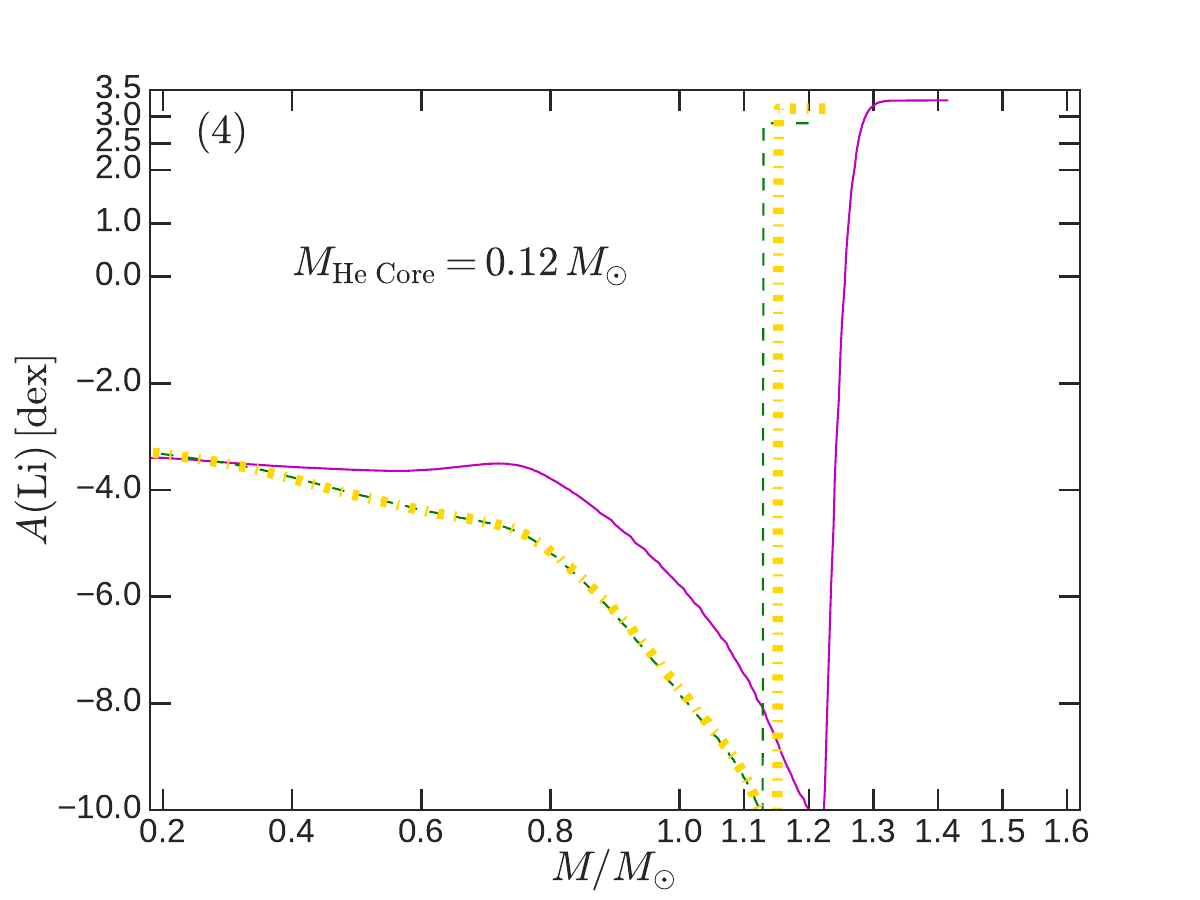}
	\includegraphics[width=5.5cm]{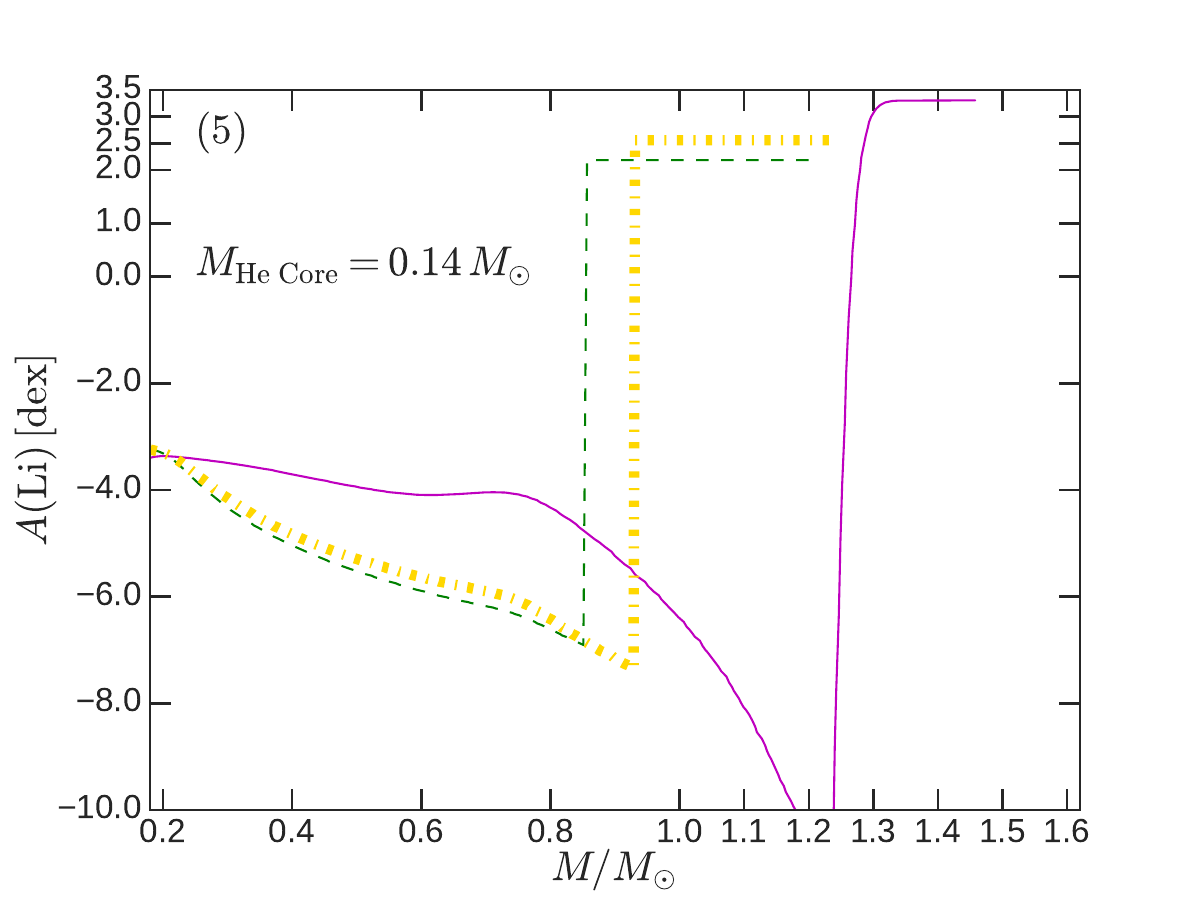}
	\includegraphics[width=5.5cm]{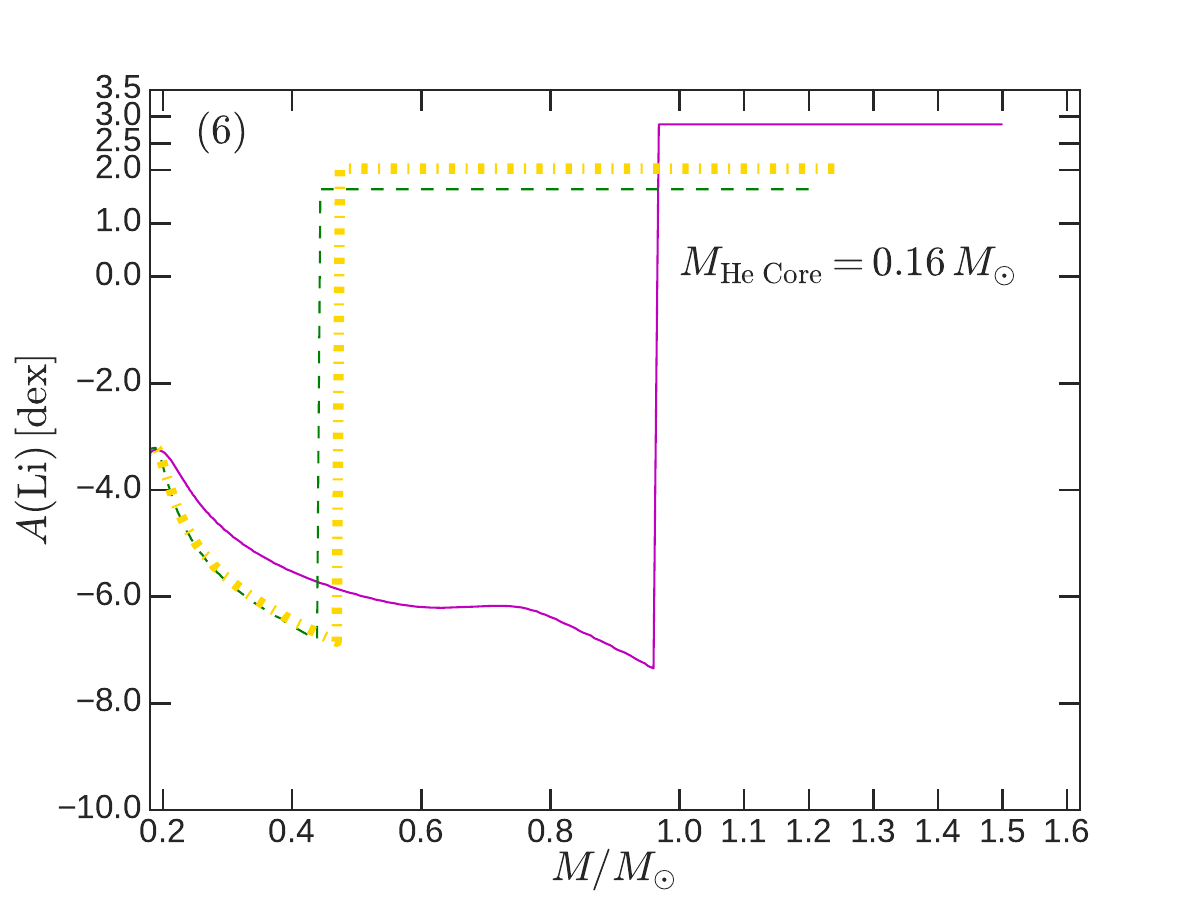}
	\includegraphics[width=5.5cm]{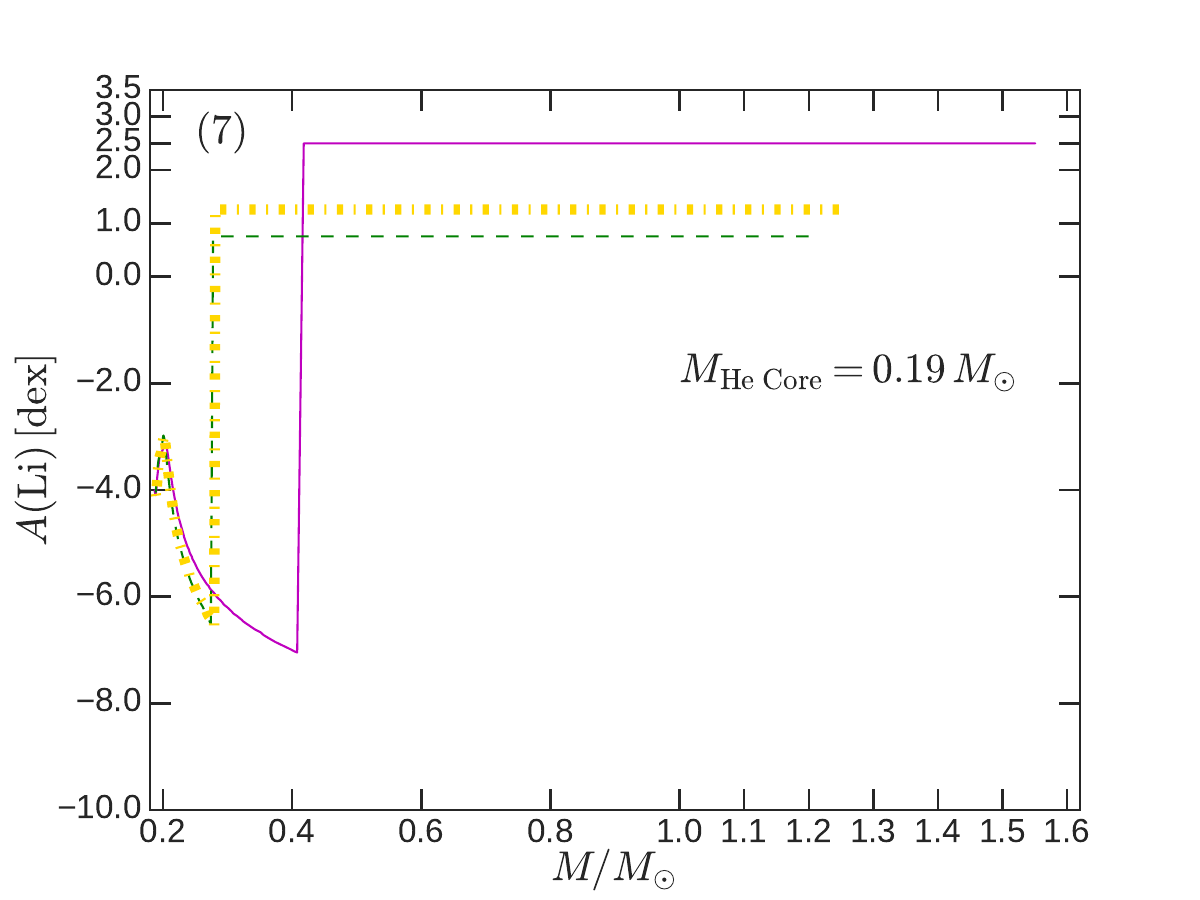}
	\includegraphics[width=5.5cm]{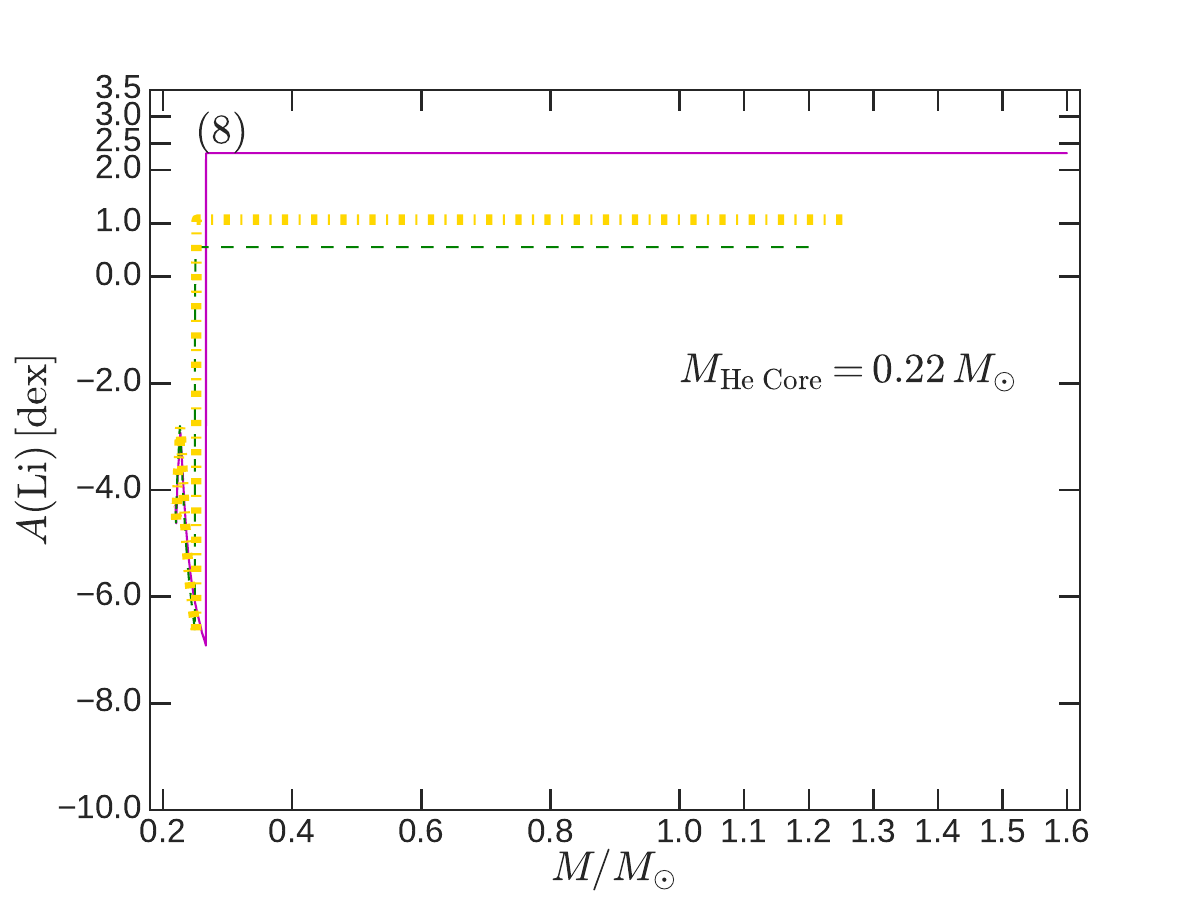}
	\includegraphics[width=5.5cm]{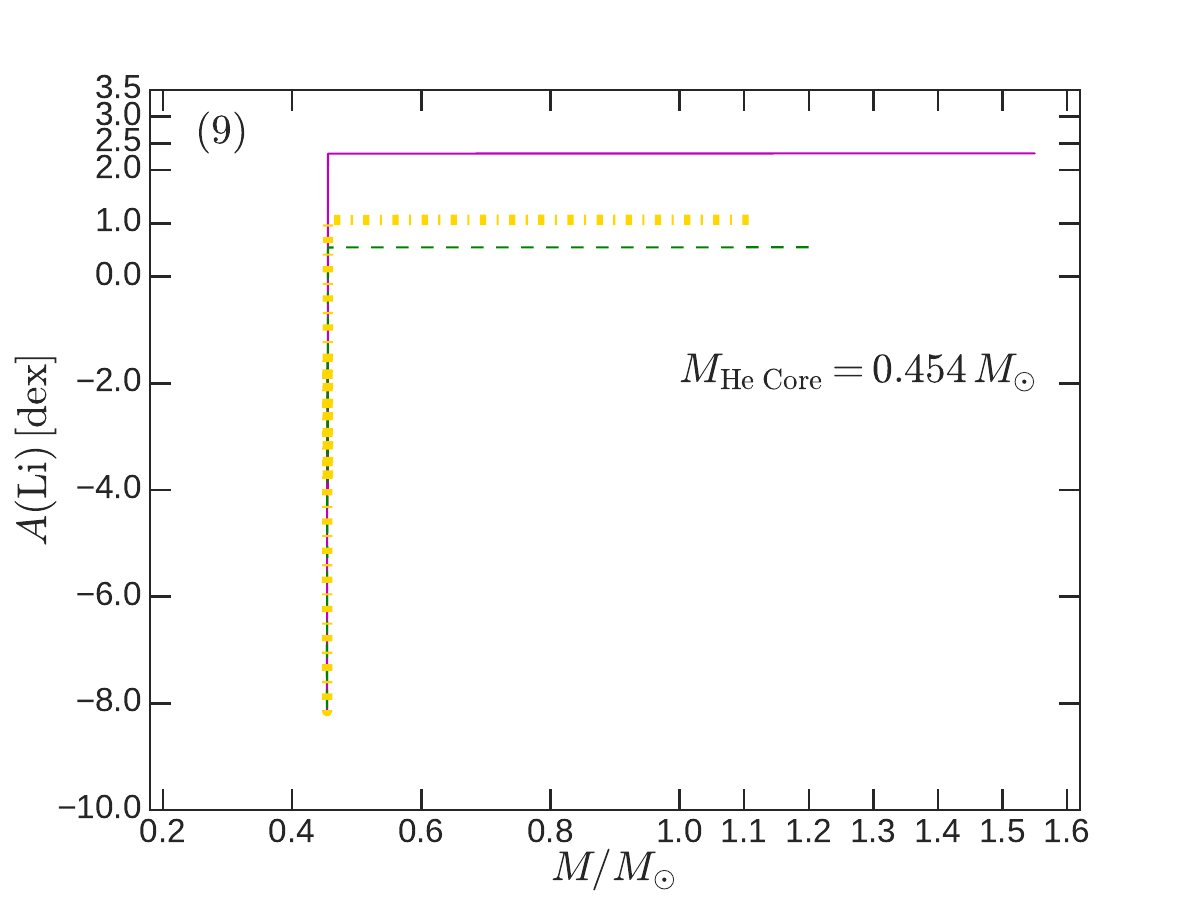}
	\includegraphics[width=9cm]{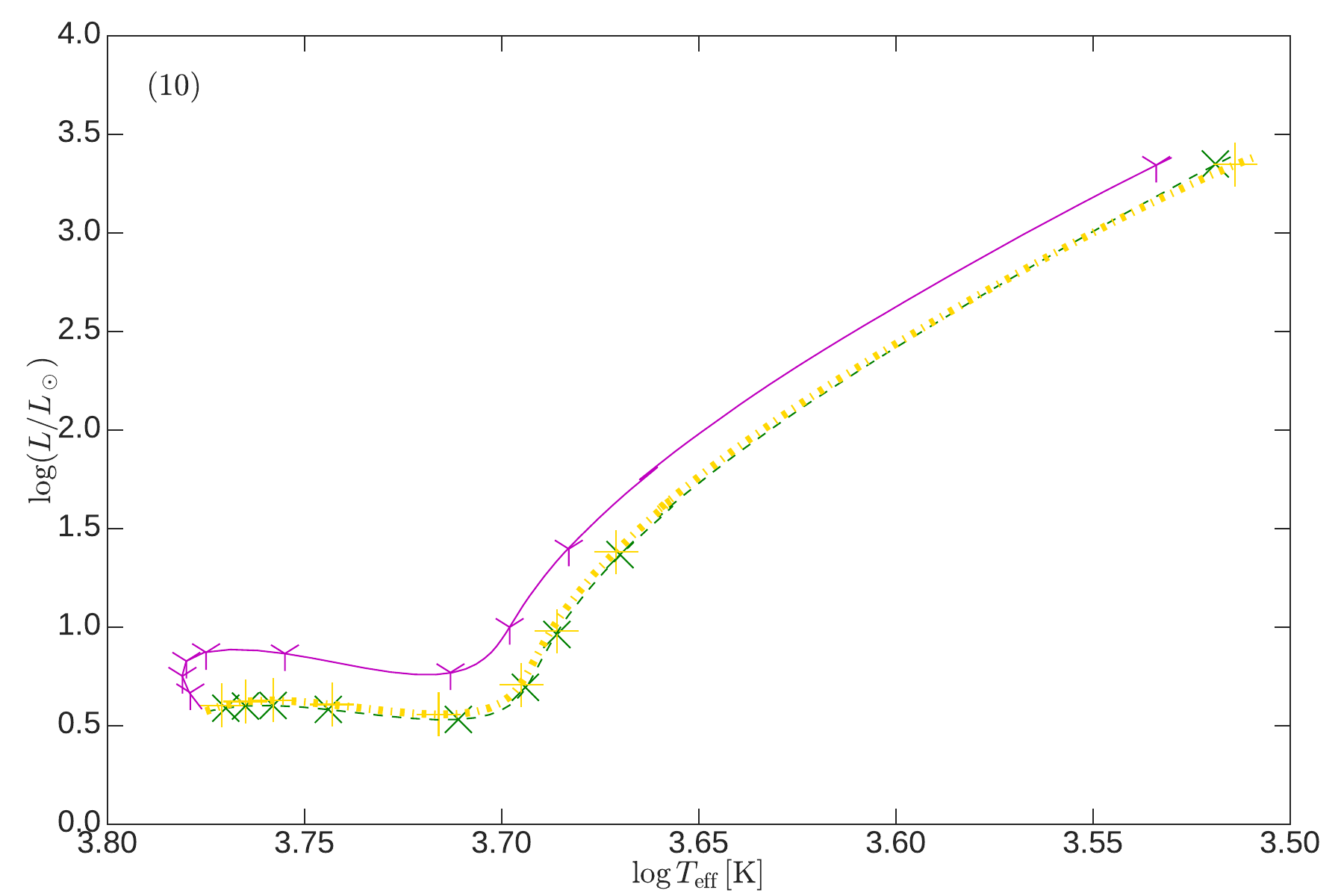}
	\caption{The radial distribution profile of Li abundance at different times. The input mass and metallicity of the stellar model are $1.2\,M_\odot$ and 0.02, respectively. The red dashed lines: The accretion rate is 0, the golden dot-dashed lines: the accretion rate is $5\times10^{-11}\,M_\odot\,\rm yr^{-1}$; the magenta solid lines: the accretion rate is $5\times10^{-10}\,M_\odot\,\rm yr^{-1}$. Panel 10 shows the positions of panels 1 to 9 on the Hertzsprung-Russell diagram, and the marked symbols along the evolutionary trajectory correspond one-to-one with the other 9 subgraphs.}
	\label{a1}%
\end{figure*}

$\rm Fig. \,\ref{a1}$ shows the distribution of Li abundance within stars at different accretion rates. The results indicate that the distribution characteristics of Li within stars are similar at different accretion rates.

\begin{figure*}[hbt]
	\centering
	\includegraphics[width=8.2cm]{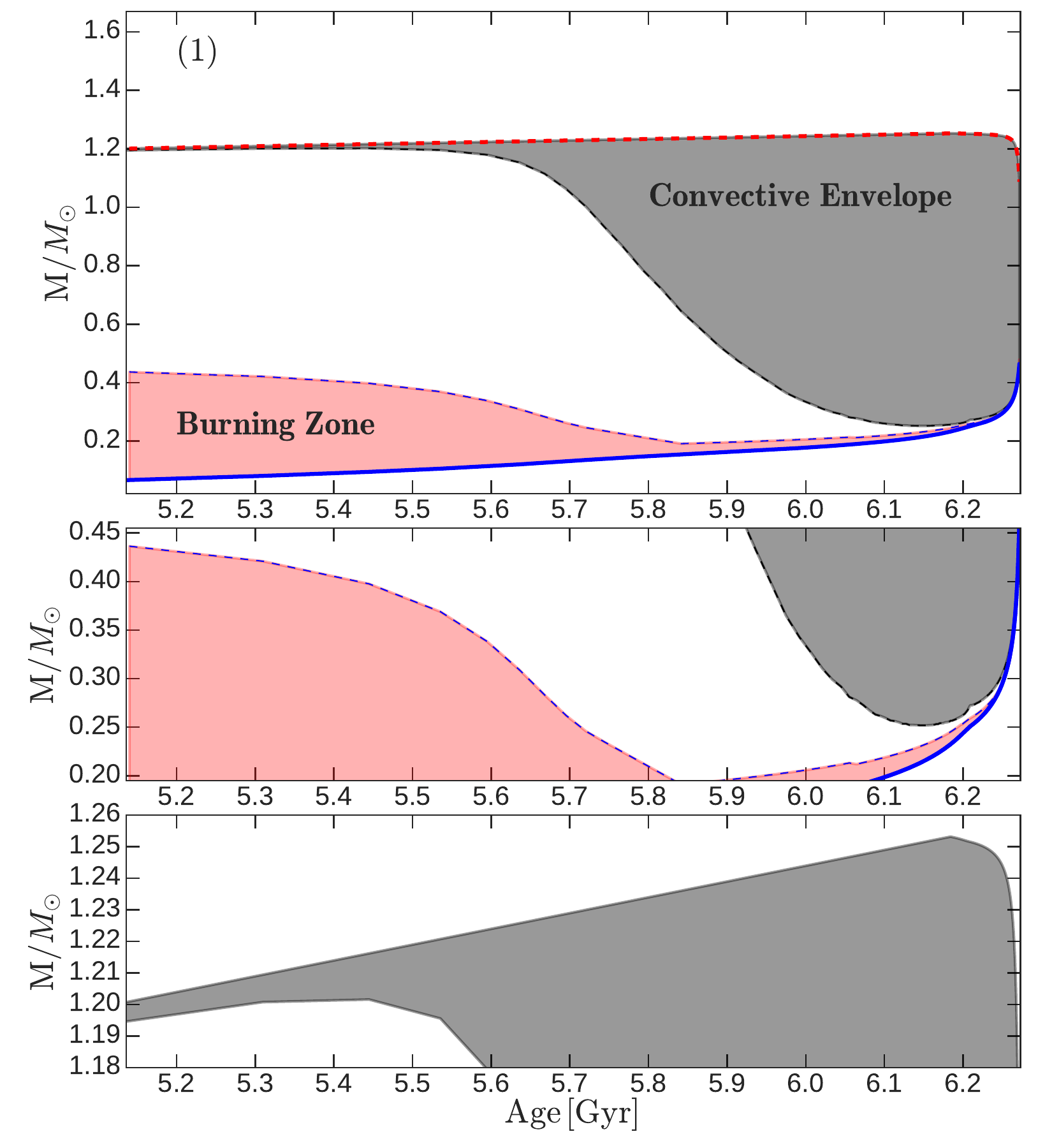}
	\includegraphics[width=8.2cm]{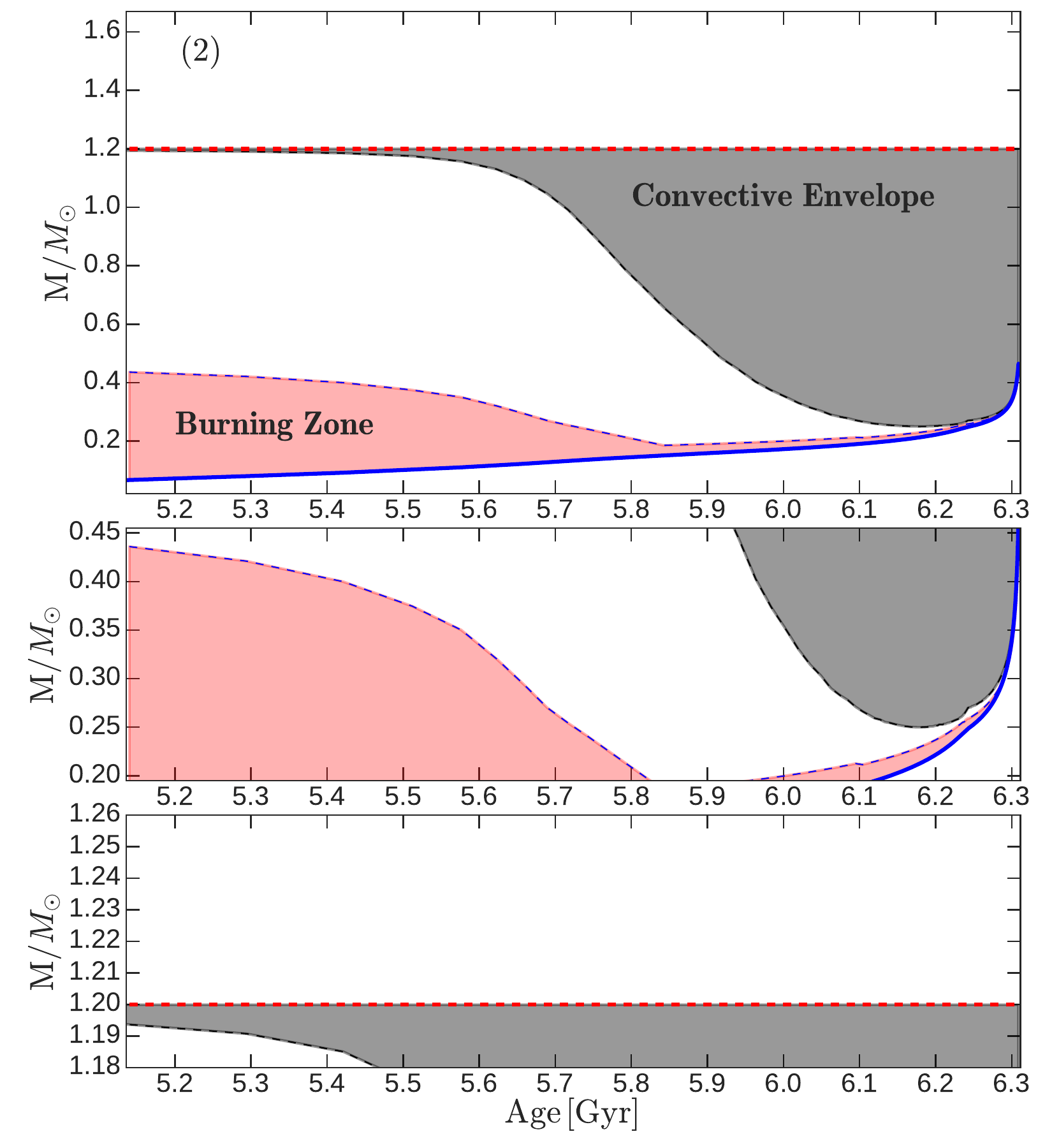}
	\includegraphics[width=8.2cm]{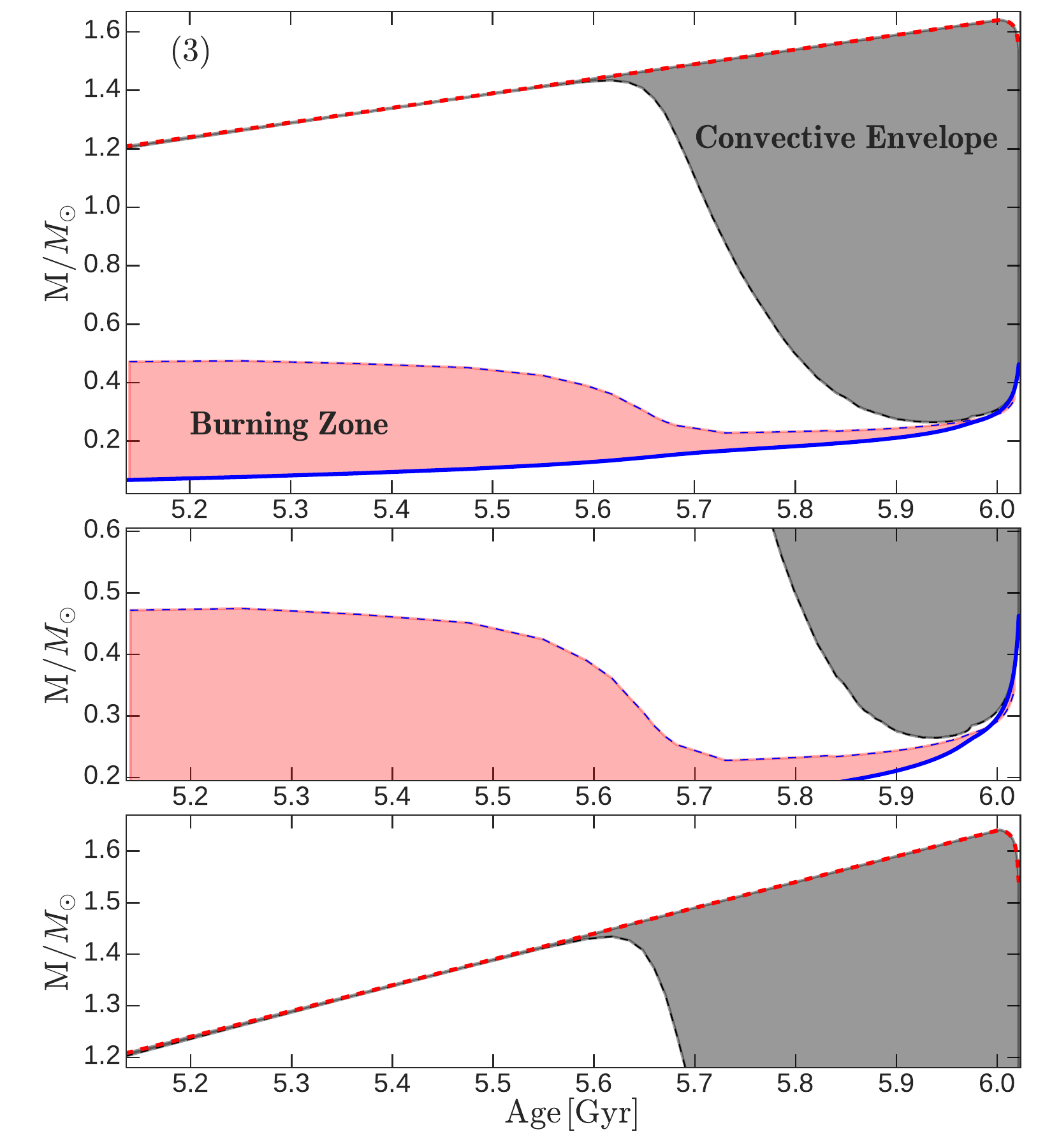}
	\includegraphics[width=8.2cm]{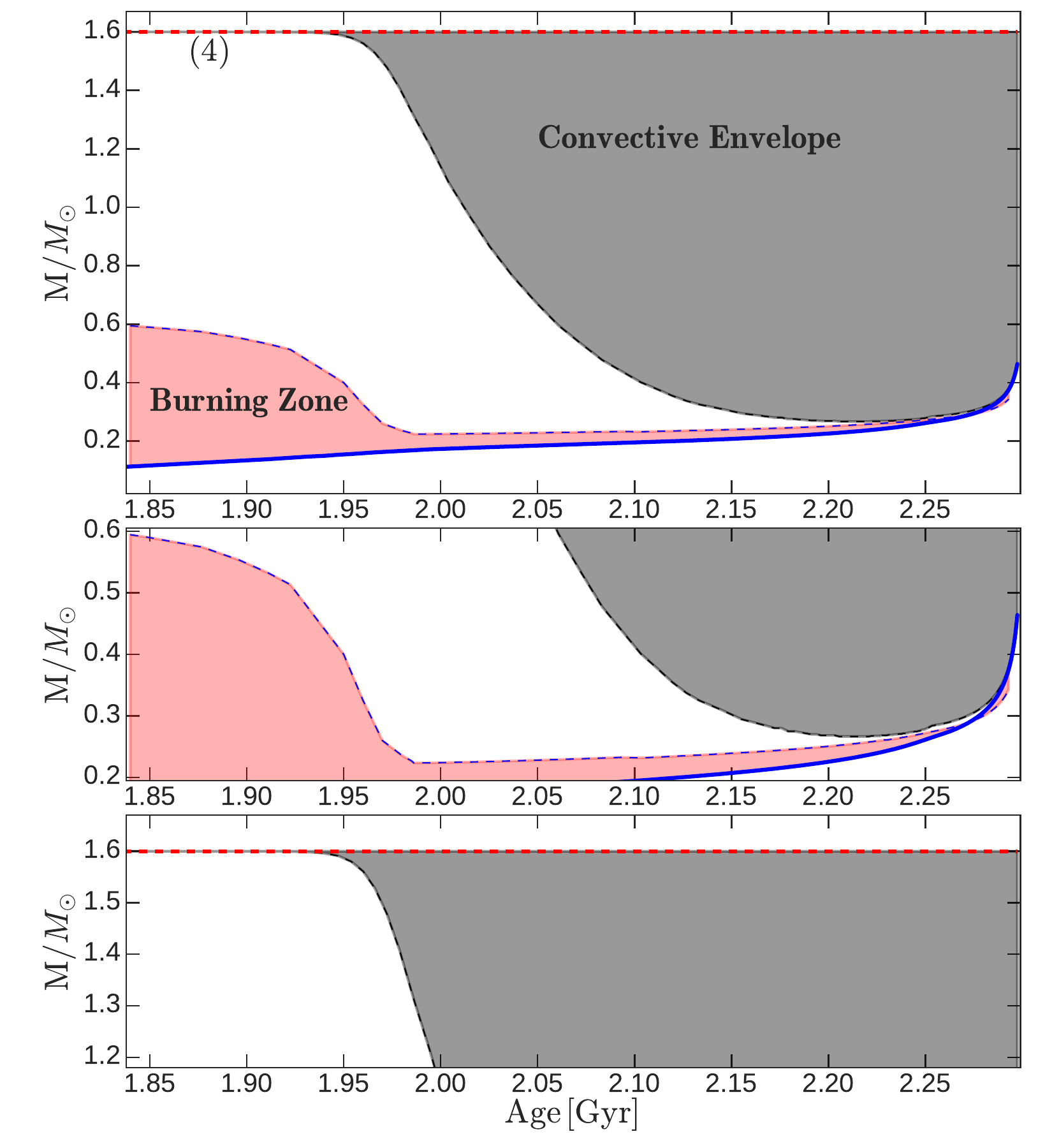}
	\caption{Kippenhahn diagram from the MS turnoff to the RGB tip. The red and black dashed lines are respectively the boundaries of the convective envelope, the blue dashed lines are the outer boundary of the burning zones, and the blue solid lines are the upper boundary of the He core. The initial masses of subfigures (1), (2), and (3) are all $1.2\,M_\odot$, but the accretion rates are $5\times10^{-11}$, 0, and $5\times10^{-10}\,M_\odot\,\rm yr^{-1}$, respectively. The initial mass of subfigure (4) is $1.6\,M_\odot$ and the accretion rate is 0. The two panels below each subfigure are local enlarged images.}
	\label{a2}%
\end{figure*}

$\rm Fig. \,\ref{a2}$ is the Kippenhahn diagram. By comparing (1) and (2), it can be found that at a lower accretion rate (with an accretion mass of $0.1\,M_\odot$), the accretion mass is basically integrated into the convective envelope. At a higher accretion rate (comparing (2) and (3)), Most of the accreted matter is also integrated into the internal region outside the burning zones. With the development of the convective zone, these areas that increase by accretion gradually become convective envelope.

\begin{figure*}[hbt]
	\centering
	\includegraphics[width=8cm]{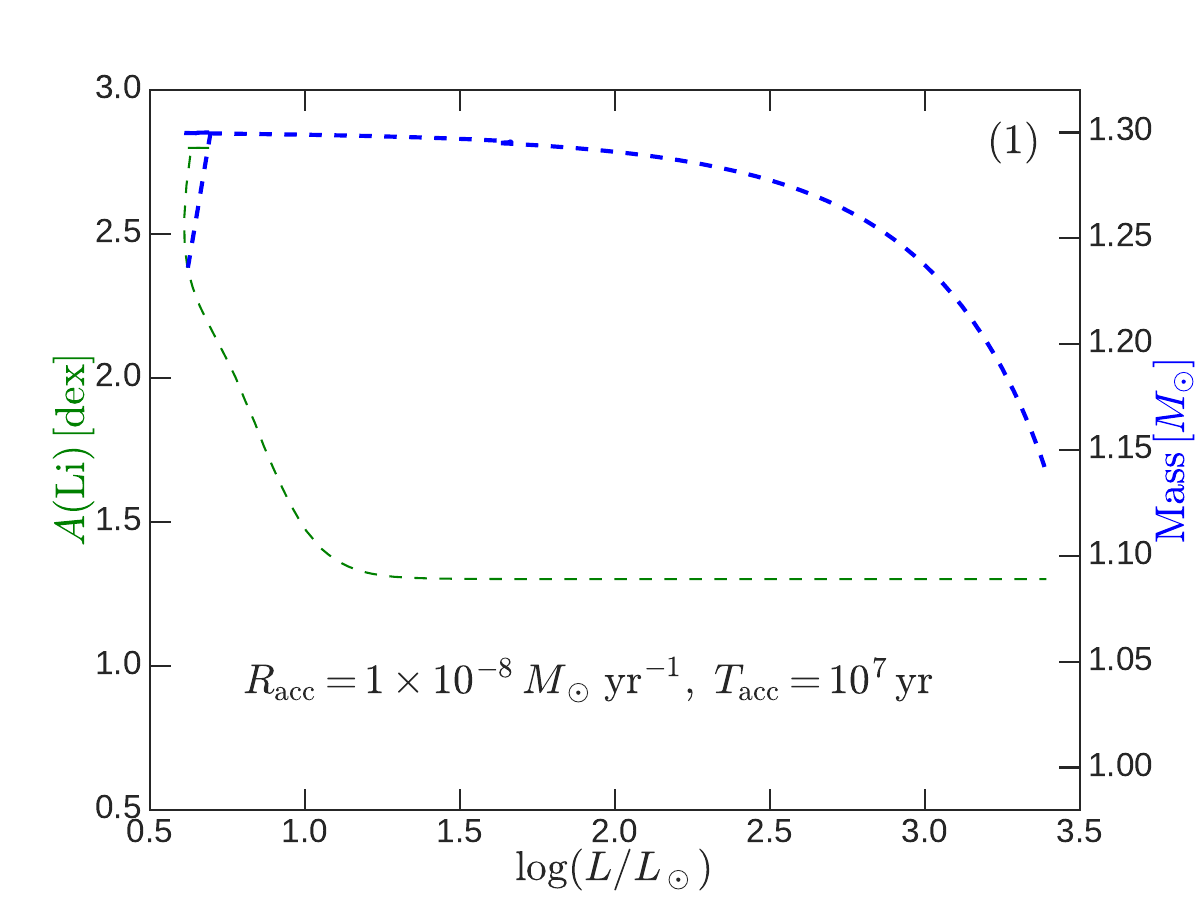}
	\includegraphics[width=8cm]{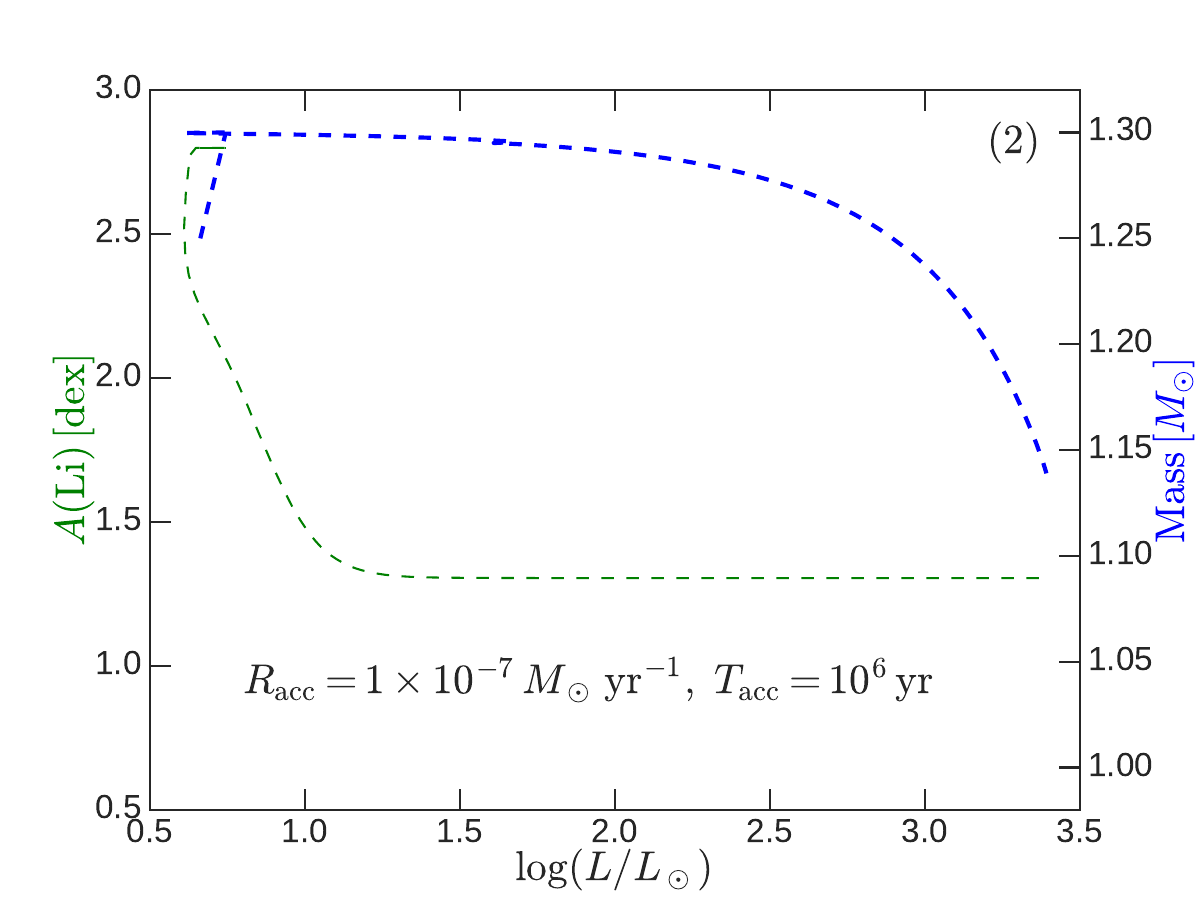}
	\includegraphics[width=8cm]{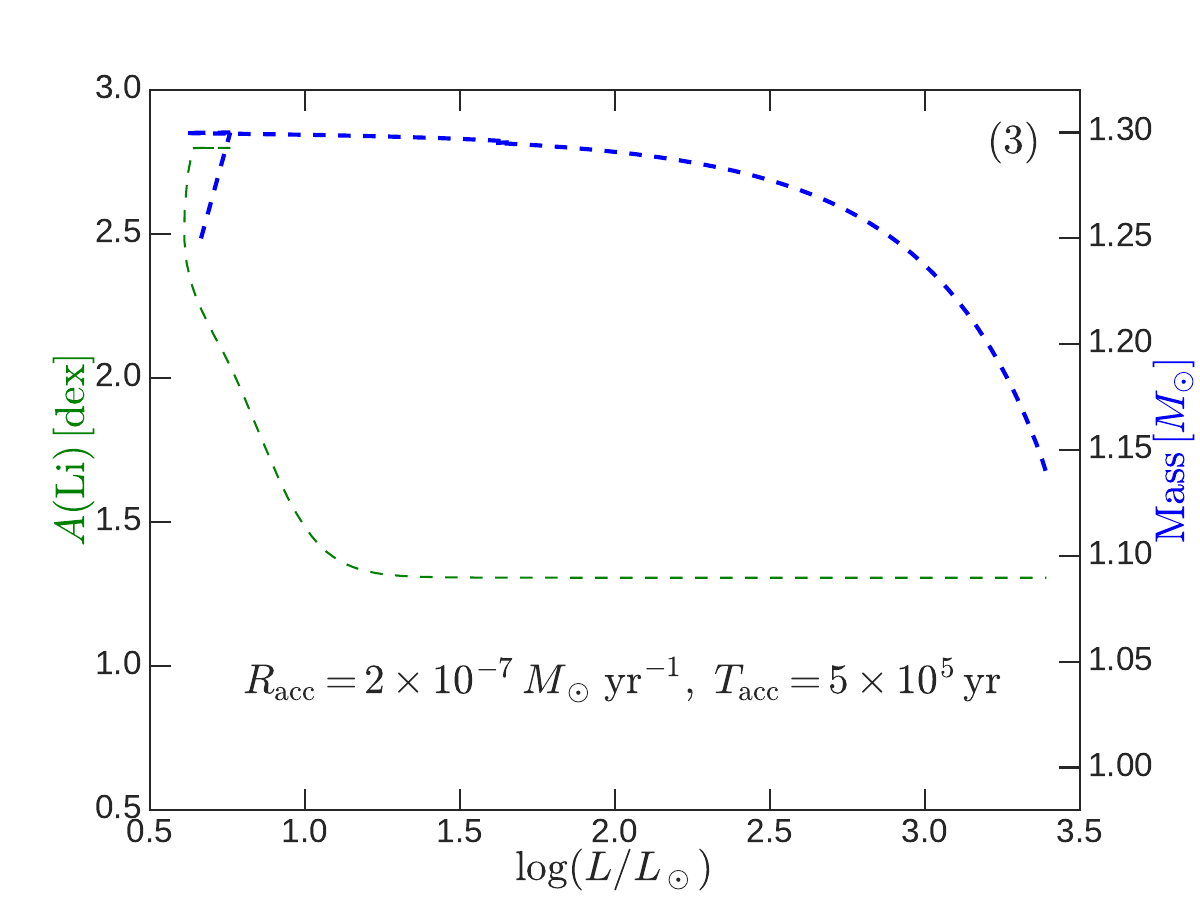}
	\includegraphics[width=8cm]{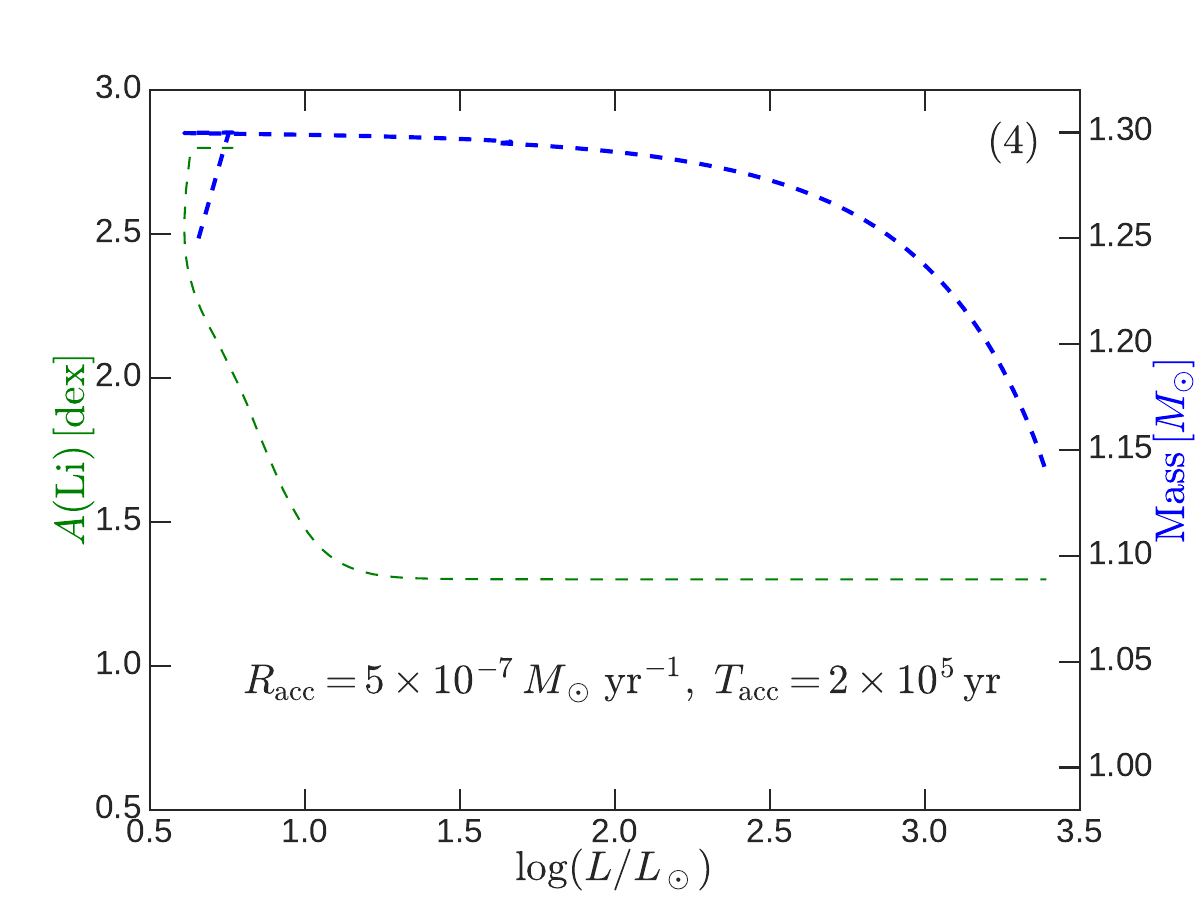}
	\caption{Similar to $\rm Fig.\,\ref{f1}$. Setting different accretion times to ensure that the accretion mass in the FDU stage is $0.1\,M_\odot$.}
	\label{a3}%
\end{figure*}

$\rm Fig. \,\ref{a3}$ simulates the evolution of Li abundance and mass with an accretion of $0.1\,M_\odot$ at different accretion times during the FDU. Given that the binary stars matter transfer is much smaller than the timescale of the FDU, in order to achieve effective accretion, we initiate the accretion process when the FDU is proceeding stably. From $X_c=10^{-9}$ to the beginning of accretion, the Li abundance decreased by approximately $0.6\,\rm dex$ due to dilution in the convection zone.

\begin{thebibliography}	

\bibitem[Aguilera-G{\'o}mez et al.(2016)]{2016ApJ...829..127A} Aguilera-G{\'o}mez, C., Chanam{\'e}, J., Pinsonneault, M.~H., et al.\ 2016, \apj, 829, 127. doi:10.3847/0004-637X/829/2/127

\bibitem[Andr{\'a}ssy \& Spruit(2015)]{2015A&A...579A.122A} Andr{\'a}ssy, R. \& Spruit, H.~C.\ 2015, \aap, 579, A122. doi:10.1051/0004-6361/201526029

\bibitem[Angulo et al.(1999)]{1999NuPhA.656....3A} Angulo, C., Arnould, M., Rayet, M., et al.\ 1999, \nphysa, 656, 1, 3. doi:10.1016/S0375-9474(99)00030-5

\bibitem[Asplund et al.(2009)]{2009ARA&A..47..481A} Asplund, M., Grevesse, N., Sauval, A.~J., et al.\ 2009, \araa, 47, 481. doi:10.1146/annurev.astro.46.060407.145222

\bibitem[Bahcall et al.(2001)]{2001ApJ...555..990B} Bahcall, J.~N., Pinsonneault, M.~H., \& Basu, S.\ 2001, \apj, 555, 990. doi:10.1086/321493

\bibitem[Balachandran et al.(2000)]{2000ApJ...542..978B} Balachandran, S.~C., Fekel, F.~C., Henry, G.~W., et al.\ 2000, \apj, 542, 978. doi:10.1086/317055

\bibitem[Baraffe et al.(2017)]{2017ApJ...845L...6B} Baraffe, I., Pratt, J., Goffrey, T., et al.\ 2017, \apjl, 845, L6. doi:10.3847/2041-8213/aa82ff

\bibitem[Baraffe \& Chabrier(2010)]{2010A&A...521A..44B} Baraffe, I. \& Chabrier, G.\ 2010, \aap, Effect of episodic accretion on the structure and the lithium depletion of low-mass stars and planet-hosting stars, 521, A44. doi:10.1051/0004-6361/201014979

\bibitem[Bastian \& Lardo(2018)]{2018ARA&A..56...83B} Bastian, N. \& Lardo, C.\ 2018, \araa, Multiple Stellar Populations in Globular Clusters, 56, 83. doi:10.1146/annurev-astro-081817-051839

\bibitem[Bharat Kumar et al.(2015)]{2015A&A...577A..10B} Bharat Kumar, Y., Reddy, B.~E., Muthumariappan, C., et al.\ 2015, \aap, 577, A10. doi:10.1051/0004-6361/201425076

\bibitem[Bloecker(1995)]{1995A&A...297..727B} Bloecker, T.\ 1995, \aap, 297, 727

\bibitem[Bozzetto et al.(2017)]{2017ApJS..230....2B} Bozzetto, L.~M., Filipovi{\'c}, M.~D., Vukoti{\'c}, B., et al.\ 2017, \apjs, 230, 2. doi:10.3847/1538-4365/aa653c

\bibitem[Brown et al.(1989)]{1989ApJS...71..293B} Brown, J.~A., Sneden, C., Lambert, D.~L., et al.\ 1989, \apjs, 71, 293. doi:10.1086/191375

\bibitem[Cai et al.(2023)]{2023AJ....165...52C} Cai, B., Kong, X., Shi, J., et al.\ 2023, \aj, 165, 52. doi:10.3847/1538-3881/aca098

\bibitem[Cao \& Pinsonneault(2025)]{2025arXiv250114867C} Cao, K. \& Pinsonneault, M.~H.\ 2025, , Modeling APOKASC-3 red giants: I. The first dredge-up and red giant branch bump, arXiv:2501.14867. doi:10.48550/arXiv.2501.14867

\bibitem[Carlberg et al.(2010)]{2010ApJ...723L.103C} Carlberg, J.~K., Smith, V.~V., Cunha, K., et al.\ 2010, \apjl, 723, L103. doi:10.1088/2041-8205/723/1/L103

\bibitem[Casey et al.(2019)]{2019ApJ...880..125C} Casey, A.~R., Ho, A.~Y.~Q., Ness, M., et al.\ 2019, \apj, 880, 125. doi:10.3847/1538-4357/ab27bf

\bibitem[Casey et al.(2016)]{2016MNRAS.461.3336C} Casey, A.~R., Ruchti, G., Masseron, T., et al.\ 2016, \mnras, 461, 3336. doi:10.1093/mnras/stw1512

\bibitem[Castilho et al.(2000)]{2000A&A...364..674C} Castilho, B.~V., Gregorio-Hetem, J., Spite, F., et al.\ 2000, \aap, 364, 674

\bibitem[Castro-Tapia et al.(2024)]{2024A&A...690A.367C} Castro-Tapia, M., Aguilera-G{\'o}mez, C., \& Chanam{\'e}, J.\ 2024, \aap, 690, A367. doi:10.1051/0004-6361/202349106

\bibitem[Caughlan \& Fowler(1988)]{1988ADNDT..40..283C} Caughlan, G.~R. \& Fowler, W.~A.\ 1988, Atomic Data and Nuclear Data Tables, 40, 283. doi:10.1016/0092-640X(88)90009-5

\bibitem[Chanam{\'e} et al.(2022)]{2022ApJ...933...58C} Chanam{\'e}, J., Pinsonneault, M.~H., Aguilera-G{\'o}mez, C., et al.\ 2022, \apj, 933, 58. doi:10.3847/1538-4357/ac70c8

\bibitem[Charbonnel et al.(1992)]{1992A&A...255..191C} Charbonnel, C., Vauclair, S., \& Zahn, J.-P.\ 1992, \aap, 255, 191

\bibitem[Charbonnel \& Balachandran(2000)]{2000A&A...359..563C} Charbonnel, C. \& Balachandran, S.~C.\ 2000, \aap, 359, 563. doi:10.48550/arXiv.astro-ph/0005280

\bibitem[Charbonnel \& Talon(2005)]{2005Sci...309.2189C} Charbonnel, C. \& Talon, S.\ 2005, Science, 309, 2189. doi:10.1126/science.1116849

\bibitem[Church et al.(2009)]{2009MNRAS.395.1127C} Church, R.~P., Dischler, J., Davies, M.~B., et al.\ 2009, \mnras, 395, 1127. doi:10.1111/j.1365-2966.2009.14619.x

\bibitem[Cox \& Giuli(1968)]{1968pss..book.....C} Cox, J.~P. \& Giuli, R.~T.\ 1968, Principles of stellar structure, by J.P. Cox and R. T. Giuli.  New York: Gordon and Breach, 1968

\bibitem[Crowther(2007)]{2007ARA&A..45..177C} Crowther, P.~A.\ 2007, \araa, 45, 177. doi:10.1146/annurev.astro.45.051806.110615

\bibitem[de La Reza et al.(1996)]{1996ApJ...456L.115D} de La Reza, R., Drake, N.~A., \& da Silva, L.\ 1996, \apjl, 456, L115. doi:10.1086/309874

\bibitem[de la Reza et al.(1997)]{1997ApJ...482L..77D} de la Reza, R., Drake, N.~A., da Silva, L., et al.\ 1997, \apjl, 482, L77. doi:10.1086/310685

\bibitem[de la Reza(2025)]{2025A&A...693A..98D} de la Reza, R.\ 2025, \aap, 693, A98. doi:10.1051/0004-6361/202451727

\bibitem[Deepak \& Reddy(2019)]{2019MNRAS.484.2000D} Deepak \& Reddy, B.~E.\ 2019, \mnras, 484, 2000. doi:10.1093/mnras/stz128

\bibitem[Deepak et al.(2020)]{2020MNRAS.494.1348D} Deepak, Lambert, D.~L., \& Reddy, B.~E.\ 2020, \mnras, 494, 1348. doi:10.1093/mnras/staa729

\bibitem[Delgado Mena et al.(2016)]{2016A&A...587A..66D} Delgado Mena, E., Tsantaki, M., Sousa, S.~G., et al.\ 2016, \aap, 587, A66. doi:10.1051/0004-6361/201527196

\bibitem[Denissenkov \& Weiss(2000)]{2000A&A...358L..49D} Denissenkov, P.~A. \& Weiss, A.\ 2000, \aap, 358, L49. doi:10.48550/arXiv.astro-ph/0005356

\bibitem[Denissenkov \& VandenBerg(2003)]{2003ApJ...593..509D} Denissenkov, P.~A. \& VandenBerg, D.~A.\ 2003, \apj, 593, 509. doi:10.1086/376410

\bibitem[Denissenkov et al.(2024)]{2024MNRAS.535.1243D} Denissenkov, P.~A., Blouin, S., Herwig, F., et al.\ 2024, \mnras, 535, 1243. doi:10.1093/mnras/stae2407

\bibitem[Ding et al.(2024)]{2024ApJS..271...58D} Ding, M.-Y., Shi, J.-R., Yan, H.-. liang ., et al.\ 2024, \apjs, 271, 58. doi:10.3847/1538-4365/ad2f28

\bibitem[Drake et al.(2002)]{2002AJ....123.2703D} Drake, N.~A., de la Reza, R., da Silva, L., et al.\ 2002, \aj, 123, 2703. doi:10.1086/339968

\bibitem[Eggenberger et al.(2012)]{2012A&A...539A..70E} Eggenberger, P., Haemmerl{\'e}, L., Meynet, G., et al.\ 2012, \aap, Impact of rotation and disc lifetime on pre-main sequence lithium depletion of solar-type stars, 539, A70. doi:10.1051/0004-6361/201118432

\bibitem[Endal \& Sofia(1981)]{1981ApJ...243..625E} Endal, A.~S. \& Sofia, S.\ 1981, \apj, 243, 625. doi:10.1086/158628

\bibitem[Fekel \& Watson(1998)]{1998AJ....116.2466F} Fekel, F.~C. \& Watson, L.~C.\ 1998, \aj, 116, 2466. doi:10.1086/300614

\bibitem[Ferlet \& Dennefeld(1984)]{1984A&A...138..303F} Ferlet, R. \& Dennefeld, M.\ 1984, \aap, 138, 303

\bibitem[Fu et al.(2018)]{2018A&A...610A..38F} Fu, X., Romano, D., Bragaglia, A., et al.\ 2018, \aap, 610, A38. doi:10.1051/0004-6361/201731677

\bibitem[Fullerton et al.(2006)]{2006ApJ...637.1025F} Fullerton, A.~W., Massa, D.~L., \& Prinja, R.~K.\ 2006, \apj, 637, 1025. doi:10.1086/498560

\bibitem[Gao et al.(2019)]{2019ApJS..245...33G} Gao, Q., Shi, J.-R., Yan, H.-L., et al.\ 2019, \apjs, 245, 33. doi:10.3847/1538-4365/ab505c

\bibitem[Gao et al.(2022)]{2022A&A...668A.126G} Gao, J., Zhu, C., Yu, J., et al.\ 2022, \aap, 668, A126. doi:10.1051/0004-6361/202243871

\bibitem[Garcia Lopez \& Spruit(1991)]{1991ApJ...377..268G} Garcia Lopez, R.~J. \& Spruit, H.~C.\ 1991, \apj, 377, 268. doi:10.1086/170356

\bibitem[Gonzalez et al.(2009)]{2009A&A...508..289G} Gonzalez, O.~A., Zoccali, M., Monaco, L., et al.\ 2009, \aap, 508, 289. doi:10.1051/0004-6361/200912469

\bibitem[Grevesse \& Sauval(1998)]{1998SSRv...85..161G} Grevesse, N. \& Sauval, A.~J.\ 1998, \ssr, 85, 161. doi:10.1023/A:1005161325181

\bibitem[Hamuy(2003)]{2003ApJ...582..905H} Hamuy, M.\ 2003, \apj, 582, 905. doi:10.1086/344689

\bibitem[Hurley et al.(2000)]{2000MNRAS.315..543H} Hurley, J.~R., Pols, O.~R., \& Tout, C.~A.\ 2000, \mnras, 315, 543. doi:10.1046/j.1365-8711.2000.03426.x

\bibitem[H{\"o}fner \& Olofsson(2018)]{2018A&ARv..26....1H} H{\"o}fner, S. \& Olofsson, H.\ 2018, \aapr, 26, 1. doi:10.1007/s00159-017-0106-5

\bibitem[Iben(1965)]{1965ApJ...141..993I} Iben, I.\ 1965, \apj, 141, 993. doi:10.1086/148193

\bibitem[Iben(1967)]{1967ApJ...147..624I} Iben, I.\ 1967, \apj, 147, 624. doi:10.1086/149040

\bibitem[Iglesias \& Rogers(1993)]{1993ApJ...412..752I} Iglesias, C.~A. \& Rogers, F.~J.\ 1993, \apj, 412, 752. doi:10.1086/172958

\bibitem[Izzo et al.(2015)]{2015ApJ...808L..14I} Izzo, L., Della Valle, M., Mason, E., et al.\ 2015, \apjl, 808, L14. doi:10.1088/2041-8205/808/1/L14

\bibitem[Iglesias \& Rogers(1996)]{1996ApJ...464..943I} Iglesias, C.~A. \& Rogers, F.~J.\ 1996, \apj, 464, 943. doi:10.1086/177381

\bibitem[Jackson \& Jeffries(2014)]{2014MNRAS.445.4306J} Jackson, R.~J. \& Jeffries, R.~D.\ 2014, \mnras, The effect of star-spots on the ages of low-mass stars determined from the lithium depletion boundary, 445, 4, 4306. doi:10.1093/mnras/stu2076

\bibitem[James et al.(2006)]{2006A&A...446..971J} James, D.~J., Melo, C., Santos, N.~C., et al.\ 2006, \aap, 446, 971. doi:10.1051/0004-6361:20053900

\bibitem[Jasniewicz et al.(1999)]{1999A&A...342..831J} Jasniewicz, G., Parthasarathy, M., de Laverny, P., et al.\ 1999, \aap, 342, 831

\bibitem[Jura(2003)]{2003ApJ...582.1032J} Jura, M.\ 2003, \apj, 582, 1032. doi:10.1086/344704

\bibitem[Kirby et al.(2016)]{2016ApJ...819..135K} Kirby, E.~N., Guhathakurta, P., Zhang, A.~J., et al.\ 2016, \apj, 819, 135. doi:10.3847/0004-637X/819/2/135

\bibitem[Knauth et al.(2003)]{2003ApJ...586..268K} Knauth, D.~C., Federman, S.~R., \& Lambert, D.~L.\ 2003, \apj, 586, 268. doi:10.1086/346264

\bibitem[Kowkabany et al.(2024)]{2024ApJ...973..125K} Kowkabany, J., Ezzeddine, R., Charbonnel, C., et al.\ 2024, \apj, Discovery of a Metal-poor Red Giant Star with the Highest Ultralithium Enhancement, 973, 2, 125. doi:10.3847/1538-4357/ad6004

\bibitem[Kumar et al.(2011)]{2011ApJ...730L..12K} Kumar, Y.~B., Reddy, B.~E., \& Lambert, D.~L.\ 2011, \apjl, 730, L12. doi:10.1088/2041-8205/730/1/L12

\bibitem[Kumar et al.(2020)]{2020NatAs...4.1059K} Kumar, Y.~B., Reddy, B.~E., Campbell, S.~W., et al.\ 2020, Nature Astronomy, 4, 1059. doi:10.1038/s41550-020-1139-7

\bibitem[Li et al.(2016)]{2016Natur.529..502L} Li, C., de Grijs, R., Deng, L., et al.\ 2016, \nat, 529, 502. doi:10.1038/nature16493

\bibitem[Li et al.(2023)]{2023ApJ...943..115L} Li, X.-F., Shi, J.-R., Li, Y., et al.\ 2023, \apj, 943, 115. doi:10.3847/1538-4357/acae9d

\bibitem[Li et al.(2024)]{2024MNRAS.529.1423L} Li, X.-F., Shi, J.-R., Li, Y., et al.\ 2024, \mnras, 529, 1423. doi:10.1093/mnras/stae639

\bibitem[Li et al.(2025)]{2025ApJ...982L...4L} Li, X.-F., Shi, J.-R., Li, Y., et al.\ 2025, \apjl, 982, L4. doi:10.3847/2041-8213/adb833

\bibitem[Li(2025)]{2025arXiv250512794L} Li, Y.\ 2025, , arXiv:2505.12794. doi:10.48550/arXiv.2505.12794

\bibitem[Lochhaas \& Thompson(2017)]{2017MNRAS.470..977L} Lochhaas, C. \& Thompson, T.~A.\ 2017, \mnras, Second-generation stars in globular clusters from rapid radiative cooling of pre-supernova massive star winds, 470, 1, 977. doi:10.1093/mnras/stx1289

\bibitem[Magrini et al.(2021)]{2021A&A...651A..84M} Magrini, L., Lagarde, N., Charbonnel, C., et al.\ 2021, \aap, 651, A84. doi:10.1051/0004-6361/202140935

\bibitem[Mallick et al.(2022)]{2022MNRAS.511.3741M} Mallick, A., Reddy, B.~E., \& Muthumariappan, C.\ 2022, \mnras, 511, 3741. doi:10.1093/mnras/stac224

\bibitem[Martell \& Shetrone(2013)]{2013MNRAS.430..611M} Martell, S.~L. \& Shetrone, M.~D.\ 2013, \mnras, 430, 611

\bibitem[Martell et al.(2021)]{2021MNRAS.505.5340M} Martell, S.~L., Simpson, J.~D., Balasubramaniam, A.~G., et al.\ 2021, \mnras, 505, 5340. doi:10.1093/mnras/stab1356

\bibitem[Moffat et al.(1988)]{1988ApJ...334.1038M} Moffat, A.~F.~J., Drissen, L., Lamontagne, R., et al.\ 1988, \apj, 334, 1038. doi:10.1086/166895

\bibitem[Monaco et al.(2011)]{2011A&A...529A..90M} Monaco, L., Villanova, S., Moni Bidin, C., et al.\ 2011, \aap, 529, A90. doi:10.1051/0004-6361/201016285

\bibitem[Mori et al.(2021)]{2021MNRAS.503.2746M} Mori, K., Kusakabe, M., Balantekin, A.~B., et al.\ 2021, \mnras, 503, 2746. doi:10.1093/mnras/stab595

\bibitem[Owocki et al.(1988)]{1988ApJ...335..914O} Owocki, S.~P., Castor, J.~I., \& Rybicki, G.~B.\ 1988, \apj, 335, 914. doi:10.1086/166977

\bibitem[Osterbart et al.(2000)]{2000A&A...357..169O} Osterbart, R., Balega, Y.~Y., Bl{\"o}cker, T., et al.\ 2000, \aap, 357, 169. doi:10.48550/arXiv.astro-ph/0003328

\bibitem[Pan et al.(2010)]{2010ApJ...715...78P} Pan, K.-C., Ricker, P.~M., \& Taam, R.~E.\ 2010, \apj, 715, 78. doi:10.1088/0004-637X/715/1/78

\bibitem[Pan et al.(2012)]{2012ApJ...750..151P} Pan, K.-C., Ricker, P.~M., \& Taam, R.~E.\ 2012, \apj, 750, 151. doi:10.1088/0004-637X/750/2/151

\bibitem[Patterson(1984)]{1984ApJS...54..443P} Patterson, J.\ 1984, \apjs, 54, 443. doi:10.1086/190940

\bibitem[Paxton et al.(2011)]{2011ApJS..192....3P} Paxton, B., Bildsten, L., Dotter, A., et al.\ 2011, \apjs, 192, 3. doi:10.1088/0067-0049/192/1/3

\bibitem[Paxton et al.(2013)]{2013ApJS..208....4P} Paxton, B., Cantiello, M., Arras, P., et al.\ 2013, \apjs, 208, 4. doi:10.1088/0067-0049/208/1/4

\bibitem[Paxton et al.(2015)]{2015ApJS..220...15P} Paxton, B., Marchant, P., Schwab, J., et al.\ 2015, \apjs, 220, 15. doi:10.1088/0067-0049/220/1/15

\bibitem[Paxton et al.(2018)]{2018ApJS..234...34P} Paxton, B., Schwab, J., Bauer, E.~B., et al.\ 2018, \apjs, 234, 34. doi:10.3847/1538-4365/aaa5a8

\bibitem[Paxton et al.(2019)]{2019ApJS..243...10P} Paxton, B., Smolec, R., Schwab, J., et al.\ 2019, \apjs, 243, 10. doi:10.3847/1538-4365/ab2241

\bibitem[Pereyra et al.(2006)]{2006A&A...449..211P} Pereyra, A., Castilho, B.~V., \& Magalh{\~a}es, A.~M.\ 2006, \aap, 449, 211. doi:10.1051/0004-6361:20054270		

\bibitem[Pinsonneault(1997)]{1997ARA&A..35..557P} Pinsonneault, M.\ 1997, \araa, 35, 557. doi:10.1146/annurev.astro.35.1.557

\bibitem[Pols et al.(1998)]{1998MNRAS.298..525P} Pols, O.~R., Schr{\"o}der, K.-P., Hurley, J.~R., et al.\ 1998, \mnras, 298, 525. doi:10.1046/j.1365-8711.1998.01658.x

\bibitem[Ramstedt et al.(2009)]{2009A&A...499..515R} Ramstedt, S., Sch{\"o}ier, F.~L., \& Olofsson, H.\ 2009, \aap, 499, 515. doi:10.1051/0004-6361/200911730

\bibitem[Rebull et al.(2015)]{2015AJ....150..123R} Rebull, L.~M., Carlberg, J.~K., Gibbs, J.~C., et al.\ 2015, \aj, 150, 123. doi:10.1088/0004-6256/150/4/123

\bibitem[Reddy et al.(2002)]{2002AJ....123.1993R} Reddy, B.~E., Lambert, D.~L., Hrivnak, B.~J., et al.\ 2002, \aj, 123, 1993. doi:10.1086/339310

\bibitem[Reddy \& Lambert(2005)]{2005AJ....129.2831R} Reddy, B.~E. \& Lambert, D.~L.\ 2005, \aj, 129, 2831. doi:10.1086/430190

\bibitem[Reddy \& Lambert(2016)]{2016A&A...589A..57R} Reddy, A.~B.~S. \& Lambert, D.~L.\ 2016, \aap, 589, A57. doi:10.1051/0004-6361/201628323

\bibitem[Reimers(1975)]{1975MSRSL...8..369R} Reimers, D.\ 1975, Memoires of the Societe Royale des Sciences de Liege, 8, 369

\bibitem[Renzo \& G{\"o}tberg(2021)]{2021ApJ...923..277R} Renzo, M. \& G{\"o}tberg, Y.\ 2021, \apj, Evolution of Accretor Stars in Massive Binaries: Broader Implications from Modeling {\ensuremath{\zeta}} Ophiuchi, 923, 2, 277. doi:10.3847/1538-4357/ac29c5

\bibitem[Ritter(1988)]{1988A&A...202...93R} Ritter, H.\ 1988, \aap, 202, 93

\bibitem[Rogers \& Nayfonov(2002)]{2002ApJ...576.1064R} Rogers, F.~J. \& Nayfonov, A.\ 2002, \apj, 576, 1064. doi:10.1086/341894

\bibitem[Romano et al.(1999)]{1999A&A...352..117R} Romano, D., Matteucci, F., Molaro, P., et al.\ 1999, \aap, 352, 117. doi:10.48550/arXiv.astro-ph/9910151

\bibitem[Ruchti et al.(2011)]{2011ApJ...743..107R} Ruchti, G.~R., Fulbright, J.~P., Wyse, R.~F.~G., et al.\ 2011, \apj, 743, 107. doi:10.1088/0004-637X/743/2/107

\bibitem[Ruiz-Lapuente et al.(2004)]{2004Natur.431.1069R} Ruiz-Lapuente, P., Comeron, F., M{\'e}ndez, J., et al.\ 2004, \nat, 431, 1069. doi:10.1038/nature03006

\bibitem[Sayeed et al.(2024)]{2024ApJ...964...42S} Sayeed, M., Ness, M.~K., Montet, B.~T., et al.\ 2024, \apj, 964, 42. doi:10.3847/1538-4357/ad1936

\bibitem[Schwab(2020)]{2020ApJ...901L..18S} Schwab, J.\ 2020, \apjl, 901, L18. doi:10.3847/2041-8213/abb45f

\bibitem[Schr{\"o}der \& Cuntz(2005)]{2005ApJ...630L..73S} Schr{\"o}der, K.-P. \& Cuntz, M.\ 2005, \apjl, 630, L73. doi:10.1086/491579

\bibitem[Siess \& Livio(1999)]{1999MNRAS.308.1133S} Siess, L. \& Livio, M.\ 1999, \mnras, 308, 1133. doi:10.1046/j.1365-8711.1999.02784.x

\bibitem[Smith \& Lambert(1990)]{1990ApJ...361L..69S} Smith, V.~V. \& Lambert, D.~L.\ 1990, \apjl, 361, L69. doi:10.1086/185829

\bibitem[Smith et al.(1995)]{1995ApJ...441..735S} Smith, V.~V., Plez, B., Lambert, D.~L., et al.\ 1995, \apj, 441, 735. doi:10.1086/175395

\bibitem[Smith(2014)]{2014ARA&A..52..487S} Smith, N.\ 2014, \araa, 52, 487. doi:10.1146/annurev-astro-081913-040025

\bibitem[Somers \& Pinsonneault(2015)]{2015ApJ...807..174S} Somers, G. \& Pinsonneault, M.~H.\ 2015, \apj, Older and Colder: The Impact of Starspots on Pre-main-sequence Stellar Evolution, 807, 2, 174. doi:10.1088/0004-637X/807/2/174

\bibitem[Somers \& Pinsonneault(2016)]{2016ApJ...829...32S} Somers, G. \& Pinsonneault, M.~H.\ 2016, \apj, 829, 32. doi:10.3847/0004-637X/829/1/32

\bibitem[Spite \& Spite(1993)]{1993A&A...279L...9S} Spite, F. \& Spite, M.\ 1993, \aap, 279, L9

\bibitem[Strassmeier et al.(2015)]{2015A&A...574A..31S} Strassmeier, K.~G., Carroll, T.~A., Weber, M., et al.\ 2015, \aap, 574, A31. doi:10.1051/0004-6361/201424130

\bibitem[Sun \& Mathieu(2023)]{2023ApJ...944...89S} Sun, M. \& Mathieu, R.~D.\ 2023, \apj, WOCS 4540: Detailed Analysis of a very Long Orbital Period Blue Straggler, 944, 1, 89. doi:10.3847/1538-4357/acacf7

\bibitem[Sundqvist \& Owocki(2013)]{2013MNRAS.428.1837S} Sundqvist, J.~O. \& Owocki, S.~P.\ 2013, \mnras, 428, 1837. doi:10.1093/mnras/sts165

\bibitem[Tayar et al.(2023)]{2023AJ....166...60T} Tayar, J., Carlberg, J.~K., Aguilera-G{\'o}mez, C., et al.\ 2023, \aj, 166, 60. doi:10.3847/1538-3881/ace25d

\bibitem[Turcotte et al.(1998)]{1998ApJ...504..559T} Turcotte, S., Richer, J., \& Michaud, G.\ 1998, \apj, 504, 559. doi:10.1086/306056

\bibitem[Villaver \& Livio(2009)]{2009ApJ...705L..81V} Villaver, E. \& Livio, M.\ 2009, \apjl, 705, L81. doi:10.1088/0004-637X/705/1/L81

\bibitem[Wallerstein \& Sneden(1982)]{1982ApJ...255..577W} Wallerstein, G. \& Sneden, C.\ 1982, \apj, A K giant with an unusually high abundance of lithium : HD 112127., 255, 577. doi:10.1086/159859

\bibitem[Weigelt et al.(1998)]{1998A&A...333L..51W} Weigelt, G., Balega, Y., Bloecker, T., et al.\ 1998, \aap, 333, L51. doi:10.48550/arXiv.astro-ph/9805022

\bibitem[Willson(2000)]{2000ARA&A..38..573W} Willson, L.~A.\ 2000, \araa, 38, 573. doi:10.1146/annurev.astro.38.1.573

\bibitem[Woosley \& Weaver(1995)]{1995ApJS..101..181W} Woosley, S.~E. \& Weaver, T.~A.\ 1995, \apjs, 101, 181. doi:10.1086/192237

\bibitem[Yan et al.(2021)]{2021NatAs...5...86Y} Yan, H.-L., Zhou, Y.-T., Zhang, X., et al.\ 2021, Nature Astronomy, 5, 86. doi:10.1038/s41550-020-01217-8

\bibitem[Yan \& Shi(2022)]{2022AcASn..63....2Y} Yan, H.~L. \& Shi, J.~R.\ 2022, Acta Astronomica Sinica, 63, 2

\bibitem[Zhang et al.(2020)]{2020ApJ...889...33Z} Zhang, X., Jeffery, C.~S., Li, Y., et al.\ 2020, \apj, 889, 33. doi:10.3847/1538-4357/ab5e89

\bibitem[Zhang et al.(2021)]{2021ApJ...919L...3Z} Zhang, J., Shi, J.-R., Yan, H.-L., et al.\ 2021, \apjl, 919, L3. doi:10.3847/2041-8213/ac224c

\bibitem[Zhou et al.(2019)]{2019ApJ...877..104Z} Zhou, Y., Yan, H., Shi, J., et al.\ 2019, \apj, 877, 104. doi:10.3847/1538-4357/ab1b4b

\bibitem[Zhou et al.(2022)]{2022ApJ...931..136Z} Zhou, Y., Wang, C., Yan, H., et al.\ 2022, \apj, 931, 136. doi:10.3847/1538-4357/ac6b3a
\end{thebibliography}
\end{document}